\documentclass[preprint,preprintnumbers,nofootinbib,amssymb,aps]{revtex4}

\pdfoutput=1

\usepackage{hyperref}
\usepackage{floatrow}
\usepackage{graphicx}
\usepackage{dcolumn}
\usepackage{amsmath}
\usepackage{epsfig}
\usepackage{subfig}
\usepackage{array}
\usepackage{epstopdf} 
\usepackage{xcolor}
\usepackage{bm}

\def\be{\begin{equation}}
\def\ee{\end{equation}}
\def\bea{\begin{eqnarray}}
\def\eea{\end{eqnarray}}
\def\NO{\nonumber}
\def\gev{\mathrm{~GeV}}

\def\md{\mathrm{d}}

\begin{document}

\normalsize

\title{QCD leading order study of the $J/\psi$ leptoproduction at HERA within the nonrelativistic QCD framework}

\author{Zhan Sun}
\affiliation{School of Science, Guizhou Minzu University, Guiyang 550025, P. R. China.}
\author{Hong-Fei Zhang}
\email{hfzhang@ihep.ac.cn (corresponding author)}
\affiliation{Department of Physics, School of Biomedical Engineering, Third Military Medical University, Chongqing 400038, China.}

\begin{abstract}
As indicated in our previous paper~\cite{Zhang:2017dia},
the existing literatures studying the $J/\psi$ production in deeply inelastic scattering (DIS) in collinear factorization are on the basis of a formalism
that will lead to wrong results when the ranges of the transverse momentum or the rapidity
of the $J/\psi$ in the laboratory frame do not cover all values possible for them.
In this paper, we present the renewed results for the $J/\psi$ production in DIS at HERA within the nonrelativistic QCD framework at QCD leading order (LO).
Three different sets of the long-distance matrix elements are employed for comparison.
The predictions via the colour-singlet (CS) model at QCD LO are generally below the experimental data
especially in the regions where perturbation theory are expected to work well,
while the colour-octet contributions are of the same order of magnitude as the CS ones, however,
in general make the agreement between theory and experiment better.
\end{abstract}

\maketitle

\section{Introduction}

Since the nonrelativistic QCD (NRQCD) framework was proposed to solve the $J/\psi$ and $\psi(2S)$ surplus puzzle in 1994~\cite{Bodwin:1994jh},
the $J/\psi$ production in various processes has been studied within this framework.
For some of these processes, QCD next-to-leading order (NLO) results have also been achieved,
including the $J/\psi$ production in $e^+e^-$ annihilation~\cite{Zhang:2005cha, Zhang:2006ay, Gong:2007db, Gong:2008ce, Ma:2008gq, Gong:2009kp, Gong:2009ng, Zhang:2009ym, Wang:2011qg, Feng:2017bdu},
the $J/\psi$ photoproduction in $e^+e^-$~\cite{Klasen:2004tz} and
$ep$~\cite{Kramer:1995nb, Maltoni:1997pt, Artoisenet:2009xh, Chang:2009uj, Li:2009fd, Butenschoen:2009zy, Butenschoen:2011ks, Bodwin:2015yma} collisions,
and the $J/\psi$ hadroproduction~\cite{Campbell:2007ws, Gong:2008sn, Gong:2008hk, Gong:2008ft, Ma:2010yw, Butenschoen:2010rq, Ma:2010jj, Butenschoen:2011yh,
Butenschoen:2012px, Chao:2012iv, Gong:2012ug, Gong:2012ah, Lansberg:2013qka, Li:2014ava, Bodwin:2014gia, Lansberg:2014swa, Shao:2014yta, Bodwin:2015iua}.
The production of some other species of charmonia,
such as $\eta_c$~\cite{Butenschoen:2014dra, Han:2014jya, Zhang:2014ybe, Gong:2016jiq} and $\chi_c$~\cite{Ma:2010vd, Li:2011yc, Shao:2012fs, Shao:2014fca, Jia:2014jfa},
was also investigated at QCD NLO level,
which provided an alternative phenomenological test of NRQCD.
Among the $J/\psi$ production processes for which experimental data are available,
deeply inelastic scattering (DIS) is an interesting yet complicated one.
Abundant data have been released by the H1~\cite{Adloff:1999zs, Adloff:2002ey, Aaron:2010gz} and ZEUS~\cite{Chekanov:2005cf} Collaborations,
however, only the QCD leading order (LO) results have been given~\cite{Baier:1981zz, Korner:1982fm, Guillet:1987xr, Merabet:1994sm, Fleming:1997fq, Yuan:2000cn, Kniehl:2001tk}.
Even at QCD LO, the existing phenomenological results cannot coincide with each other (see e.g.~\cite{Brambilla:2004wf}).
However, the $J/\psi$ production in deeply inelastic scattering is an excellent laboratory for the study of the $J/\psi$ production mechanism.
For one thing, perturbation theory works better and the resolved photon contributions are less important in large $Q^2$ region,
relative to the $J/\psi$ photoproduction.
For another, multiple distributions are measured, which can provide reference to distinguish different models.

In Reference~\cite{Zhang:2017dia} we pointed out that most of the existing calculations on the $J/\psi$ production in DIS
are based on a formalism that is not valid when the ranges of the physical observables,
such as the transverse momentum ($p_t$) or the rapidity $y_\psi$ of the $J/\psi$,
in the laboratory frame do not cover all their possible values.
Actually, In early 1980s, the azimuthal dependence of the $J/\psi$ production in DIS has already been studied
in both the colour-evaporation~\cite{Duke:1980sq} and the colour-singlet (CS)~\cite{Korner:1982fm} models.
Unfortunately, some of the succeeding investigations were not aware of this, and abused the azimuthal symmetry under an unsuitable circumstance.
Besides, after the most recent phenomenological studies, three independent measurements were published.
All these facts suggest that an up-to-date theoretical study of the $J/\psi$ production in DIS in the NRQCD framework is needed.

This paper is devoted to the phenomenological investigation of the $J/\psi$ production in DIS within the NRQCD framework at QCD LO.
By exploiting the equations presented in Reference~\cite{Zhang:2017dia},
we obtain the comprehensive analytical formalism for the calculation of the $J/\psi$ production in DIS,
which is also valid for calculating the $p_t^2$ and $y_\psi$ distributions of the $J/\psi$ production cross sections.
One of the crucial issue of the NRQCD research is the determination of the long-distance matrix elements (LDMEs),
of which there are several independent extractions, with different strategy, having obtained different results.
In order to study the aftermath of the LDME uncertainties,
three typical sets of these parameters are employed to present our numerical results.
The rest of this paper is organised as follows.
In Section \ref{sec:anaframe}, we briefly describe the analytic formalism of our computation,
following which the numerical results are given in Section \ref{sec:numresults}.
At the end of this paper is a concluding remark in Section \ref{sec:summary}.

\section{Analytical Framework\label{sec:anaframe}}

\subsection{General Formalism\label{subsec:genform}}

\begin{figure}
\includegraphics[scale=0.5]{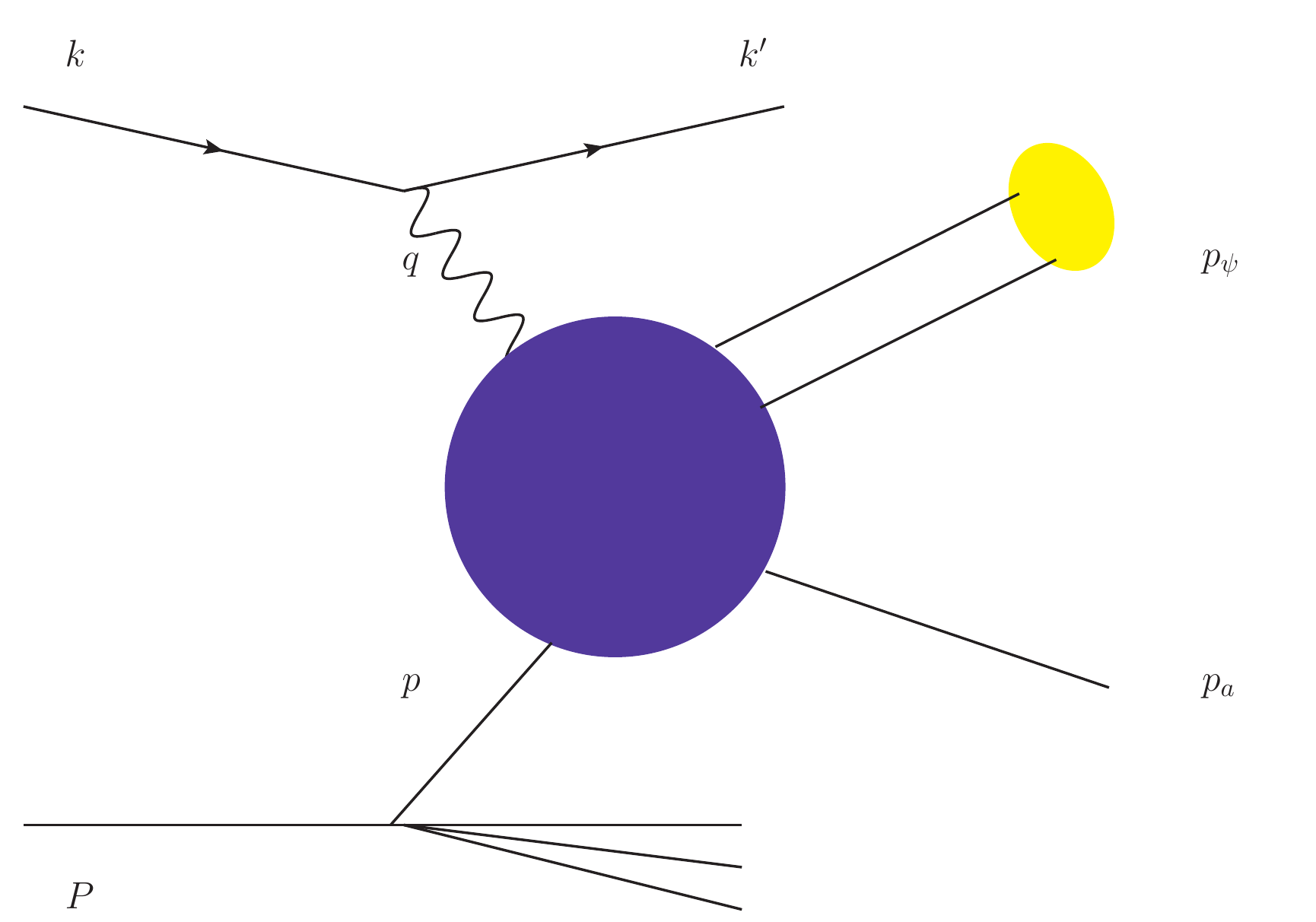}
\caption{\label{fig:diag}
The illustrative diagram for the $J/\psi$ production in DIS.
}
\end{figure}

In electron-proton ($ep$) DIS, the kinematics of the scattered lepton can be described in terms of any two of the following variables,
\bea
&&Q^2=-q^2\equiv-(k-k')^2,~~~~W^2=(P+q)^2, \NO \\
&&x_B=\frac{Q^2}{2P\cdot q},~~~~y=\frac{P\cdot q}{P\cdot k}, \label{eqn:lepvar}
\eea
where $P$, $k$, $k'$ and $q$ are, as illustrated in Figure~\ref{fig:diag}, the momenta of the initial proton,
the initial and final lepton, and the virtual boson, respectively.
Throughout this paper, we work in the limit $m_l\rightarrow 0$ and $m_p\rightarrow 0$,
where $m_l$ and $m_p$ are the mass of the initial lepton and proton, respectively.
To describe the kinematics of the $J/\psi$, we need to introduce two additional variables,
namely the $J/\psi$ transverse momentum ($p_t$) and the inelasticity coefficient,
\be
z=\frac{P\cdot p_\psi}{P\cdot q}, \label{eqn:z}
\ee
where $p_\psi$ is the momentum of the $J/\psi$ meson.

In collinear factorization, the $J/\psi$ producton cross section at QCD LO in $ep$ DIS can be expressed as
\bea
\md\sigma(e+p\rightarrow J/\psi+e+X)=\int\md x\sum_af_{a/p}(x,\mu_f)\md\sigma(e+a\rightarrow J/\psi+e+a), \label{eqn:cs}
\eea
where $a$ runs over all the species of partons, the mass of which is below the $c$-quark mass,
namely $g$, $u$, $d$, $s$ and the corresponding anti-quarks.
$f_{a/p}(x,\mu_f)$ is the parton distribution function (PDF) evaluated at the factorization scale $\mu_f$,
where the momentum of the parton is $p=xP$.
Note that only for the inclusive DIS at QCD LO, when the invariant mass of the hadronic final states is zero,
$x$ is identical to the Bjorken-$x$, $x_B$, defined in Equation~\ref{eqn:lepvar}.

Within the NRQCD framework, the partonic cross section can be further factorized,
accordingly, the cross section defined in Equation~\ref{eqn:cs} can be written as~\cite{Bodwin:1994jh}
\bea
&&\md\sigma(e+p\rightarrow J/\psi+e+X) \NO \\
&&~~~~=\int\md x\sum_{a,n}f_{a/p}(x,\mu_f)\md\hat{\sigma}(e+a\rightarrow c\bar{c}[n]+e+a)\langle O^{J/\psi}(n)\rangle, \label{eqn:csNRQCD}
\eea
where $\langle O^{J/\psi}(n)\rangle$ is the LDME,
which describes the hadronisation of a $c\bar{c}$ pair with quantum number $n$,
and $\md\hat{\sigma}$ is the corresponding short-distance coefficient (SDC).
For the $J/\psi$ production, $n$ can be $^3S_1^{[1]}$, $^1S_0^{[8]}$, $^3S_1^{[8]}$ and $^3P_J^{[8]}$ up to the order of $v^4$,
where $v$ denotes the typical relative velocity of the $c\bar{c}$ pair inside the $J/\psi$ meson.
Taking advantage of the relations $\langle O^{J/\psi}(^3P_2^{[8]})\rangle=5\langle O^{J/\psi}(^3P_0^{[8]})\rangle$,
and $\langle O^{J/\psi}(^3P_1^{[8]})\rangle=3\langle O^{J/\psi}(^3P_0^{[8]})\rangle$,
we can synthesise the three SDCs for $n=^3P_0^{[8]}$, $n=^3P_1^{[8]}$, and $n=^3P_2^{[8]}$ by defining
\bea
&&\md\hat{\sigma}(e+a\rightarrow c\bar{c}[^3P_J^{[8]}]+e+a)\equiv\md\hat{\sigma}(e+a\rightarrow c\bar{c}[^3P_0^{[8]}]+e+a) \NO \\
&&~~~~+3\md\hat{\sigma}(e+a\rightarrow c\bar{c}[^3P_1^{[8]}]+e+a)+5\md\hat{\sigma}(e+a\rightarrow c\bar{c}[^3P_2^{[8]}]+e+a). \label{eqn:sdc3pj8def}
\eea
The corresponding LDME for this synthesised SDC thus should be $\langle O^{J/\psi}(^3P_0^{[8]})\rangle$.

The partonic SDCs can in general be written as
\bea
&&\md\hat{\sigma}(e+a\rightarrow e+c\bar{c}[n]+a) \NO \\
&&~~~~=\frac{1}{4xP\cdot k}\frac{1}{N_cN_s}\frac{1}{(Q^2)^2}L_{\mu\nu}H^{\mu\nu}[n]\md\Phi, \label{eqn:sdc}
\eea
where $1/(N_cN_s)$ is the color and spin average factor, $L_{\mu\nu}$ and $H^{\mu\nu}[n]$ are the leptonic and hadronic tensors, respectively, and
\bea
\md\Phi&=&(2\pi)^4\delta^4(p+q-p_\psi-p_a)\frac{\md^3k'}{(2\pi)^32k'_0}\frac{\md^3p_\psi}{(2\pi)^32p_{\psi0}}\frac{\md^3p_a}{(2\pi)^32p_{a0}}, \label{eqn:ps}
\eea
where $p_a$ is the momentum of the final state parton.
Note that, here, the LDMEs have been eliminated from the hadronic tensors.

In the following subsections, we will present the explicit form of the elements needed in our calculation in Equation~\ref{eqn:sdc}.

\subsection{The Parameterisation of the Physical Variables\label{subsec:para}}

In this subsection, we provide the expressions for the physical variables needed in our calculation.

The squared colliding energy at the lepto-hadronic level and the partonic level are defined as
\bea
S=(P+k)^2=2P\cdot k,~~~~\hat{s}=(p+q)^2, \label{eqn:ss}
\eea
respectively.
However, in most of the cases, $\hat{s}+Q^2$ emerges as a unity, thus, we define
\be
s=\hat{s}+Q^2=2p\cdot q. \label{eqn:s}
\ee

Our calculations are carried out in the virtual-boson-proton ($\gamma p$) centre-of-mass frame.
All the quantities measured in this frame are labeled by a superscript $\star$ in this paper.
Assigning the $J/\psi$ mass as $M$, the momenta involved can be parameterised as
\bea
&&p^\mu=(xE_p^\star,~0,~0,~-xE_p^\star), \NO \\
&&q^\mu=(q_0^\star,~0,~0,~E_p^\star), \NO \\
&&p_\psi^\mu=(\frac{zW^2+m_t^{\star2}/z}{2W},~p_t^\star,~0,~\frac{zW^2-m_t^{\star2}/z}{2W}), \NO \\
&&k^\mu=(E_k^\star,~k_t^\star\mathrm{cos}\psi^\star,~k_t^\star\mathrm{sin}\psi^\star,~k_l^\star), \label{eqn:momenta}
\eea
where
\bea
&&E_p^\star=\frac{W^2+Q^2}{2W},~~~~q_0^\star=\frac{W^2-Q^2}{2W}, \NO \\
&&m_t^\star=\sqrt{p_t^{\star2}+M^2},~~~~x=\frac{s}{W^2+Q^2}, \NO \\
&&E_k^\star=\frac{S-Q^2}{2W},~~~~k_l^\star=\frac{1}{2W}(Q^2+\frac{W^2-Q^2}{W^2+Q^2}S),~~~~k_t^\star=\frac{Q}{y}\sqrt{1-y}. \label{eqn:commom}
\eea
The momentum of the $J/\psi$ can be obtained through the following relation:
\be
z=\frac{m_t^\star e^{y_\psi^\star}}{W}, \label{eqn:rapidity}
\ee
where $y_\psi^\star$ is the rapidity of the $J/\psi$ measured in the $\gamma p$ rest frame.
Regarding the sign of the longitudinal component of $q$,
the forward $z$ direction is defined as that of the incident virtual photon,
which is consistent with the HERA convention.

For some of the available data, the ranges of the transverse momentum ($p_t$) or the rapidity ($y_\psi$) of the $J/\psi$ meson in the laboratory frame are also specified,
thus, we need to find the relations between these two variables and $\psi^\star$\footnote{
The relation between $p_t$ and $\psi^\star$ can also be obtained in the proton rest frame by applying a rotation.
However, here we introduce an alternative approach, which can avoid implementing the limit $m_p\rightarrow 0$.
}.
In the laboratory frame, the rapidity of the $J/\psi$ is calculated with respect to the incident proton direction.
Accordingly, we can obtain the following equations:
\bea
&&m_t^2\equiv p_t^2+M^2=\frac{4(P\cdot p_\psi)(k\cdot p_\psi)}{S}=2yz(k\cdot p_\psi), \NO \\
&&z=\frac{2E_pm_t}{W^2+Q^2}e^{-y_\psi}, \label{eqn:mty}
\eea
where $E_p$ is the energy of the incident proton in the laboratory frame.
After a short calculation, we arrive at
\bea
k\cdot p_\psi=\frac{1}{2yz}[m_t^{\star2}+(1-y)z^2Q^2-2z\sqrt{1-y}Qp_t^\star\mathrm{cos}\psi^\star]. \label{eqn:kppsi}
\eea
Then we can obtain the following relations:
\bea
&&p_t^2=p_t^{\star2}+z^2(1-y)Q^2-2z\sqrt{1-y}Qp_t^\star\mathrm{cos}\psi^\star, \NO \\
&&y_\psi=\frac{1}{2}\mathrm{ln}[\frac{m_t^{\star2}+z^2(1-y)Q^2-2z\sqrt{1-y}Qp_t^\star\mathrm{cos}\psi^\star}{4y^2z^2E_l^2}], \label{eqn:ptypsi}
\eea
where $E_l$ is the energy of the incident lepton in the laboratory frame.

\subsection{The Calculation of the Leptonic and Hadronic Tensors\label{subsec:tensor}}

In HERA experimental condition, $Q$ is much smaller than the mass of the $Z_0$ boson,
thus, we can neglect the contributions from the $Z_0$ propagator.
The leptonic tensor can be obtained as
\bea
L_{\mu\nu}&=&8\pi \alpha Q^2(-g_{\mu\nu}+\frac{4k_\mu k_\nu-2k_\mu q_\nu-2q_\mu k_\nu}{Q^2}) \NO \\
&\equiv&8\pi \alpha Q^2 l_{\mu\nu}. \label{eqn:lepcurrent}
\eea
The normalised leptonic tensor, $l_{\mu\nu}$,
can be decomposed into the linear combination of four independent Lorentz invariant structures as~\cite{Zhang:2017dia}
\bea
l^{\mu\nu}=A_g(-g^{\mu\nu}-\frac{q^\mu q^\nu}{Q^2})+A_L\epsilon_L^\mu\epsilon_L^\nu
+A_{LT}(\epsilon_L^\mu\epsilon_T^\nu+\epsilon_T^\mu\epsilon_L^\nu)+A_T\epsilon_T^\mu\epsilon_T^\nu, \label{eqn:lepreduced}
\eea
where\footnote{We have replaced $P$ in the formalism presented in Reference~\cite{Zhang:2017dia} by $p$.}
\bea
&&\epsilon_L=\frac{1}{Q}(q+\frac{2Q^2}{s}p), \NO \\
&&\epsilon_T=\frac{1}{p^{\star}_t}(p_\psi-\rho p-zq), \NO \\
&&\rho=\frac{m_t^{\star2}/z+zQ^2}{s}, \label{eqn:epLT}
\eea
and
\bea
&&A_g=1+\frac{2(1-y)}{y^2}-\frac{2(1-y)}{y^2}\cos(2\psi^{\star}), \NO \\
&&A_L=1+\frac{6(1-y)}{y^2}-\frac{2(1-y)}{y^2}\cos(2\psi^{\star}), \NO \\
&&A_{LT}=\frac{2(2-y)}{y^2}\sqrt{1-y}\cos(\psi^{\star}), \NO \\
&&A_T=\frac{4(1-y)}{y^2}\cos(2\psi^{\star}). \label{eqn:coefa}
\eea
The leptonic tensor can be modified to be more convenient for our computation as
\bea
l^{\mu\nu}=C_1(-g^{\mu\nu})+C_2p^\mu p^\nu+C_3\frac{p^\mu p_\psi^\nu+p_\psi^\mu p^\nu}{2}
+C_4p_\psi^\mu p_\psi^\nu, \label{eqn:lepredc}
\eea
where
\bea
&&C_1=A_g, \NO \\
&&C_2=\frac{4Q^2}{s^2}(A_L-2\beta A_{LT}+\beta^2A_T), \NO \\
&&C_3=\frac{4Q}{p_t^\star s}(A_{LT}-\beta A_T), \NO \\
&&C_4=\frac{1}{p_t^{\star2}}A_T,
\eea
with
\bea
\beta=\frac{m^{\star2}_t/z+zQ^2}{2p_t^\star Q}. \label{eqn:beta}
\eea
Then the contraction of the leptonic tensor with the hadronic one can be expressed as
\bea
L_{\mu\nu}H^{\mu\nu}[n]=(4\pi)^4\alpha^2\alpha_s^2Q^2(C_1H_1[n]+C_2H_2[n]
+C_3H_3[n]+C_4H_4[n]), \label{eqn:contraction}
\eea
where
\bea
&&H_1[n]=-g^{\mu\nu}h_{\mu\nu}[n],~~~~H_2[n]=p^\mu p^\nu h_{\mu\nu}[n], \NO \\
&&H_3[n]=\frac{p^\mu p_\psi^\nu+p_\psi^\mu p^\nu}{2}h_{\mu\nu}=p^\mu p_\psi^\nu h_{\mu\nu}[n],~~~~H_4[n]=p_\psi^\mu p_\psi^\nu h_{\mu\nu}[n], \label{eqn:hele}
\eea
and
\bea
H_{\mu\nu}[n]=32\pi^3h_{\mu\nu}[n]. \label{eqn:defh}
\eea
In the derivation of the above equations, we have exploited the relations, $q_\mu H^{\mu\nu}=0$ and $H_{\mu\nu}=H_{\nu\mu}$.

The analytical results for $H_i[n]$ ($i=$1, 2, 3, 4) are presented in Appendix~\ref{app:h}.

\subsection{Phase Space\label{subsec:ps}}

The only missing element for calculating the SDCs in Equation~\ref{eqn:sdc} is the expressions for the phase space in Equation~\ref{eqn:ps}.

The phase space for the scattered lepton can be obtained as
\bea
\md\Phi_L\equiv\frac{\md^3k'}{(2\pi)^32k'_0}=\frac{1}{32\pi^3S}\md Q^2\md W^2\md\psi, \label{eqn:pslep}
\eea
while that for the hadrons is
\bea
\md\Phi_H&\equiv&(2\pi)^4\delta^4(p+q-p_\psi-p_a)\frac{\md^3p_\psi}{(2\pi)^32p_{\psi0}}\frac{\md^3p_{a0}}{(2\pi)^32p_{a0}} \NO \\
&=&\frac{1}{16\pi p_{a0}}\delta(p_0+q_0-p_{\psi0}-p_{a0})\md p_t^2\frac{\md z}{z}. \label{eqn:pshadron}
\eea
Note that Equation~\ref{eqn:pshadron} is valid in any frame,
thus, we do not restrict it in the $\gamma p$ rest frame, and erase all the superscript $\star$.
To keep the Lorentz invariance of our formalism, we integrate over $\md x$ to eliminate the last dimension of the $\delta$ function and obtain
\bea
f_{a/p}(x,\mu_f)\md x\md\Phi_H=\frac{1}{8\pi(W^2+Q^2)z(1-z)}f_{a/p}(x,\mu_f)\md p_t^2\md z,
\eea
where the value of $x$ can be found in Equation~\ref{eqn:commom}.
Then we arrive at the final expression of the phase space,
\bea
f_{a/p}(x,\mu_f)\md x\md\Phi&=&\frac{1}{(4\pi)^4S(W^2+Q^2)z(1-z)}f_{a/p}(x,\mu_f)\md Q^2\md W^2\md p_t^2\md z\md\psi \NO \\
&=&\frac{1}{(4\pi)^4z(1-z)}f_{a/p}(x,\mu_f)\md x_B\md y\md p_t^2\md z\md\psi. \label{eqn:psfinal}
\eea

\subsection{The Cross Section}

The cross section for the $J/\psi$ leptoproduction can be obtained by combining Eqs.~(\ref{eqn:csNRQCD}),
(\ref{eqn:sdc}), (\ref{eqn:contraction}) and (\ref{eqn:psfinal}), and reduces to
\bea
&&\md\sigma(e+p\rightarrow e+J/\psi+X) \NO \\
&&~~=\frac{\alpha_s^2\alpha^2}{2S^2N_cN_s}\sum_{a,n}\langle\mathcal{O}^{J/\psi}(n)\rangle\frac{1}{x}f_{a/p}(x,\mu_f)
\sum_{i=1}^4C_iH_i[n]\frac{\md Q^2}{Q^2}\frac{\md W^2}{W^2+Q^2}\md p_t^2\frac{\md z}{z(1-z)}\md\psi \NO \\
&&~~=\frac{\alpha_s^2\alpha^2}{2S^2N_cN_s}\sum_{a,n}\langle\mathcal{O}^{J/\psi}(n)\rangle\frac{1}{x}f_{a/p}(x,\mu_f)
\sum_{i=1}^4C_iH_i[n]\frac{\md x_B}{x_B}\frac{\md y}{y}\md p_t^2\frac{\md z}{z(1-z)}\md\psi. \label{eqn:csfinal}
\eea

\section{Numerical Results\label{sec:numresults}}

Before our work, three papers~\cite{Fleming:1997fq, Yuan:2000cn, Kniehl:2001tk} studying the $J/\psi$ production in DIS within the NRQCD framework have already been published.
It is worth noting that they are mutually incompatible with each other.
With the same leptonic tensors, our results agree with those in Reference~\cite{Kniehl:2001tk}.
However, as we pointed out in Reference~\cite{Zhang:2017dia},
Reference~\cite{Kniehl:2001tk} used a form of the leptonic tensor identical to
\bea
l_{\mu\nu}=\frac{2-2y+y^2}{y^2}(-g_{\mu\nu}-\frac{q_\mu q_\nu}{Q^2})+\frac{6-6y+y^2}{y^2}\frac{1}{Q^2}(q+\frac{Q^2}{p\cdot q}p)_\mu(q+\frac{Q^2}{p\cdot q}p)_\nu, \label{eqn:lepred}
\eea
which will lead to wrong results when $p_t$ or $y_\psi$ in the laboratory frame do not cover all their possible values.
Another important issue to address is that, to calculate the polarisation of the $J/\psi$ meson,
an additional momentum other than $q$ and $p$ will also emerge in the hadronic tensor,
thus, with the leptonic tensor in Equation~\ref{eqn:lepred}, one will also obtain wrong results.
To illuminate the difference between Equation~\ref{eqn:lepredc} and Equation~\ref{eqn:lepred},
we present the comparison of our results with those in Table 1 of Reference~\cite{Kniehl:2001tk}.
Applying the same parameter choices, namely, $m_c=1.5\gev$, $E_l=27.5\gev$, $E_p=820\gev$, $30\gev<W<150\gev$, $\alpha=1/137$, $\mu_r=\mu_f=\sqrt{Q^2+M^2}$,
$\langle O^{J/\psi}(^3S_1^{[1]})\rangle=1.1\gev^3$, $\langle O^{J/\psi}(^1S_0^{[8]})\rangle=1\times10^{-2}\gev^3$,
$\langle O^{J/\psi}(^3S_1^{[8]})\rangle=1.12\times10^{-2}\gev^3$, $\langle O^{J/\psi}(^3P_0^{[8]})\rangle/m_c^2=5\times10^{-3}\gev^3$,
the LO proton PDF given in Reference~\cite{Gluck:1994uf},
and the one-loop $\alpha_s$ running with $n_f=4$ and $\Lambda^{(4)}=0.13\gev$,
we present the corresponding results in Table~\ref{tab:comparison},
where our results for employing the leptonic tensors presented in Equation~\ref{eqn:lepred}
and Equation~\ref{eqn:lepredc} are labeled by I and II, respectively.
The slight difference between our results I and those in Reference~\cite{Kniehl:2001tk}
might be due to the difference in the precision of the parameter choices kept in the two independent evaluations.
However, when the $p_t^2$ cut is applied, the difference between our result II and that in Reference~\cite{Kniehl:2001tk} is significant,
which manifests the necessity of the inclusion of the azimuthal-asymmetric terms in the calculation.

\begin{table}
\centering
\caption{\label{tab:comparison}
Comparison of our results with those of Reference~\cite{Kniehl:2001tk}.
}
\begin{tabular}{lcccccc}
\hline
Type & Cuts & Reference~\cite{Kniehl:2001tk} & Our results I & Our results II \\
CS & $Q^2>4\gev^2$ & 107pb & 103.4pb & 103.4pb \\
CS & $Q^2$, $p_t^2>4\gev^2$ & 62pb & 60.0pb & 54.5pb \\
CS & $Q^2>4\gev^2$, $p_t^{\star2}>2\gev^2$, $z<0.8$ & 24pb & 23.5pb & 23.5pb \\
CO & $Q^2>4\gev^2$, $p_t^{\star2}>2\gev^2$, $z<0.8$ & 16pb & 15pb & 15pb \\
\hline
\end{tabular}
\end{table}

Now we are in a position to confront our phenomenological results with the H1~\cite{Adloff:1999zs, Adloff:2002ey, Aaron:2010gz} and ZEUS~\cite{Chekanov:2005cf} data.
In the rest part of this paper, we take $m_c=1.5$ GeV, $\alpha=1/137$ and $E_l=27.5\gev$.
For the experimental conditions in Reference~\cite{Adloff:1999zs}, $E_p=820\gev$,
while for those in References~\cite{Adloff:2002ey, Chekanov:2005cf, Aaron:2010gz}, $E_p=920\gev$.
The one-loop $\alpha_s$ running is employed, where its value at the $Z_0$ boson mass is set to be $\alpha_s(M_Z)=0.13$.
Correspondingly, we employ CTEQ6L1~\cite{Pumplin:2002vw} as the PDF for the proton.
Our default set of the renormalisation and factorisation scales is $\mu_r=\mu_f=\mu_0\equiv\sqrt{Q^2+M^2}$.
To present the uncertainties of our theoretical predictions,
the bands in the plots cover all the results for simultaneously varying these two scales from $0.5\mu_0$ to $2\mu_0$.
The NRQCD predictions suffer severely from the uncertainties of the LDMEs, thus,
we will adopt three sets of the LDMEs on the market, the detailed values of which can be found in Table~\ref{tab:LDME},
to present our numerical results\footnote{
With the recent data of the hadroporduction of $\eta_c$ from the LHCb Collaboration,
the authors of Reference~\cite{Han:2014jya} have imposed a strong upper limit on $\langle O^{J/\psi}(^1S_0^{[8]})\rangle$.
Since the values of their new set of LDMEs almost coincide with the ones in Reference~\cite{Zhang:2014ybe},
in our current paper, only the latter set is adopted.
For the same reason, the LDMEs presented in Reference~\cite{Sun:2015pia},
which is an update of Reference~\cite{Zhang:2014ybe} to adapt to the $J/\psi$ polarization measurements,
are also ignored in the presentation of our results.
To understand the $J/\psi$ polarisation puzzle, the authors of Reference~\cite{Chao:2012iv} proposed the $^1S_0^{[8]}$ dominance picture.
There are some other works~\cite{Gong:2012ug, Faccioli:2014cqa, Bodwin:2015iua} studying the $J/\psi$ polarisation,
which, although with different fitting strategies, also arrived at similar results,
namely the $^1S_0^{[8]}$ LDME being almost one order of magnitude larger than the other two.
Applying these LDMEs, the results for the $J/\psi$ production in DIS at HERA do not differ significantly from those obtained by employing the LDMEs in Reference~\cite{Chao:2012iv}.
Therefore, we just adopt the three representative sets of the LDMEs presenting our results.}.
The other origin of the uncertainties in our results is those arising from the LDMEs.
Note that the $c$-quark mass ($m_c$) dependence of the SDCs can generally be balanced out by that of the LDMEs (see e.g.~\cite{Ma:2010jj}).
In this paper, we do not vary the value of $m_c$.

\begin{table}
\centering
\caption{\label{tab:LDME}Three sets of the LDMEs used in our calculation.}
\begin{tabular}{lccccc}
\hline
& $\langle O^{J/\psi}(^3S_1^{[1]})\rangle$ & $\langle O^{J/\psi}(^1S_0^{[8]})\rangle$ & $\langle O^{J/\psi}(^3S_1^{[8]})\rangle$ & $\langle O^{J/\psi}(^3P_0^{[8]})\rangle$ \\
& $\gev^3$ & $\times10^{-2}\gev^3$ & $\times10^{-2}\gev^3$ & $\times10^{-2}\gev^5$ \\
Reference~\cite{Chao:2012iv} & $1.16$ & $8.9\pm0.98$ & $0.3\pm0.12$ & $1.26\pm0.47$ \\
Reference~\cite{Butenschoen:2011yh} & $1.32$ & $3.04\pm0.35$ & $0.168\pm0.046$ & $-0.908\pm0.161$ \\
Reference~\cite{Zhang:2014ybe} & $0.645\pm0.405$ & $0.785\pm0.42$ & $1.0\pm0.3$ & $3.8\pm1.1$ \\
\hline
\end{tabular}
\end{table}

We notice that the data for the $J/\psi$ leptoproduction at HERA is plentiful.
In some of the experimental conditions, the perturbative calculation based on the collinear factorisation is not expected to work well,
therefore, we need first to address the ranges of the variables for good perturbative expansion.
The cross sections diverge at $Q^2=0$, thus, when $Q^2$ is too small,
the results will suffer from a divergence factor, $1/Q^2$.
The newest HERA measurements applied the cut, $Q^2<2.5\gev^2$, for photoproduction,
according to which, $Q^2=2.5\gev^2$ can be considered as a moderate boundary to separate regions for good and bad perturbative calculations.
For a specific value of $p_t^{\star2}$ and $p_t^2$,
the minimum values of $x$ are approximately $[(m_t^\star+p_t^\star)^2+Q^2]/(W^2+Q^2)$ and $m_t^2/[z(W^2+Q^2)]$, respectively.
Thus, in small $p_t$ and $p_t^\star$ regions, the gluon saturation effect might ruin our results.
The perturbation theory works better in the region, $p_t^2>6.4\gev^2$ or $p_t^{\star2}>6.4\gev^2$,
where the value of $x$ is generally larger than 0.001.
Since the selection rules do not forbid the processes $\gamma g\rightarrow c\bar{c}[n]$ for $n=^1S_0^{[8]}$ and $n=^3P_J^{[8]}$,
both soft and collinear singularities will arise for the $c\bar{c}[^1S_0^{[8]}]$ and $c\bar{c}[^3P_J^{[8]}]$ production at $z=1$.
One will suffer from a divergence factor, $1/(1-z)$, in the region, $z\rightarrow1$,
as a result, only when $z$ is much smaller than 1, the NRQCD predictions are expected to be reliable.
In the region, $z>0.6$, this divergence factor is larger than 2.5,
which might significantly enhances the terms which ought to be small.
For this reason, we consider the perturbative calculation in the region, $z>0.6$, to be not reliable.
In addition, around $z=1$, the diffractively produced $J/\psi$ will dominate the prompt $J/\psi$ production,
which, however, cannot be effectively eliminated from the $J/\psi$ inelastic production events.
In the region $M_X>10\gev$, the value of $z$ can be as large as 0.997,
thus, we completely omit the data corresponding to this condition.

\begin{figure}
\includegraphics[width=0.32\textwidth]{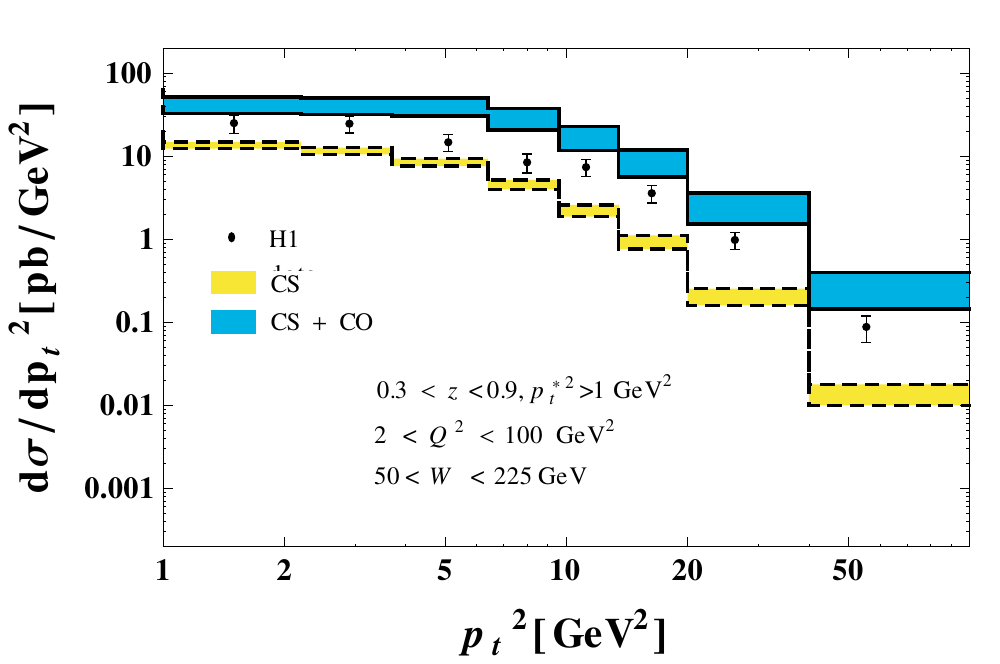}
\includegraphics[width=0.32\textwidth]{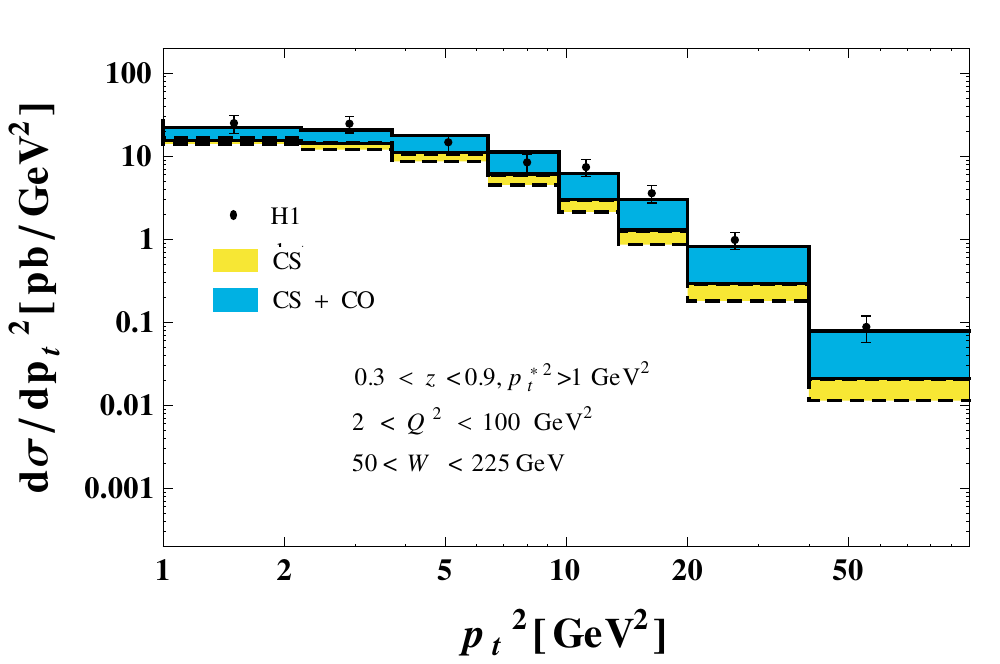}
\includegraphics[width=0.32\textwidth]{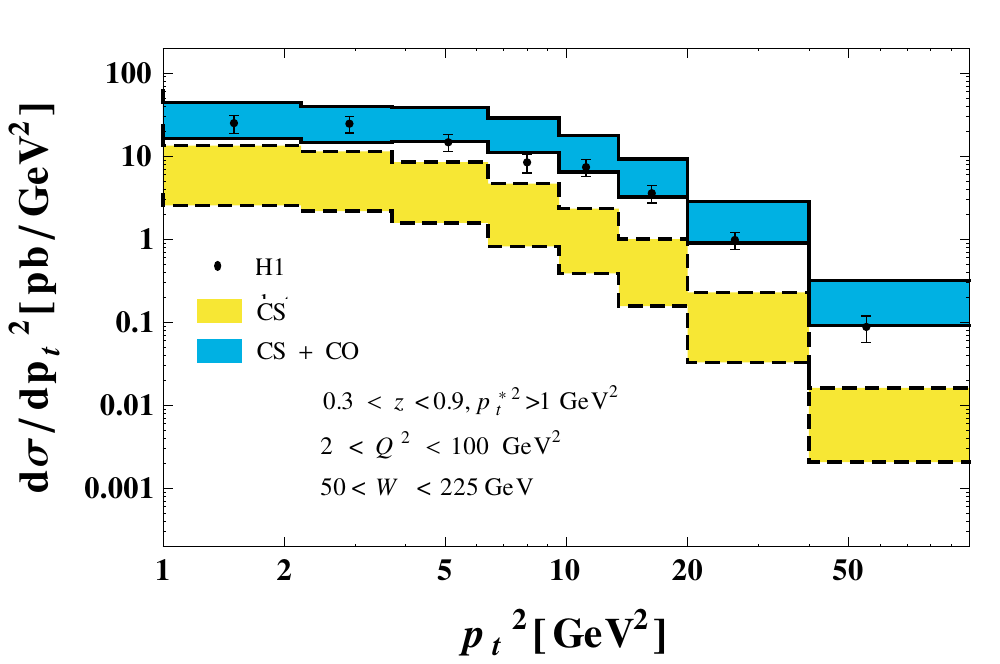}\\
\includegraphics[width=0.32\textwidth]{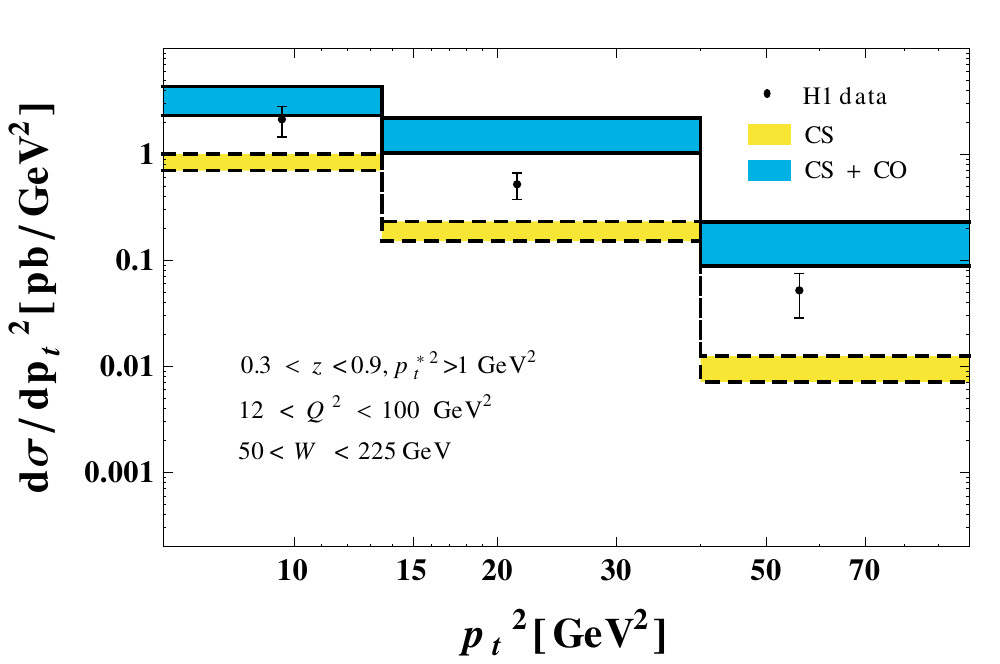}
\includegraphics[width=0.32\textwidth]{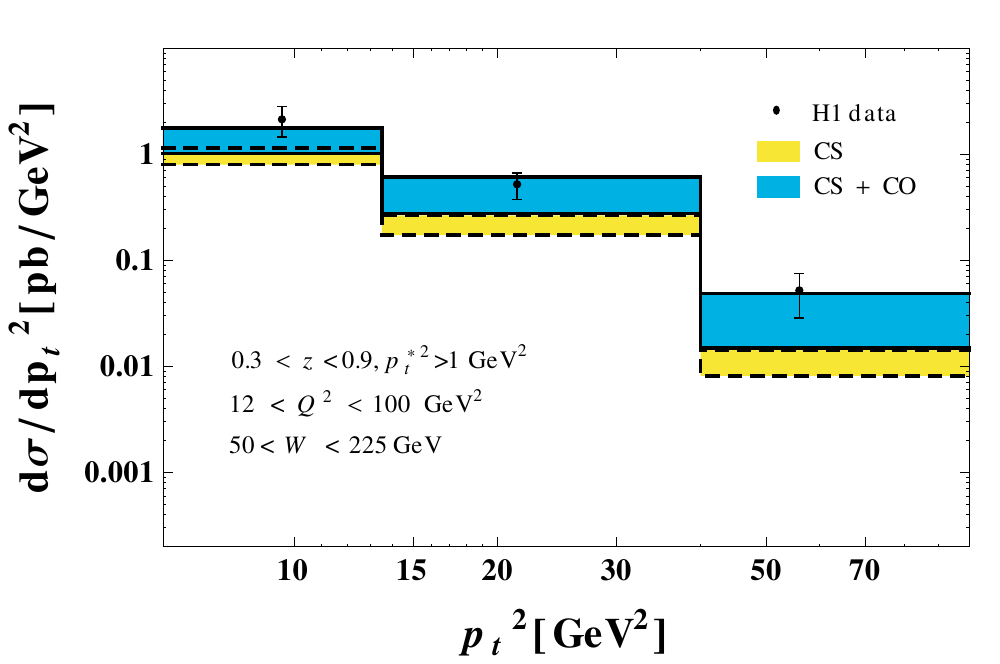}
\includegraphics[width=0.32\textwidth]{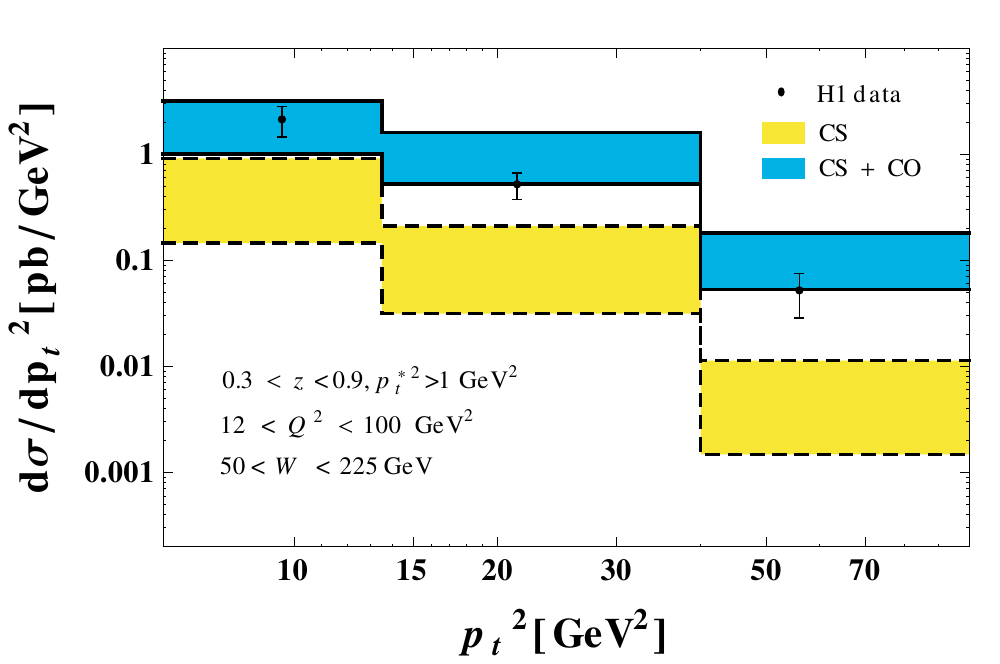}
\caption{\label{fig:pt2-2002}
The differential cross sections for the $J/\psi$ production in DIS with respect to $p_t^2$.
The experimental data are taken from Reference~\cite{Adloff:2002ey}.
The l.h.s., mid, and r.h.s plots correspond to the LDMEs taken in References~\cite{Chao:2012iv},~\cite{Butenschoen:2011yh}, and~\cite{Zhang:2014ybe}, respectively.
}
\end{figure}

\begin{figure}
\includegraphics[width=0.32\textwidth]{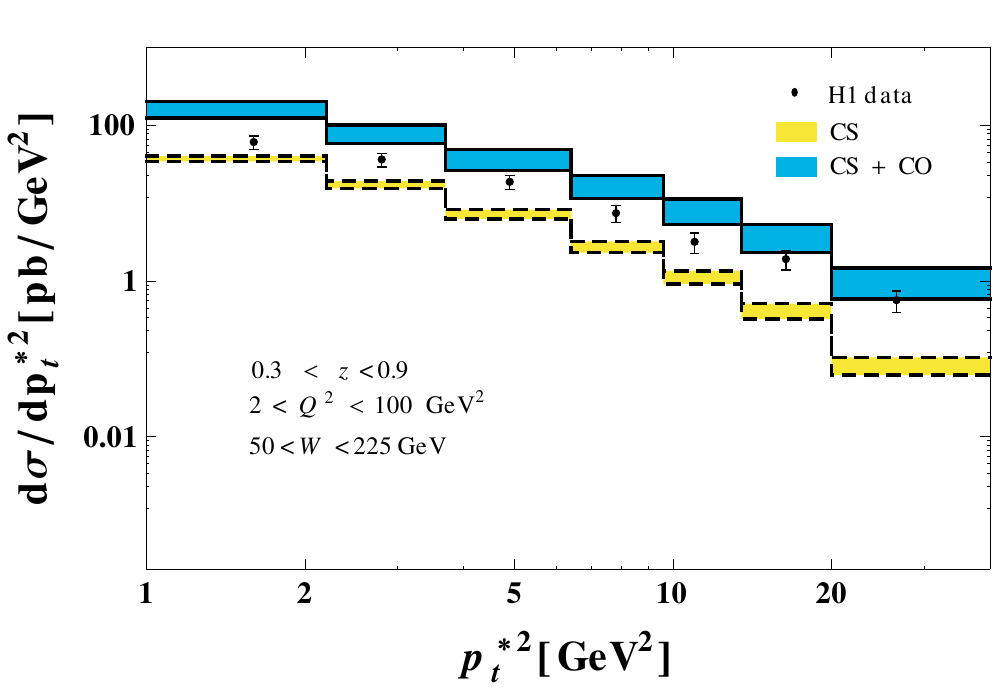}
\includegraphics[width=0.32\textwidth]{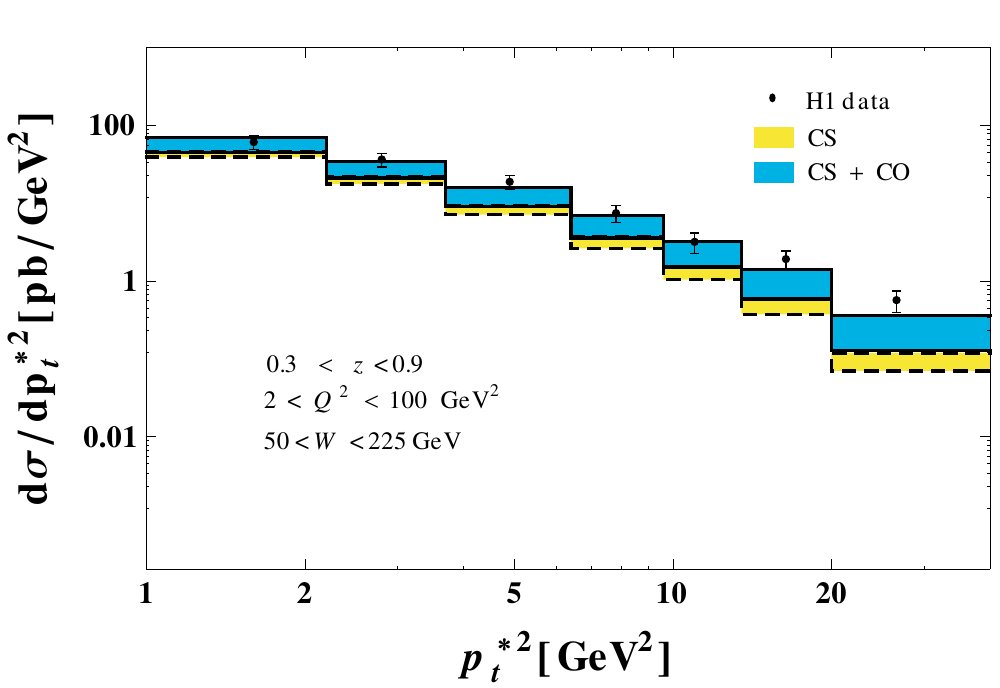}
\includegraphics[width=0.32\textwidth]{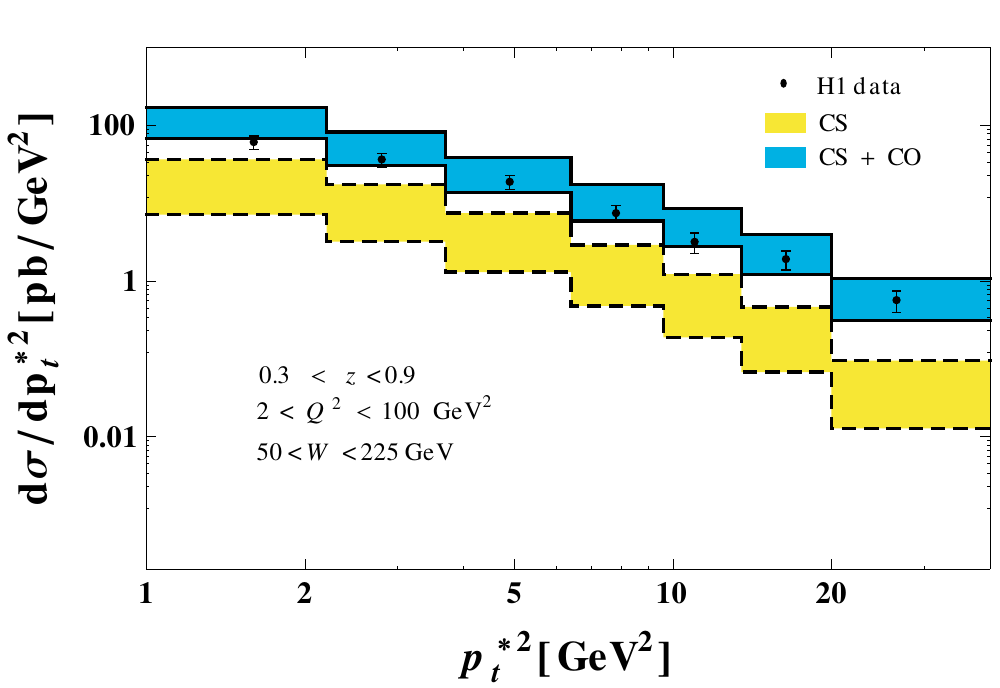}\\
\includegraphics[width=0.32\textwidth]{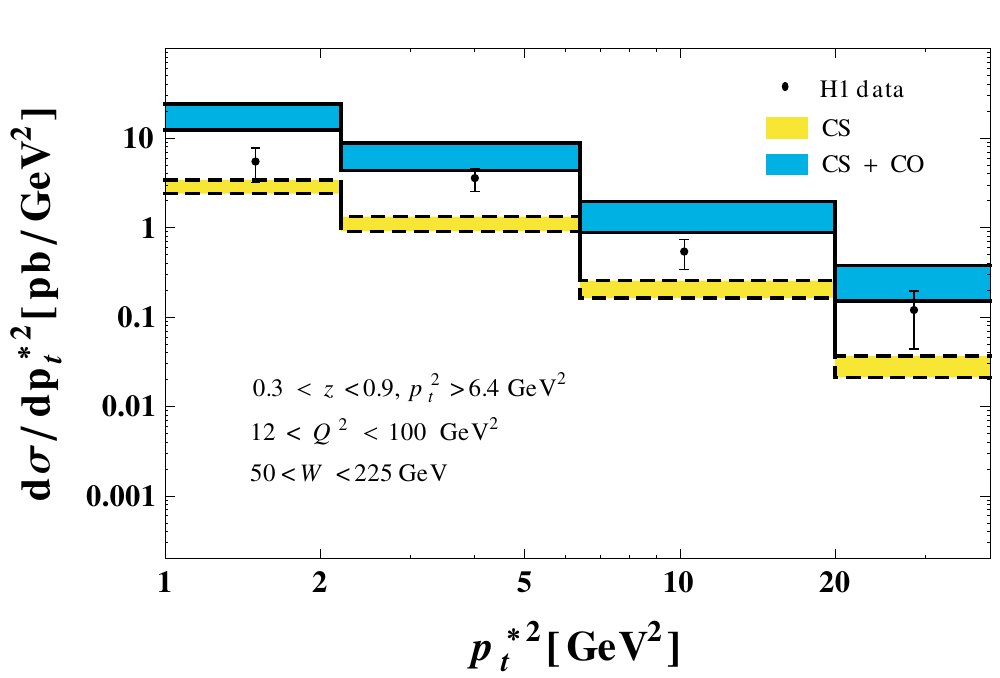}
\includegraphics[width=0.32\textwidth]{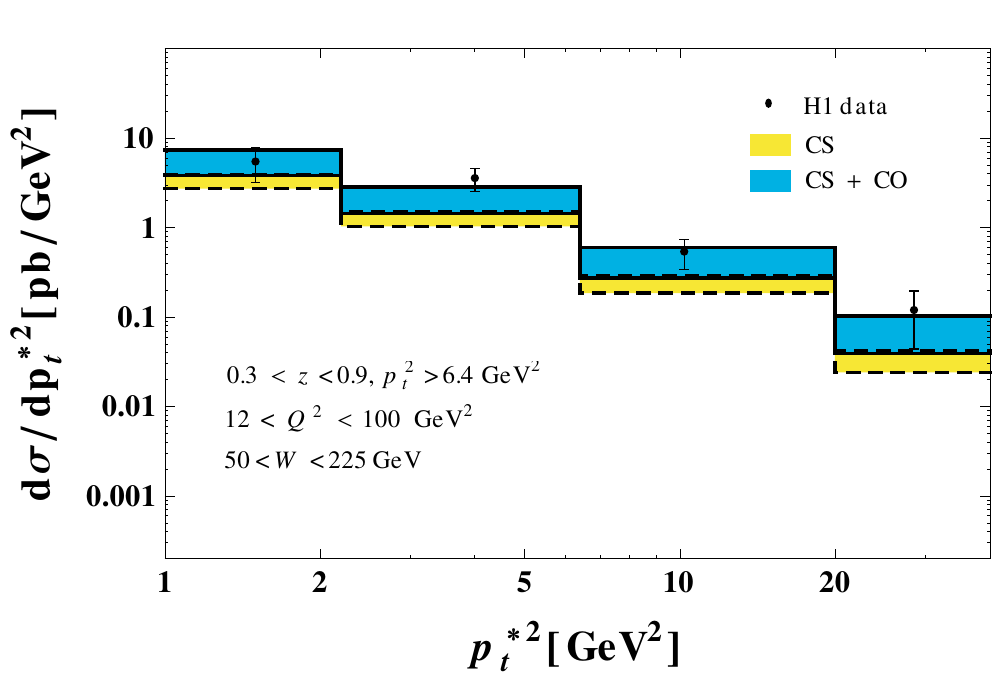}
\includegraphics[width=0.32\textwidth]{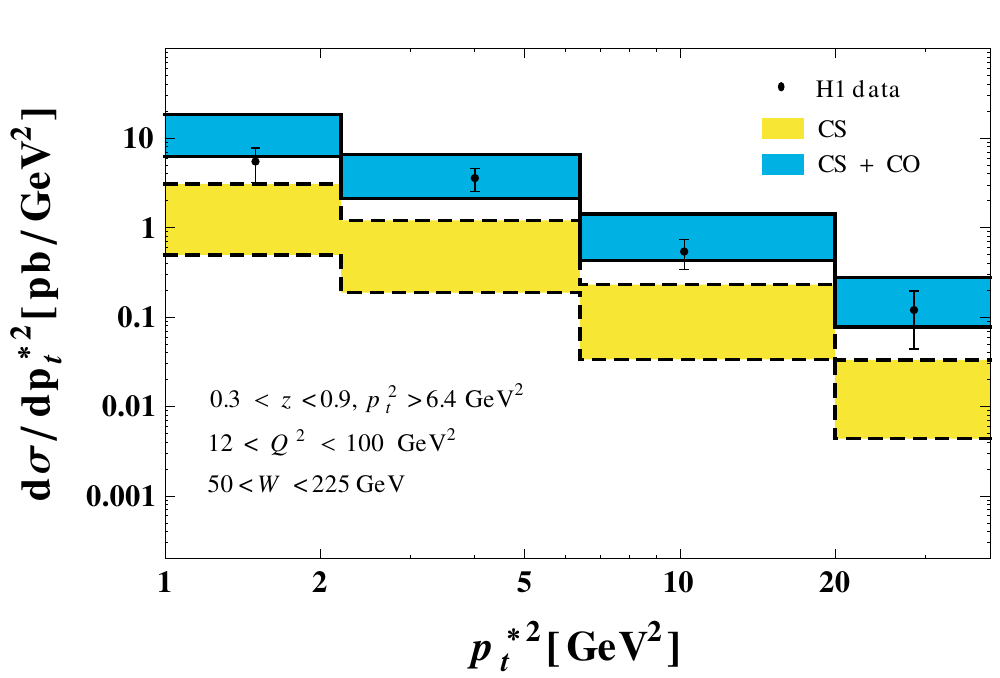}
\caption{\label{fig:pts2-2002}
The differential cross sections for the $J/\psi$ production in DIS with respect to $p_t^{\star2}$.
The experimental data are taken from Reference~\cite{Adloff:2002ey}.
The l.h.s., mid, and r.h.s plots correspond to the LDMEs taken in References~\cite{Chao:2012iv},~\cite{Butenschoen:2011yh}, and~\cite{Zhang:2014ybe}, respectively.
}
\end{figure}

\begin{figure}
\includegraphics[width=0.32\textwidth]{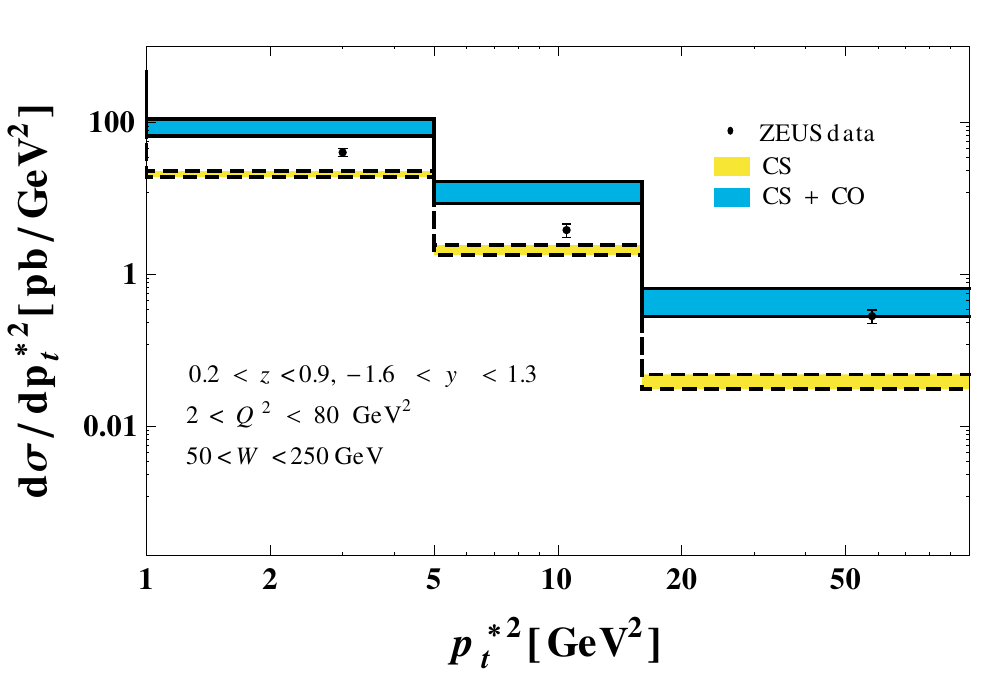}
\includegraphics[width=0.32\textwidth]{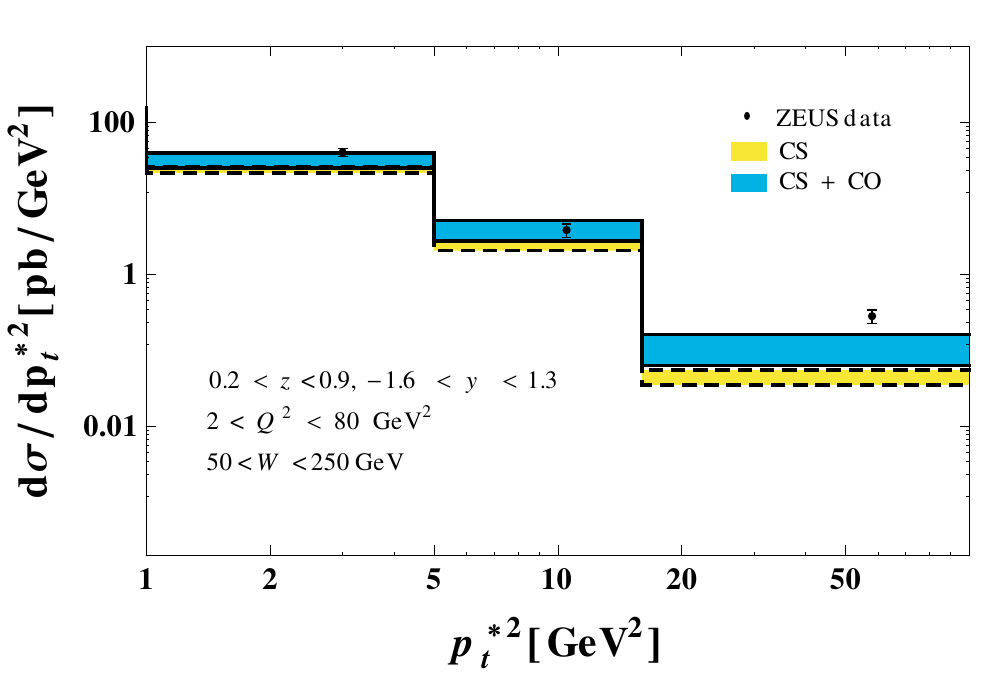}
\includegraphics[width=0.32\textwidth]{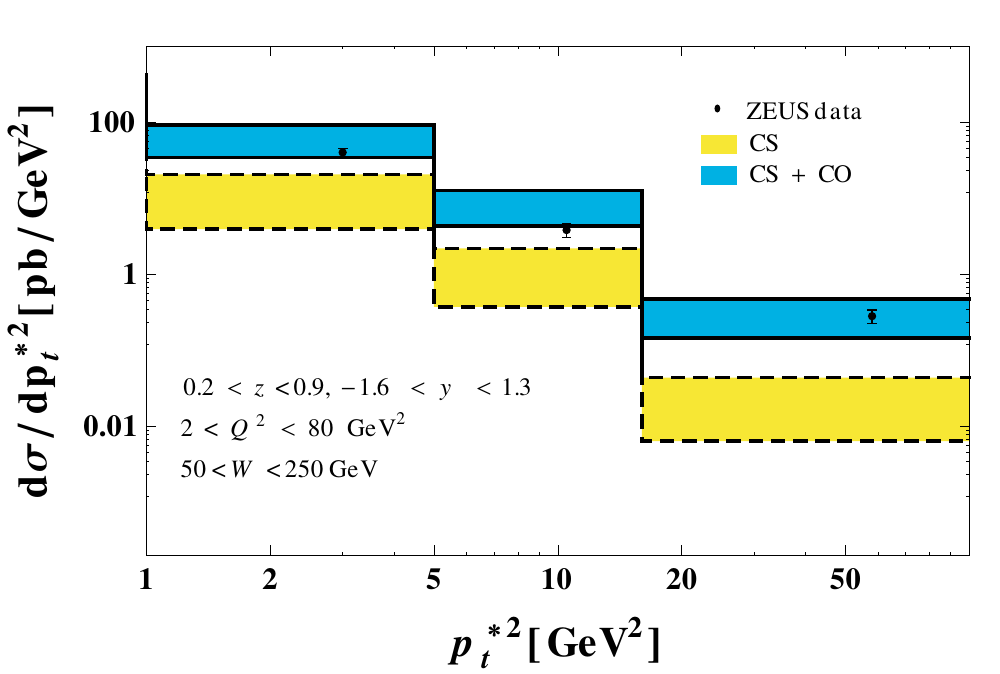}\\
\includegraphics[width=0.32\textwidth]{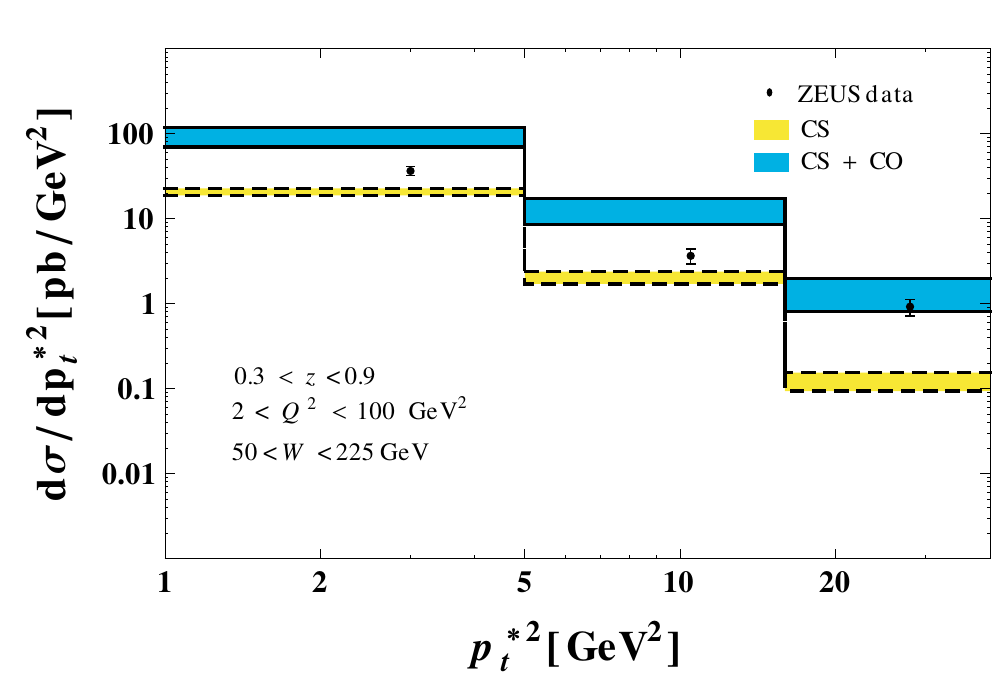}
\includegraphics[width=0.32\textwidth]{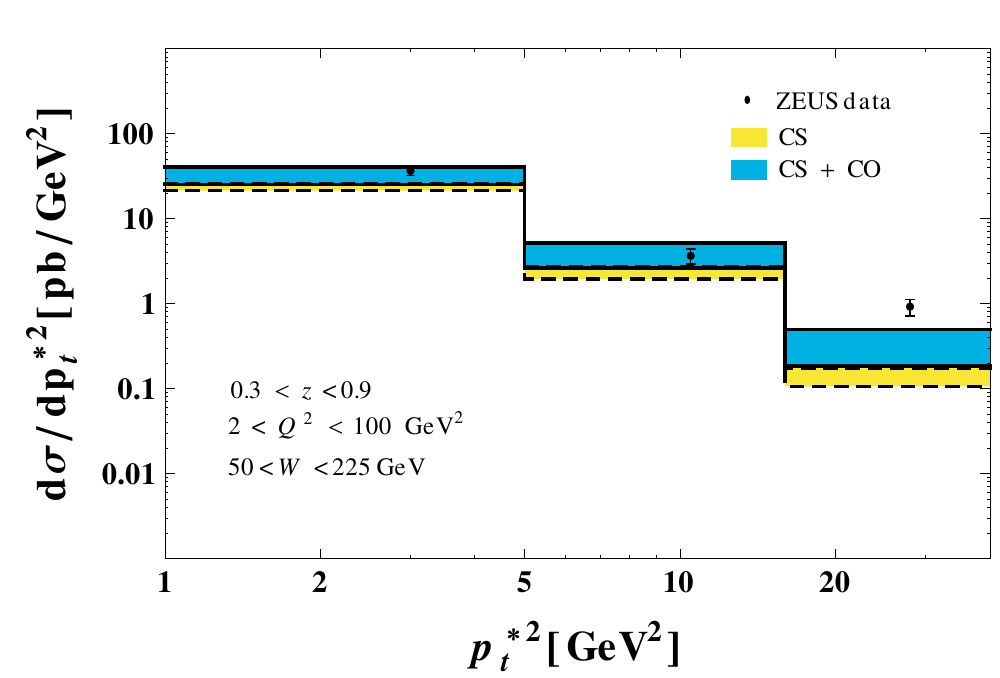}
\includegraphics[width=0.32\textwidth]{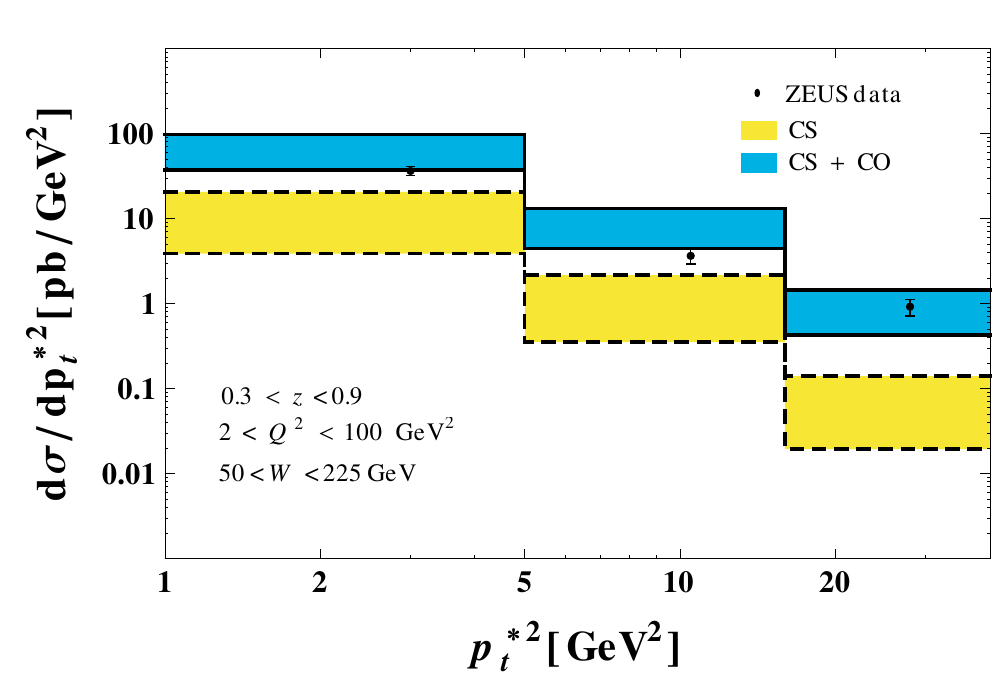}
\caption{\label{fig:pts2-2005}
The differential cross sections for the $J/\psi$ production in DIS with respect to $p_t^{\star2}$.
The experimental data are taken from Reference~\cite{Chekanov:2005cf}.
The l.h.s., mid, and r.h.s plots correspond to the LDMEs taken in References~\cite{Chao:2012iv},~\cite{Butenschoen:2011yh}, and~\cite{Zhang:2014ybe}, respectively.
}
\end{figure}

\begin{figure}
\includegraphics[width=0.32\textwidth]{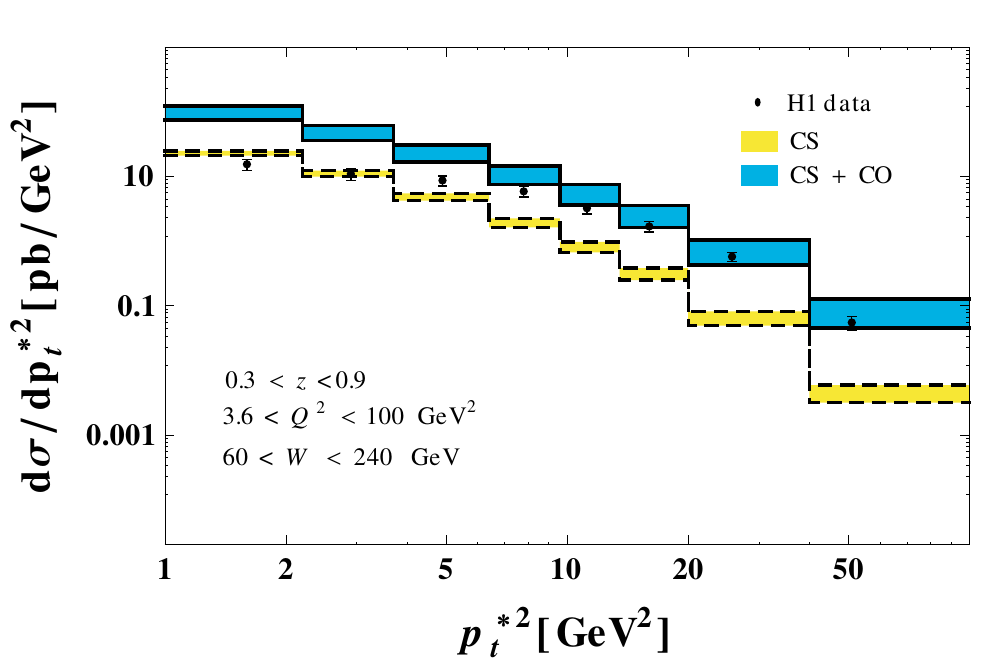}
\includegraphics[width=0.32\textwidth]{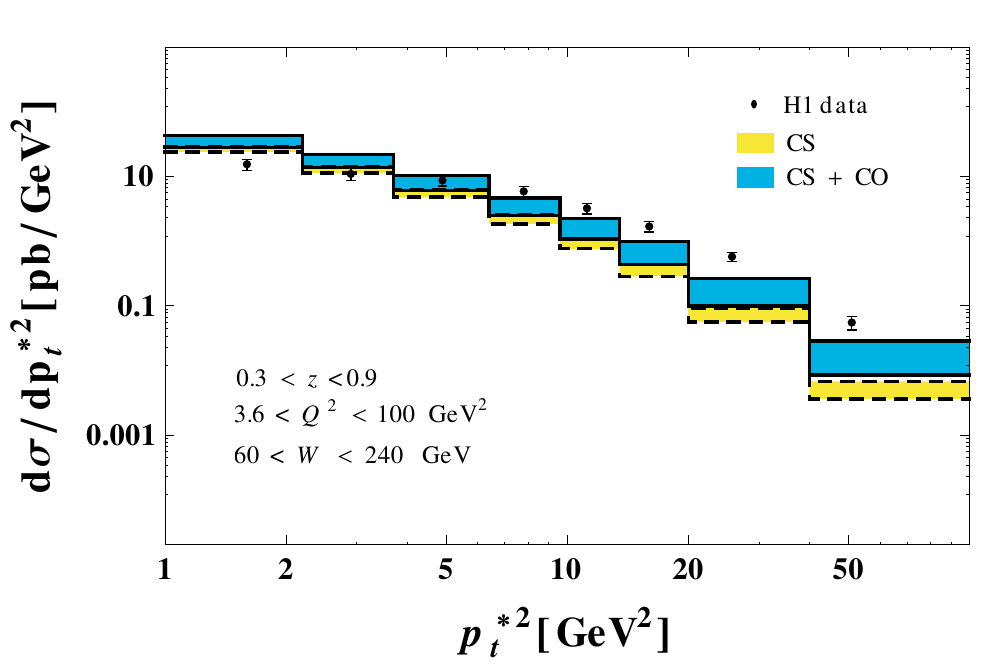}
\includegraphics[width=0.32\textwidth]{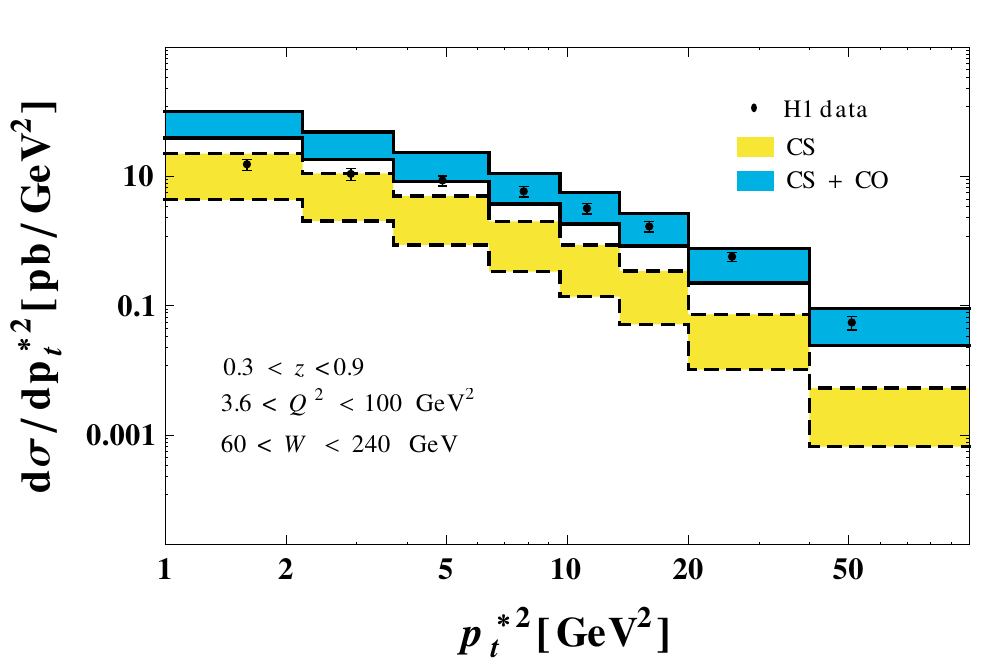}\\
\includegraphics[width=0.32\textwidth]{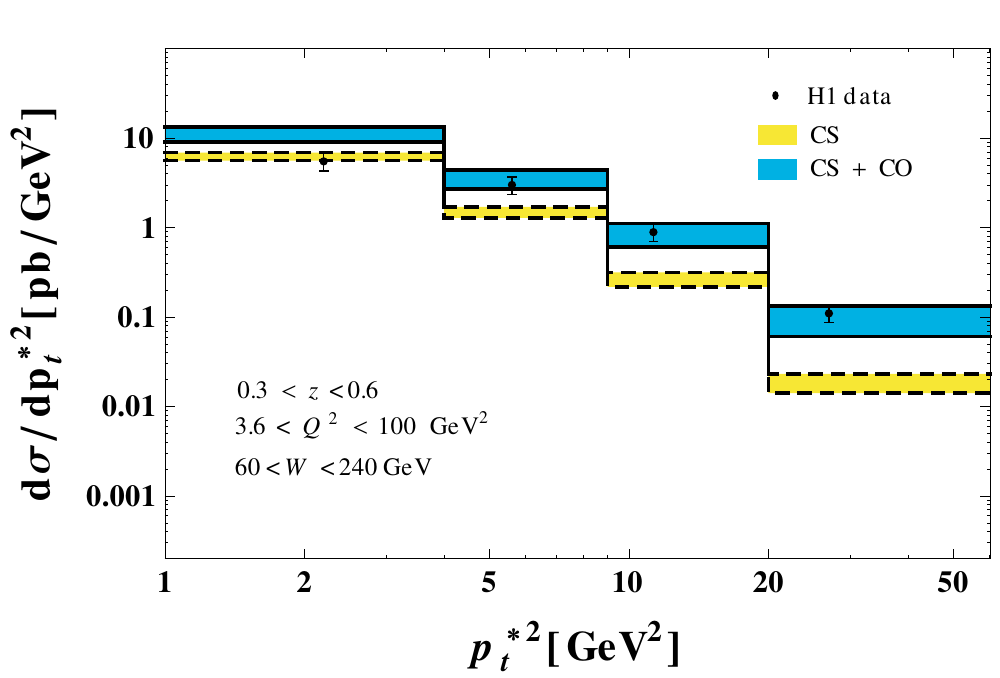}
\includegraphics[width=0.32\textwidth]{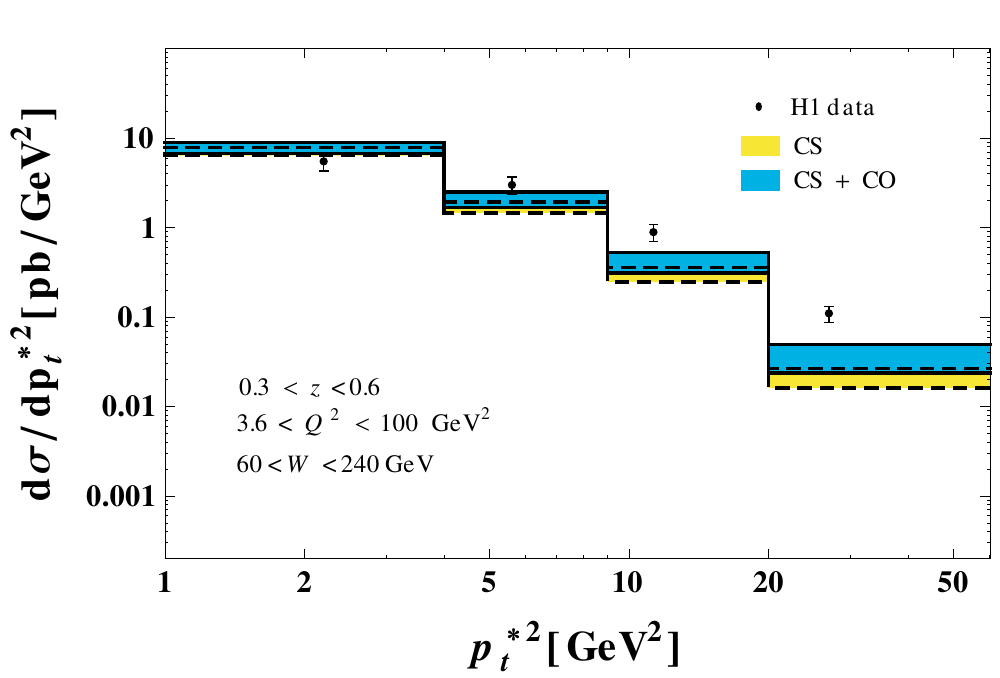}
\includegraphics[width=0.32\textwidth]{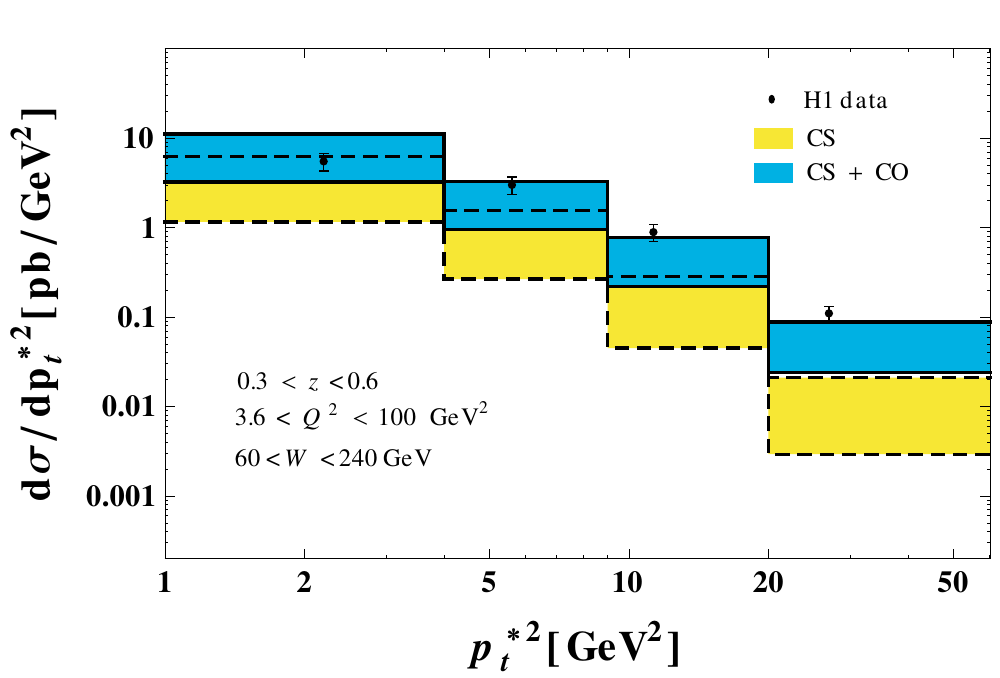}\\
\includegraphics[width=0.32\textwidth]{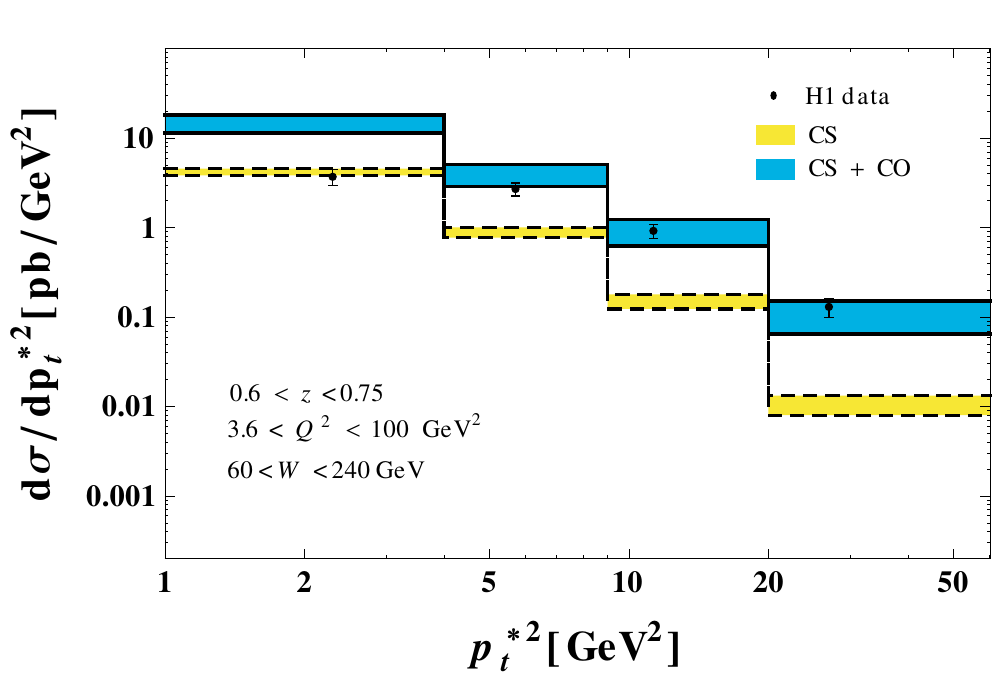}
\includegraphics[width=0.32\textwidth]{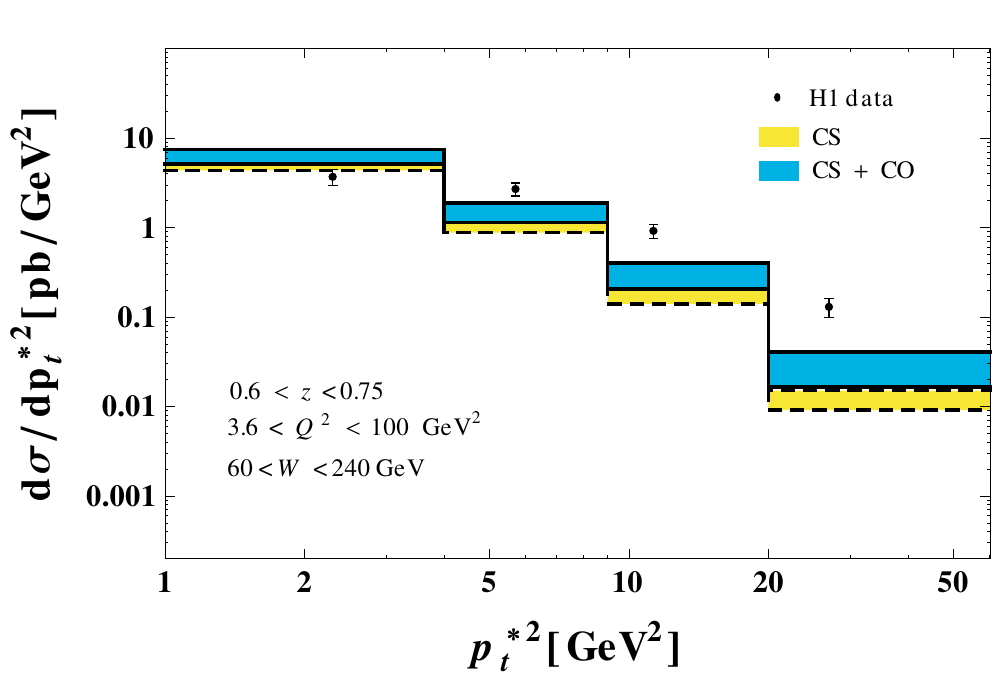}
\includegraphics[width=0.32\textwidth]{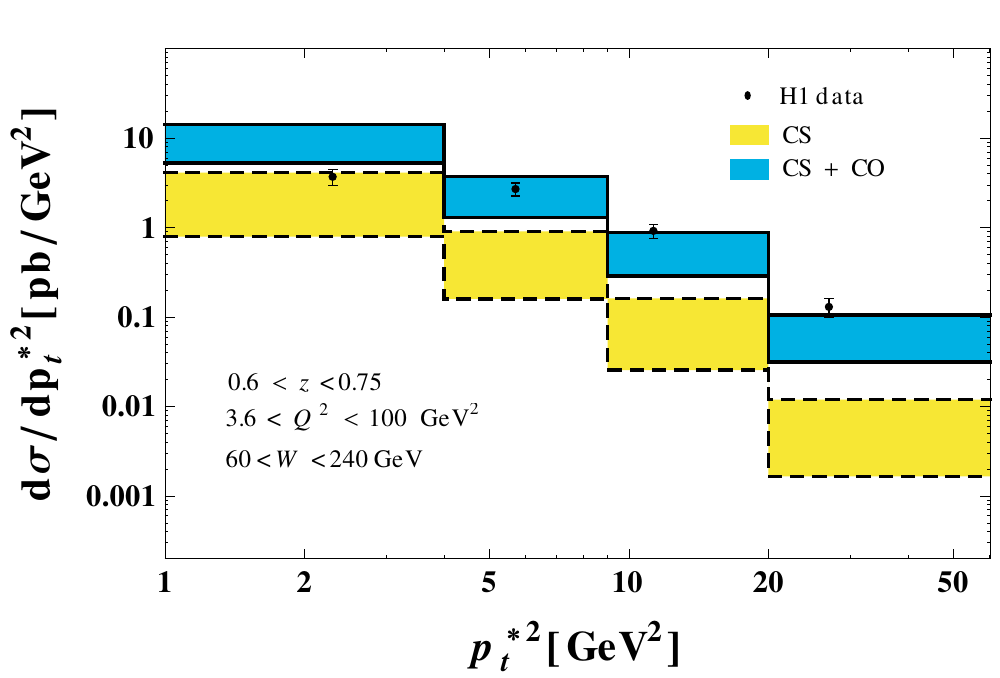}\\
\includegraphics[width=0.32\textwidth]{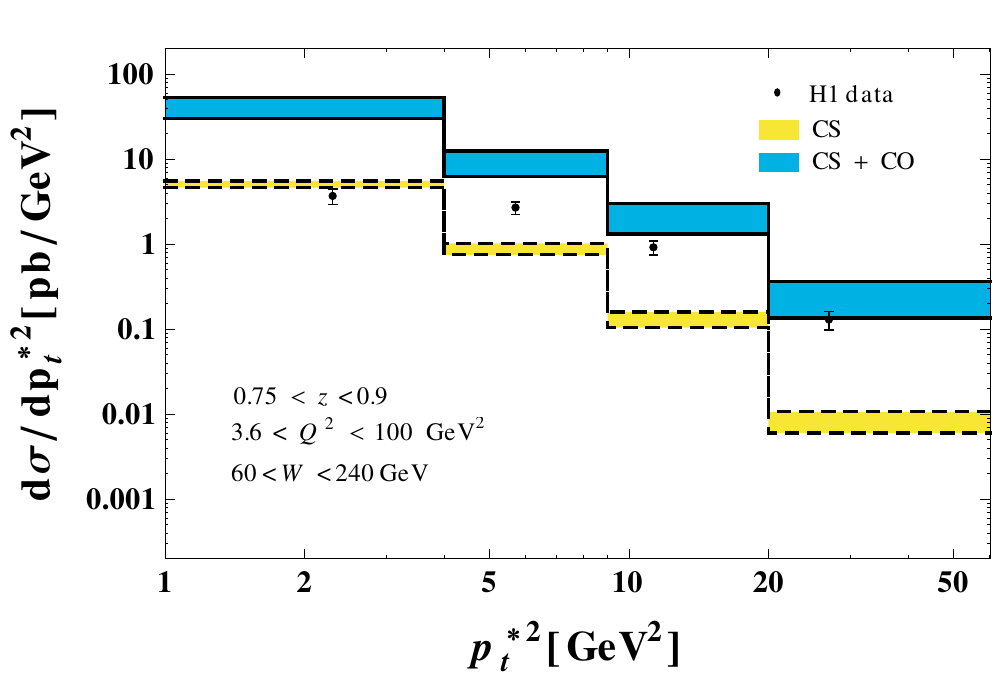}
\includegraphics[width=0.32\textwidth]{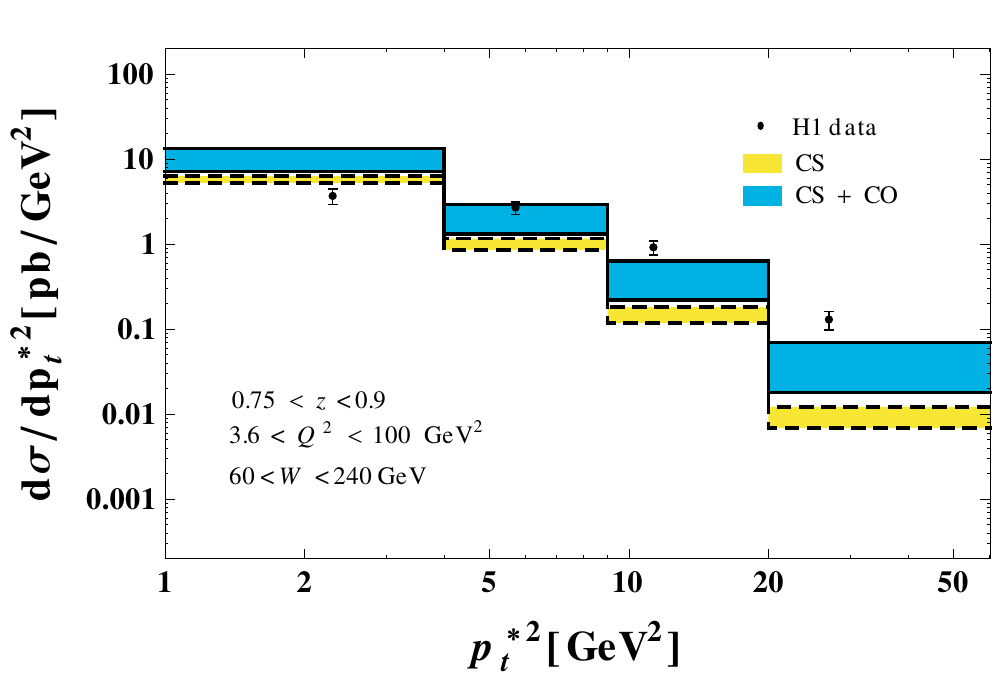}
\includegraphics[width=0.32\textwidth]{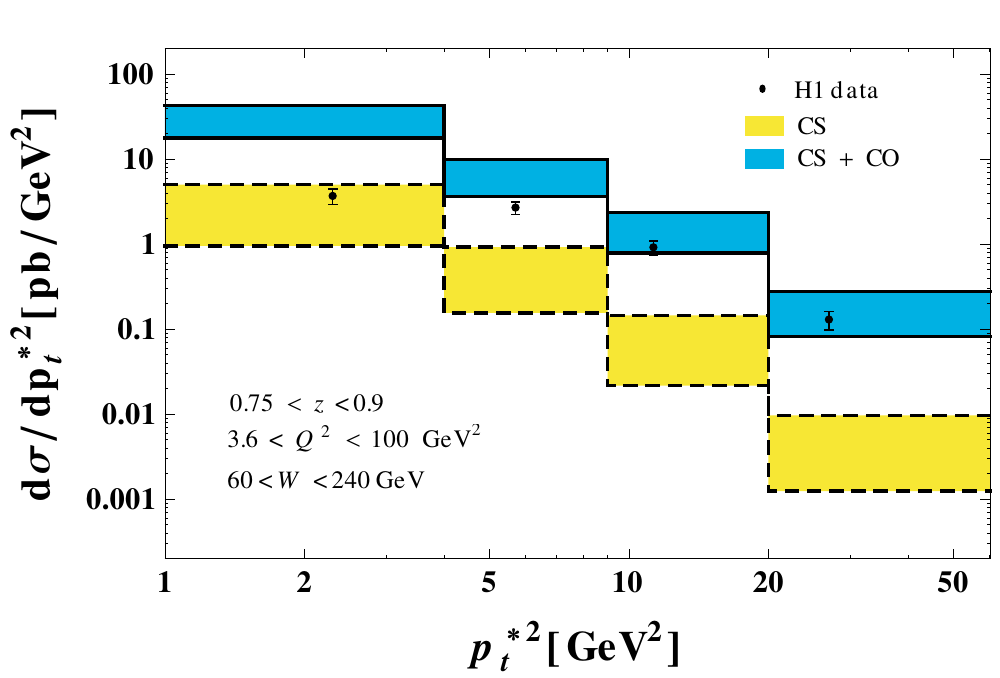}
\caption{\label{fig:pts2-2010}
The differential cross sections for the $J/\psi$ production in DIS with respect to $p_t^{\star2}$.
The experimental data are taken from Reference~\cite{Aaron:2010gz}.
The l.h.s., mid, and r.h.s plots correspond to the LDMEs taken in References~\cite{Chao:2012iv},~\cite{Butenschoen:2011yh}, and~\cite{Zhang:2014ybe}, respectively.
}
\end{figure}

\begin{figure}
\includegraphics[width=0.32\textwidth]{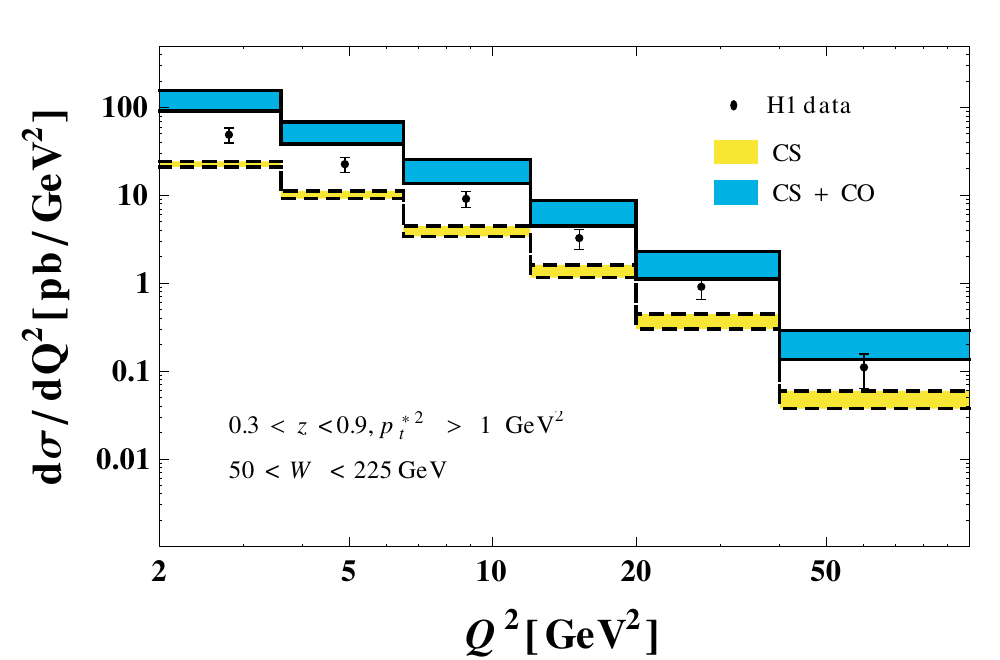}
\includegraphics[width=0.32\textwidth]{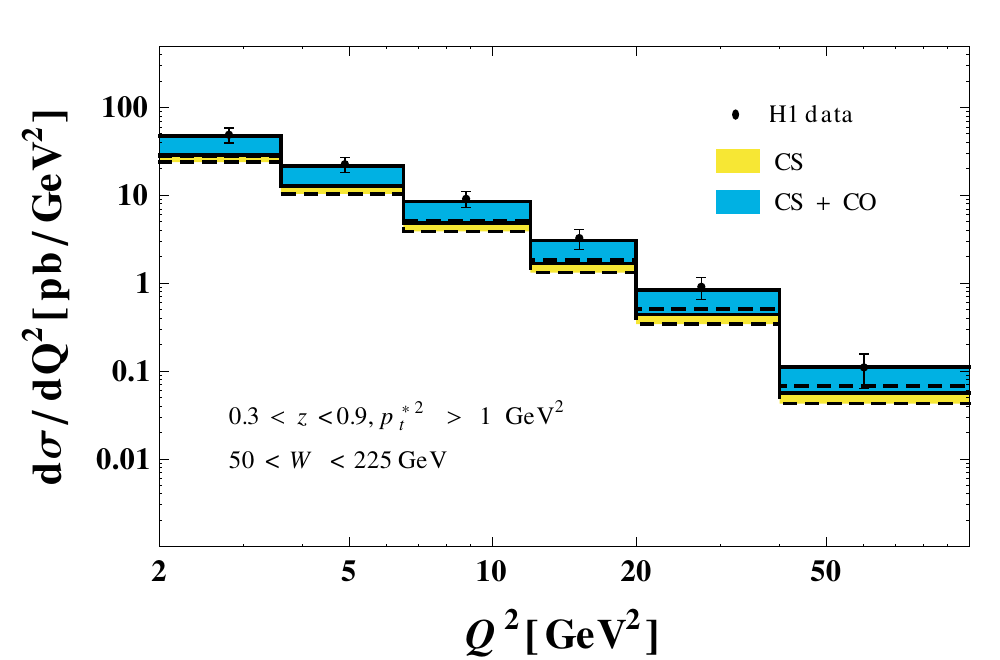}
\includegraphics[width=0.32\textwidth]{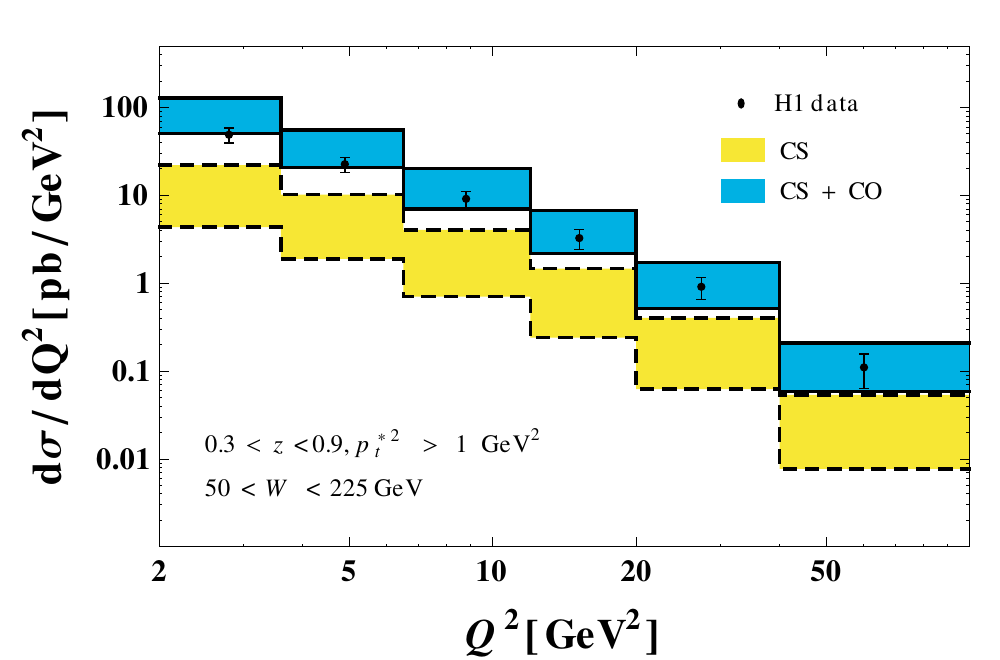}\\
\includegraphics[width=0.32\textwidth]{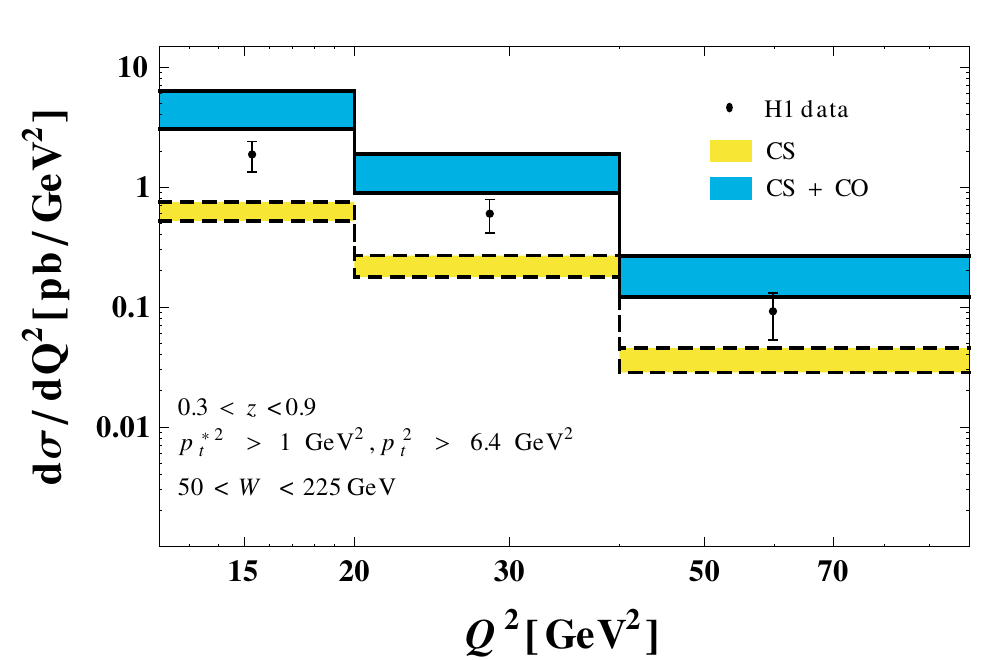}
\includegraphics[width=0.32\textwidth]{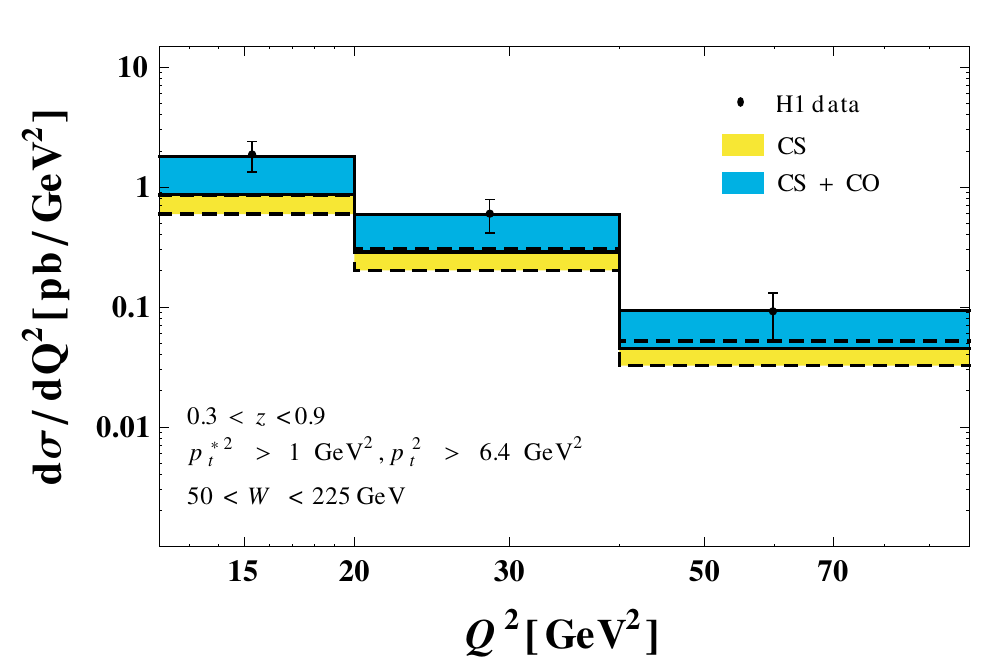}
\includegraphics[width=0.32\textwidth]{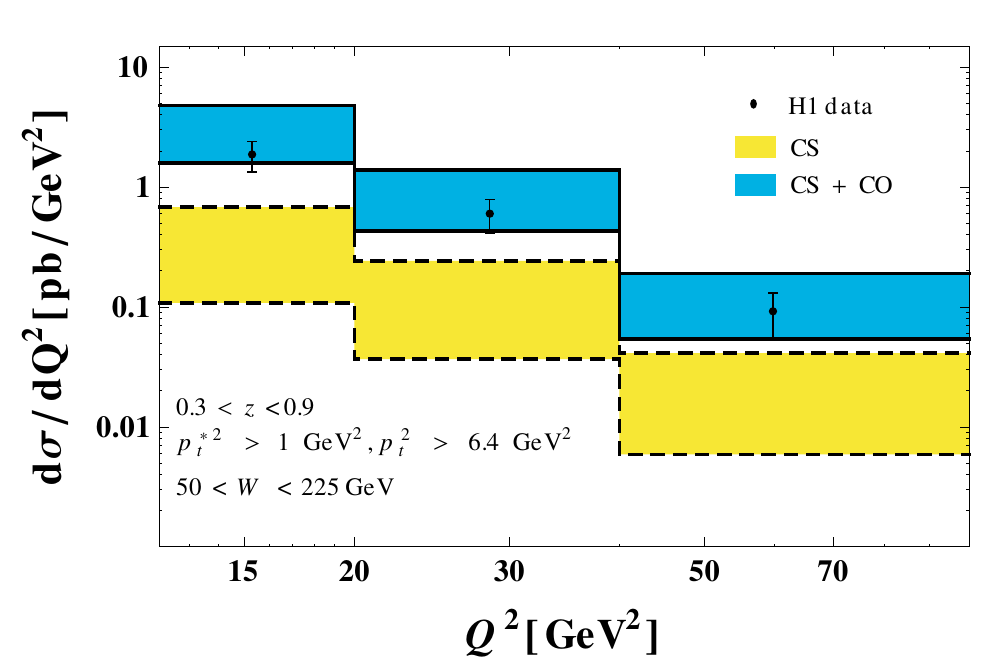}
\caption{\label{fig:q2-2002}
The differential cross sections for the $J/\psi$ production in DIS with respect to $Q^2$.
The experimental data are taken from Reference~\cite{Adloff:2002ey}.
The l.h.s., mid, and r.h.s plots correspond to the LDMEs taken in References~\cite{Chao:2012iv},~\cite{Butenschoen:2011yh}, and~\cite{Zhang:2014ybe}, respectively.
}
\end{figure}

\begin{figure}
\includegraphics[width=0.32\textwidth]{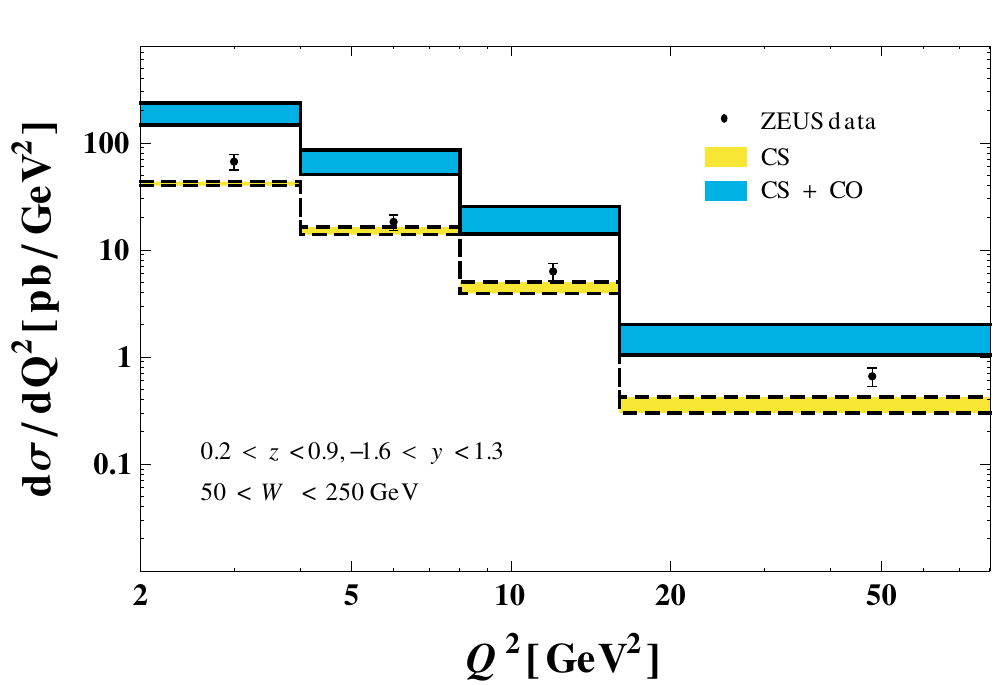}
\includegraphics[width=0.32\textwidth]{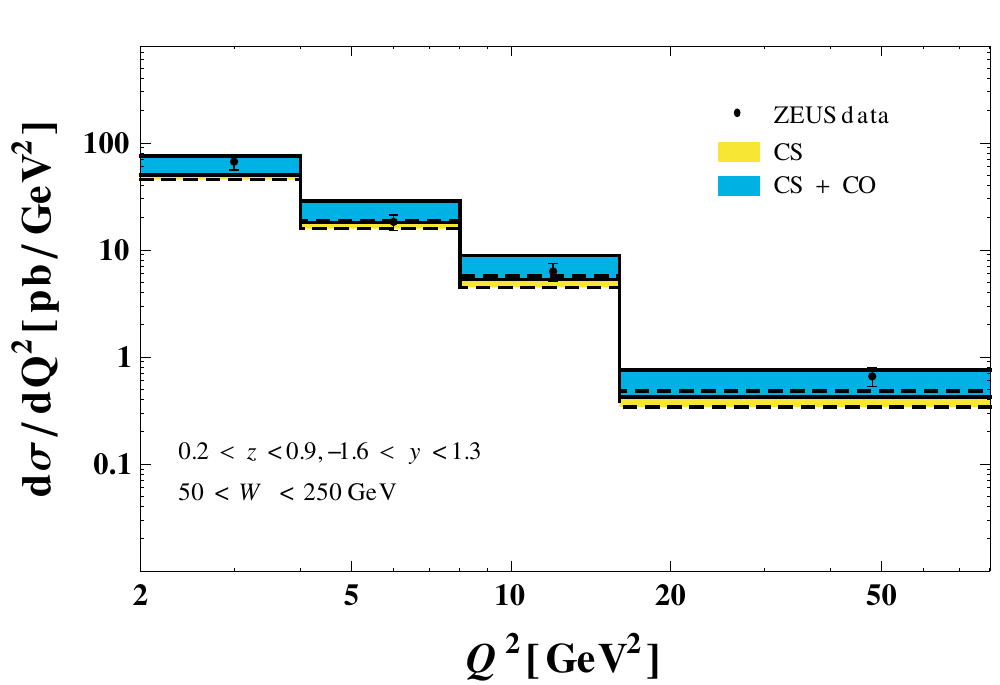}
\includegraphics[width=0.32\textwidth]{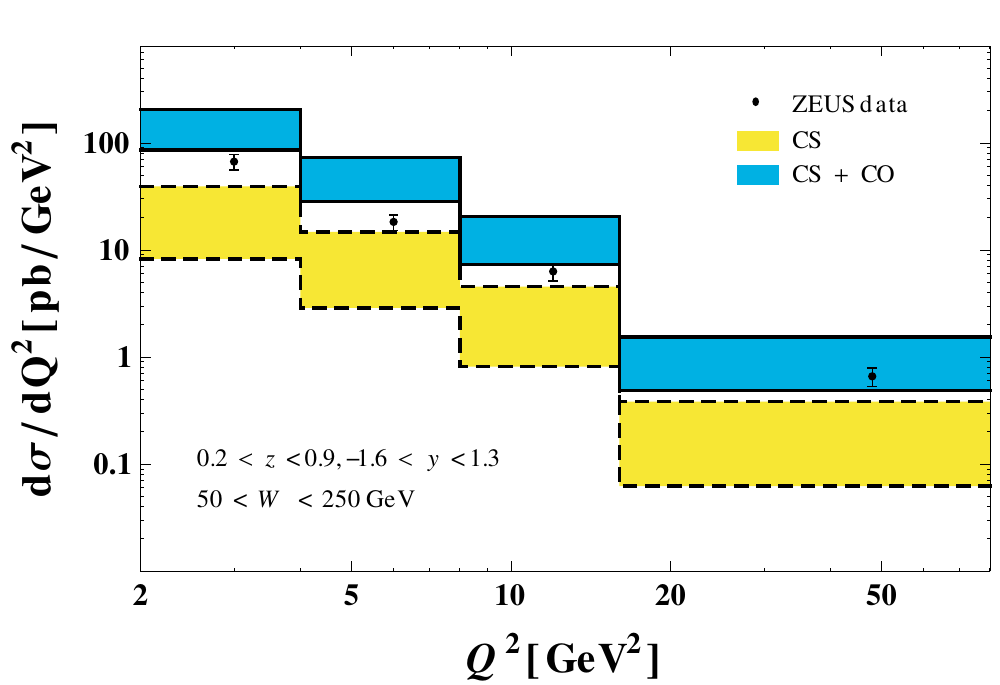}
\caption{\label{fig:q2-2005}
The differential cross sections for the $J/\psi$ production in DIS with respect to $Q^2$.
The experimental data are taken from Reference~\cite{Chekanov:2005cf}.
The l.h.s., mid, and r.h.s plots correspond to the LDMEs taken in References~\cite{Chao:2012iv},~\cite{Butenschoen:2011yh}, and~\cite{Zhang:2014ybe}, respectively.
}
\end{figure}

\begin{figure}
\includegraphics[width=0.32\textwidth]{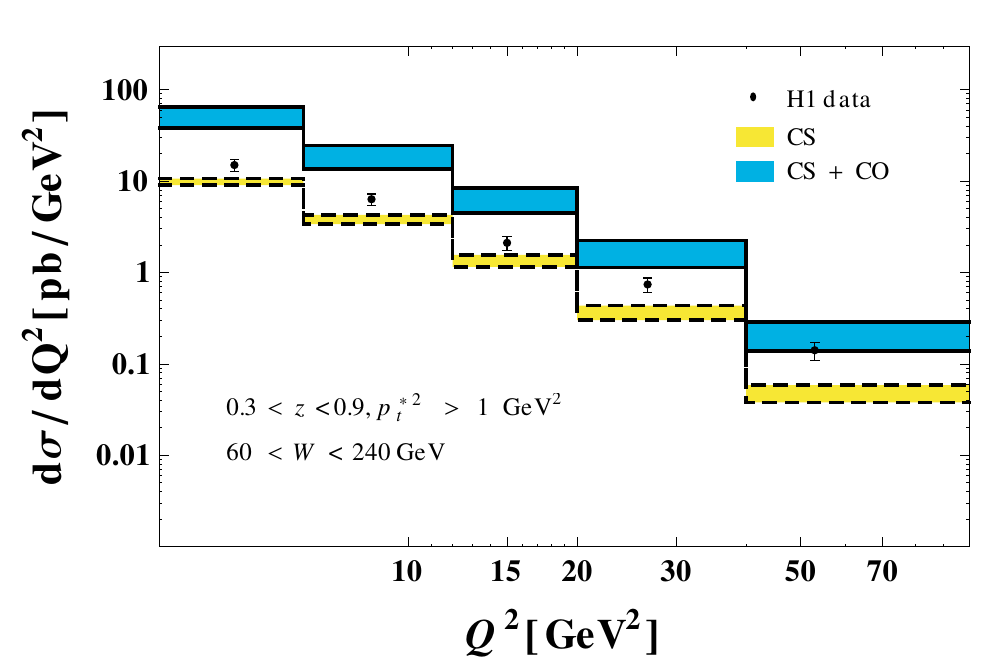}
\includegraphics[width=0.32\textwidth]{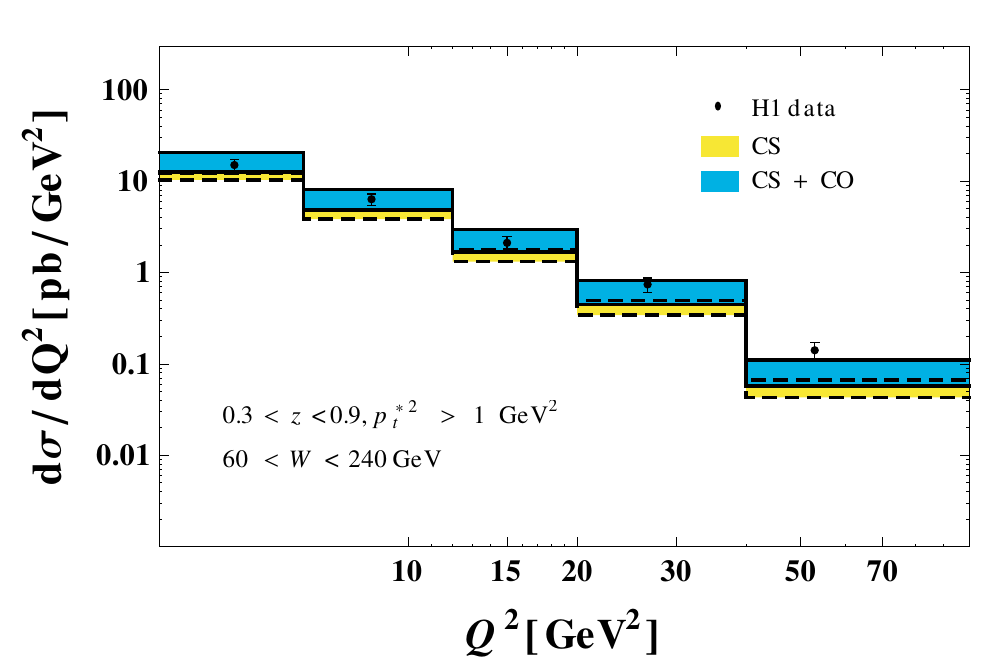}
\includegraphics[width=0.32\textwidth]{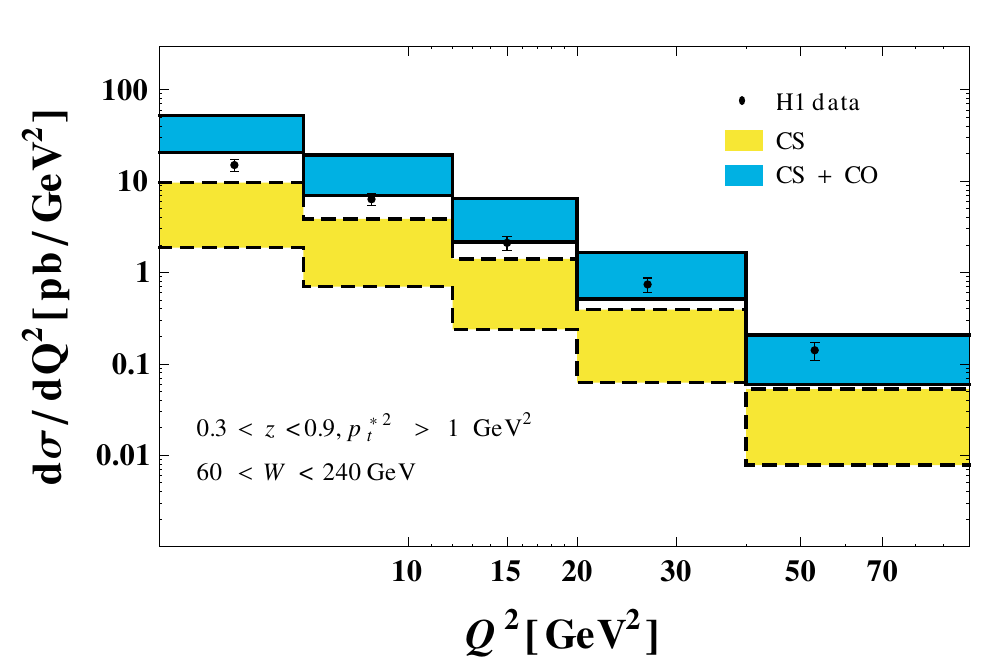}
\caption{\label{fig:q2-2010}
The differential cross sections for the $J/\psi$ production in DIS with respect to $Q^2$.
The experimental data are taken from Reference~\cite{Aaron:2010gz}.
The l.h.s., mid, and r.h.s plots correspond to the LDMEs taken in References~\cite{Chao:2012iv},~\cite{Butenschoen:2011yh}, and~\cite{Zhang:2014ybe}, respectively.
}
\end{figure}

\begin{figure}
\includegraphics[width=0.32\textwidth]{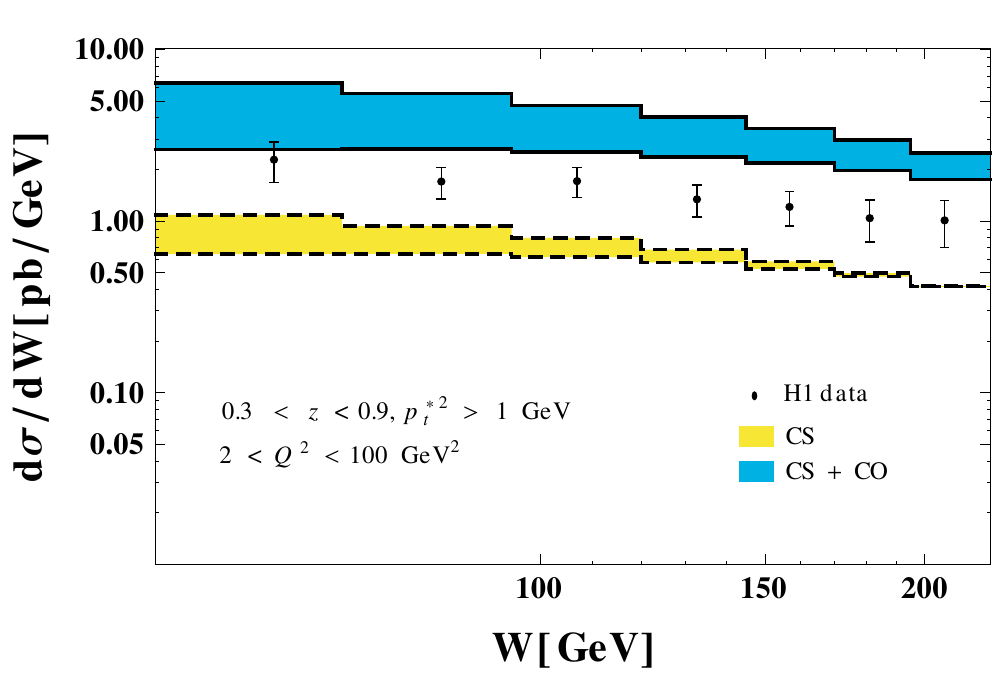}
\includegraphics[width=0.32\textwidth]{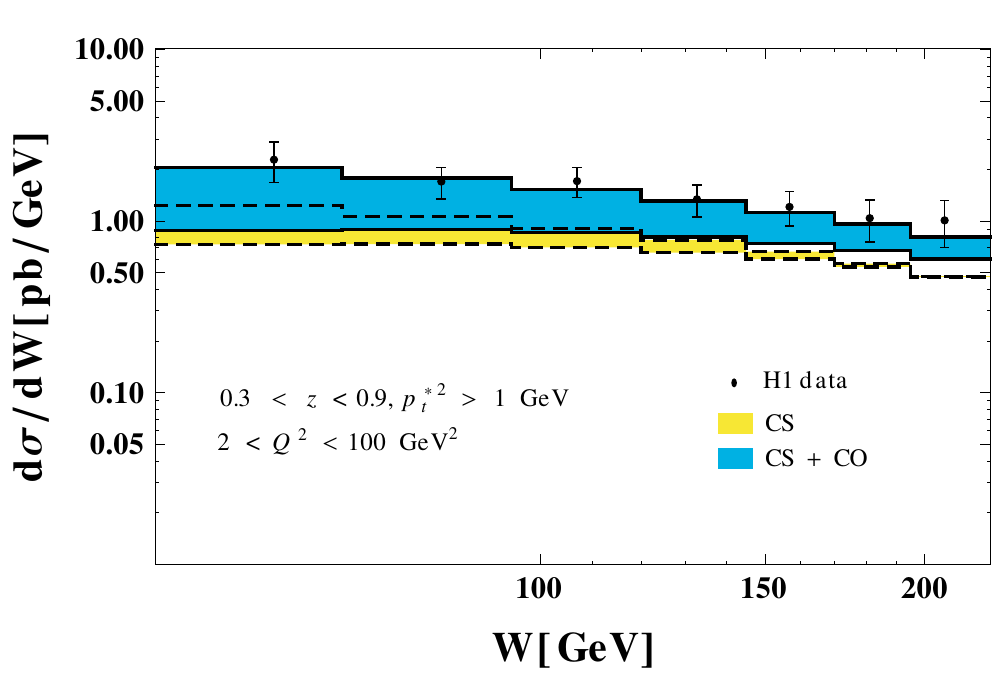}
\includegraphics[width=0.32\textwidth]{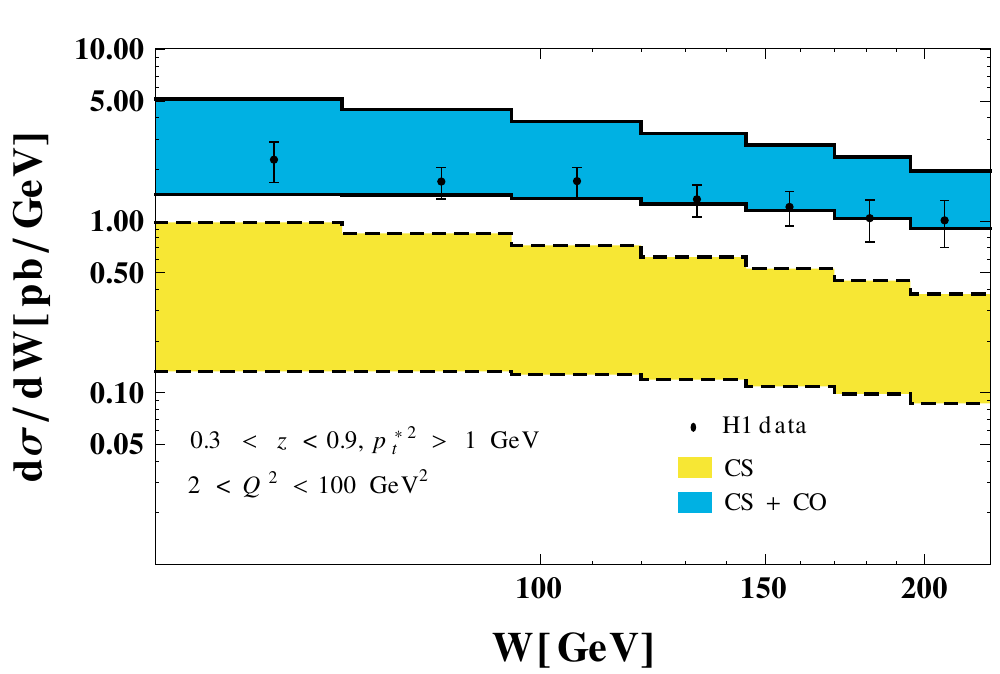}\\
\includegraphics[width=0.32\textwidth]{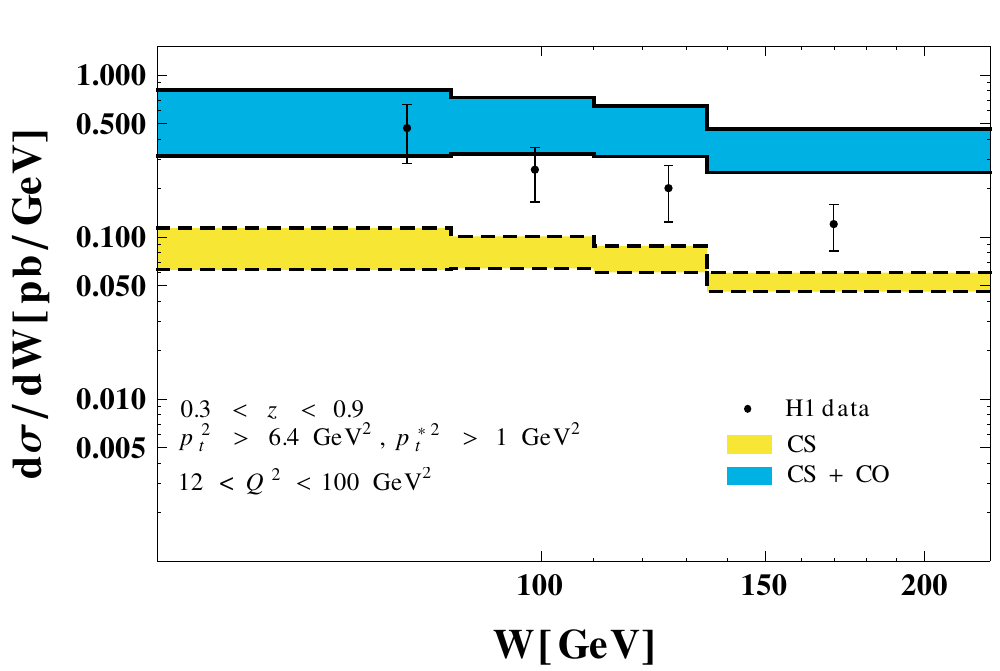}
\includegraphics[width=0.32\textwidth]{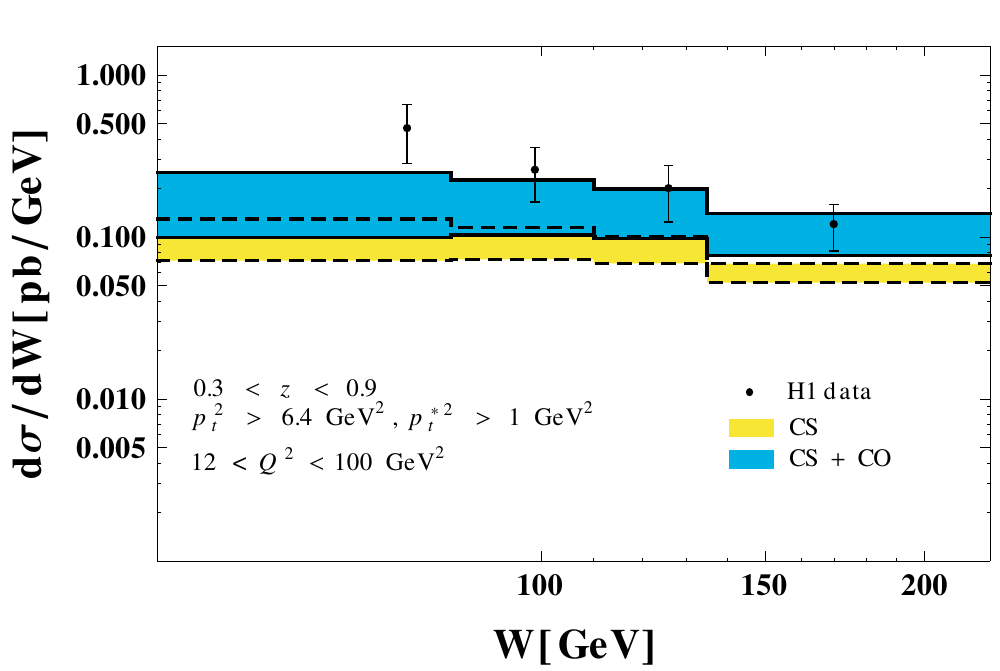}
\includegraphics[width=0.32\textwidth]{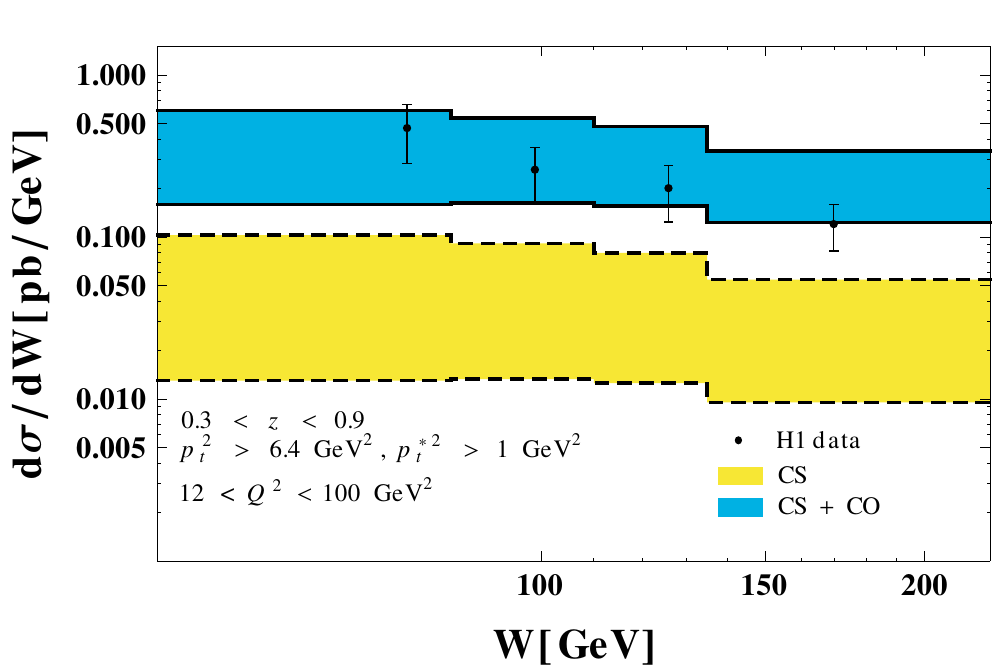}
\caption{\label{fig:w-2002}
The differential cross sections for the $J/\psi$ production in DIS with respect to $W$.
The experimental data are taken from Reference~\cite{Adloff:2002ey}.
The l.h.s., mid, and r.h.s plots correspond to the LDMEs taken in References~\cite{Chao:2012iv},~\cite{Butenschoen:2011yh}, and~\cite{Zhang:2014ybe}, respectively.
}
\end{figure}

\begin{figure}
\includegraphics[width=0.32\textwidth]{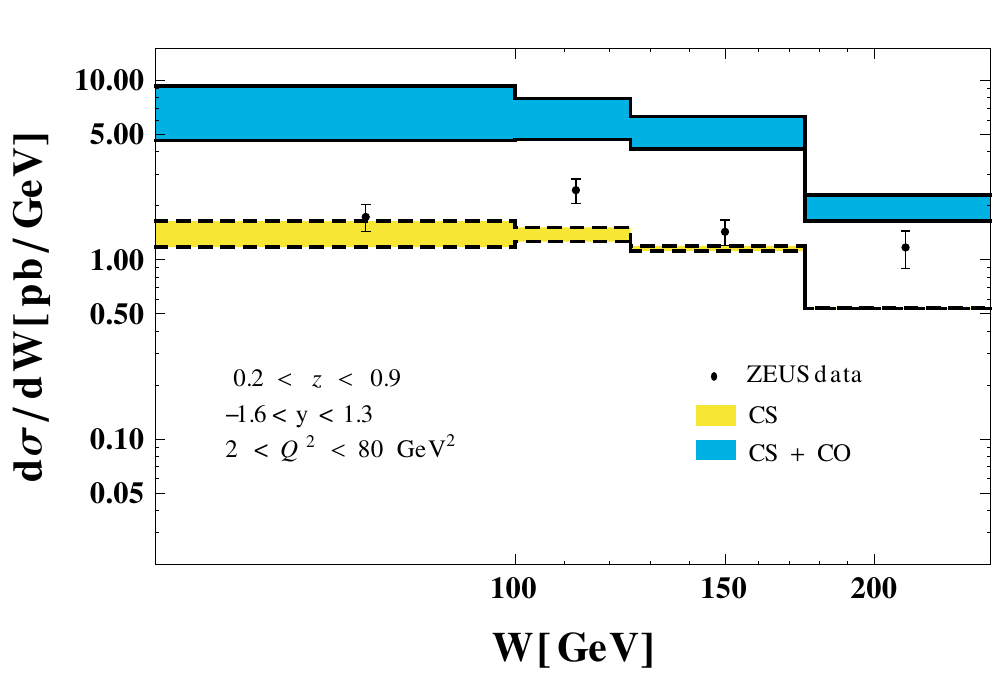}
\includegraphics[width=0.32\textwidth]{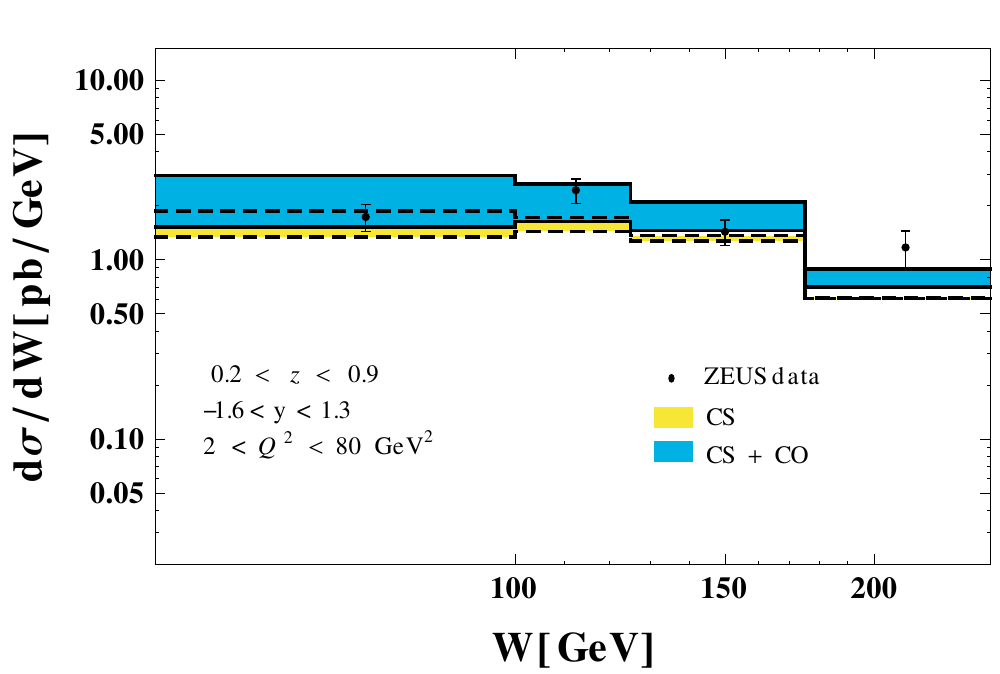}
\includegraphics[width=0.32\textwidth]{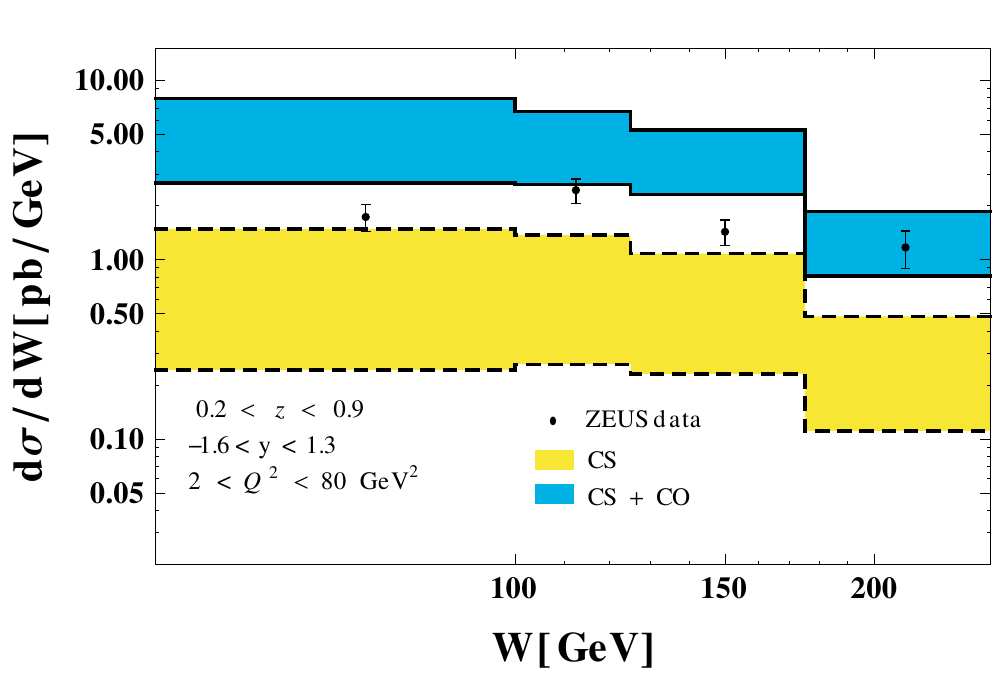}
\caption{\label{fig:w-2005}
The differential cross sections for the $J/\psi$ production in DIS with respect to $W$.
The experimental data are taken from Reference~\cite{Chekanov:2005cf}.
The l.h.s., mid, and r.h.s plots correspond to the LDMEs taken in References~\cite{Chao:2012iv},~\cite{Butenschoen:2011yh}, and~\cite{Zhang:2014ybe}, respectively.
}
\end{figure}

\begin{figure}
\includegraphics[width=0.32\textwidth]{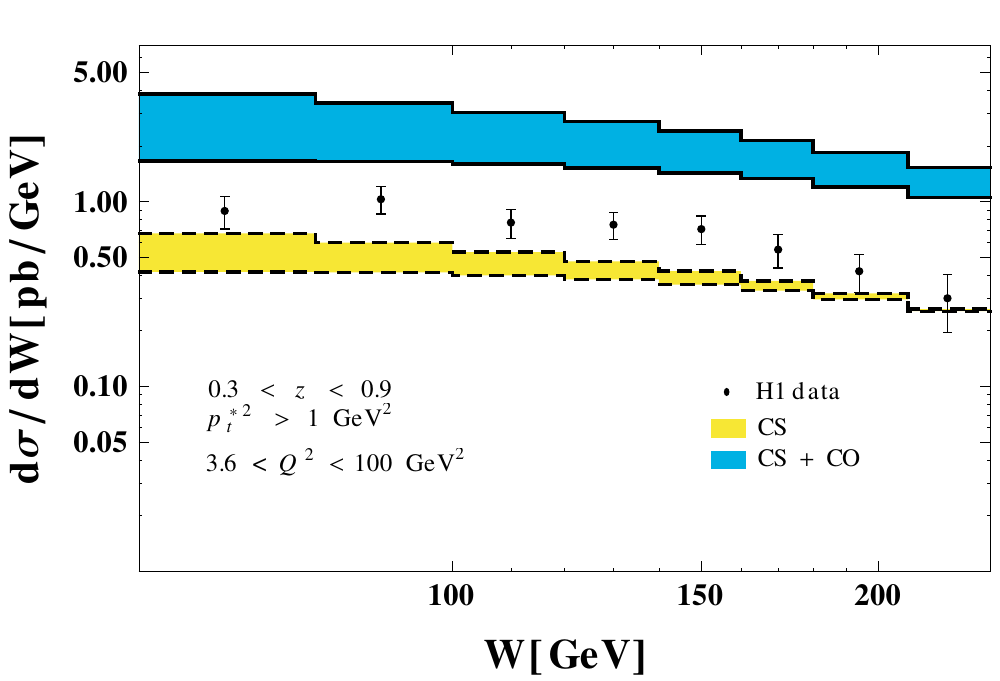}
\includegraphics[width=0.32\textwidth]{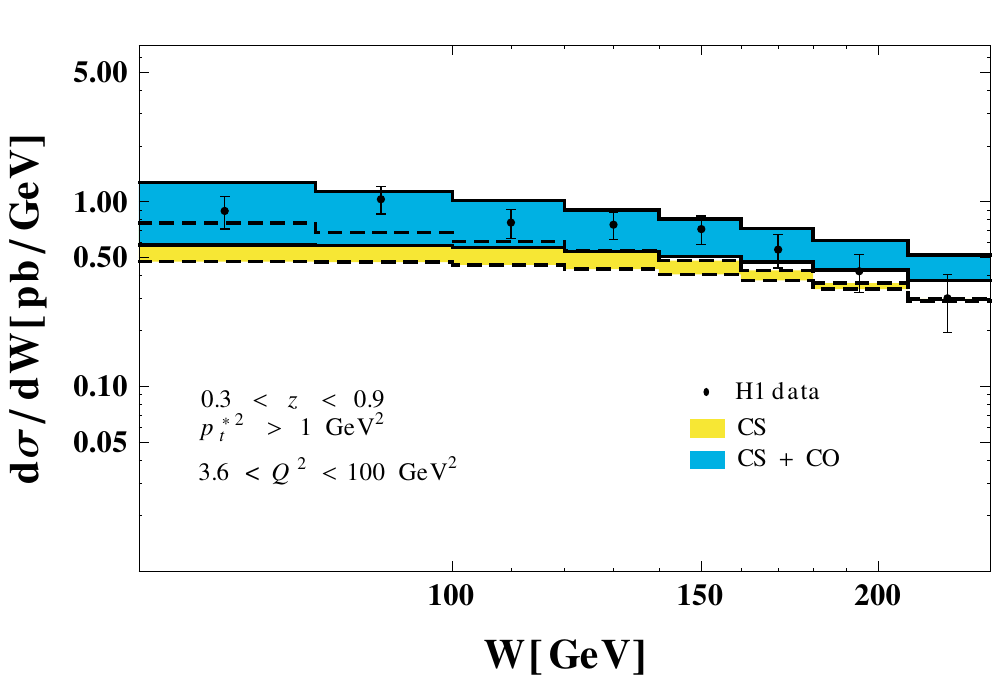}
\includegraphics[width=0.32\textwidth]{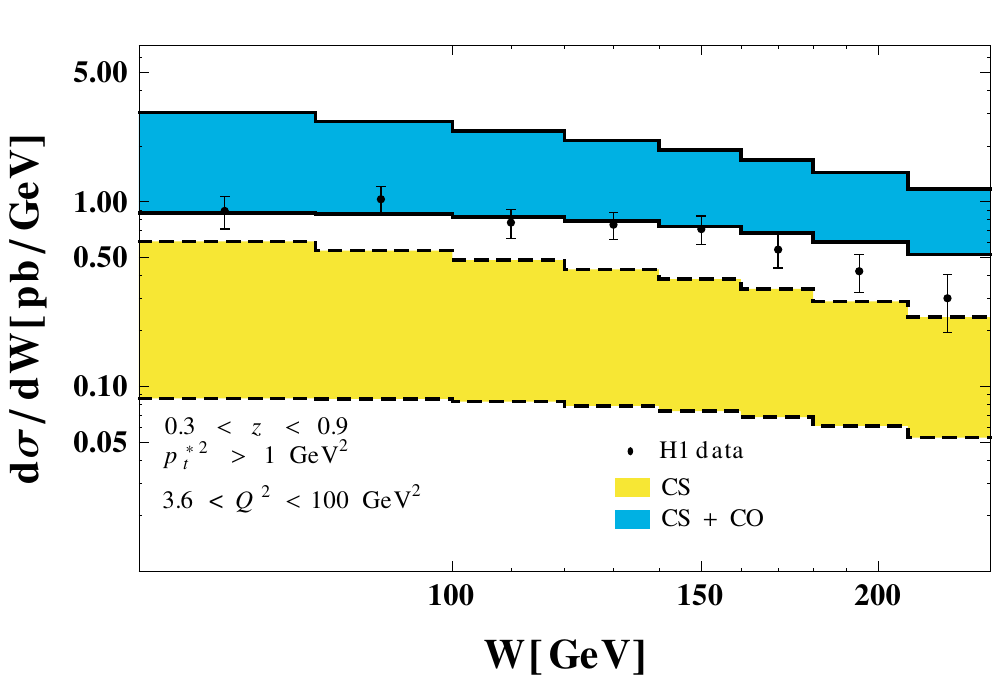}
\caption{\label{fig:w-2010}
The differential cross sections for the $J/\psi$ production in DIS with respect to $W$.
The experimental data are taken from Reference~\cite{Aaron:2010gz}.
The l.h.s., mid, and r.h.s plots correspond to the LDMEs taken in References~\cite{Chao:2012iv},~\cite{Butenschoen:2011yh}, and~\cite{Zhang:2014ybe}, respectively.
}
\end{figure}

\begin{figure}
\includegraphics[width=0.32\textwidth]{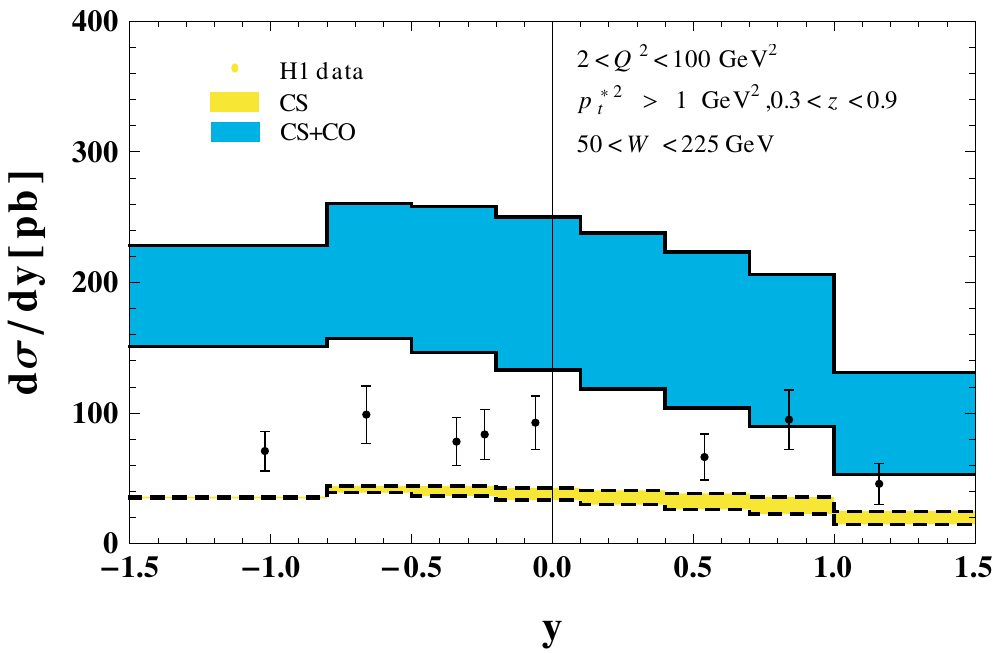}
\includegraphics[width=0.32\textwidth]{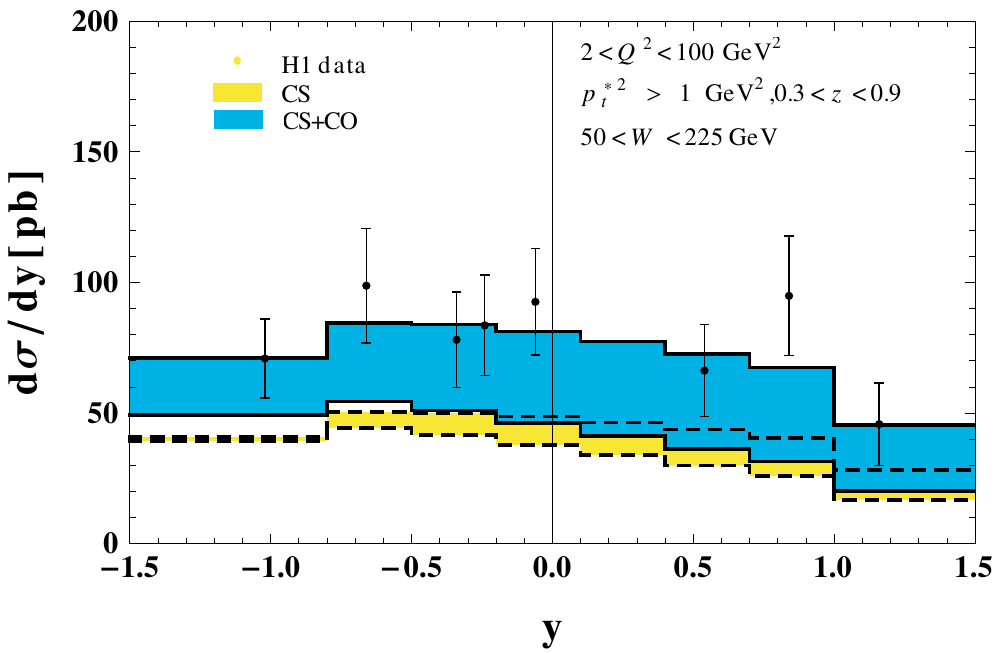}
\includegraphics[width=0.32\textwidth]{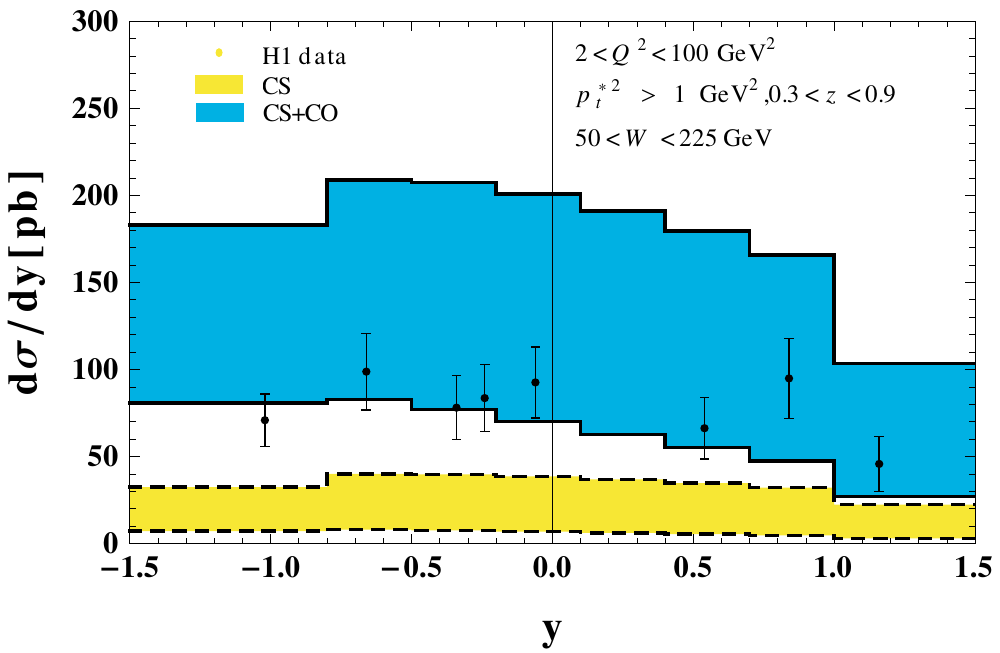}
\caption{\label{fig:y-2002}
The differential cross sections for the $J/\psi$ production in DIS with respect to $y_\psi$.
The experimental data are taken from Reference~\cite{Adloff:2002ey}.
The l.h.s., mid, and r.h.s plots correspond to the LDMEs taken in References~\cite{Chao:2012iv},~\cite{Butenschoen:2011yh}, and~\cite{Zhang:2014ybe}, respectively.
}
\end{figure}

\begin{figure}
\includegraphics[width=0.32\textwidth]{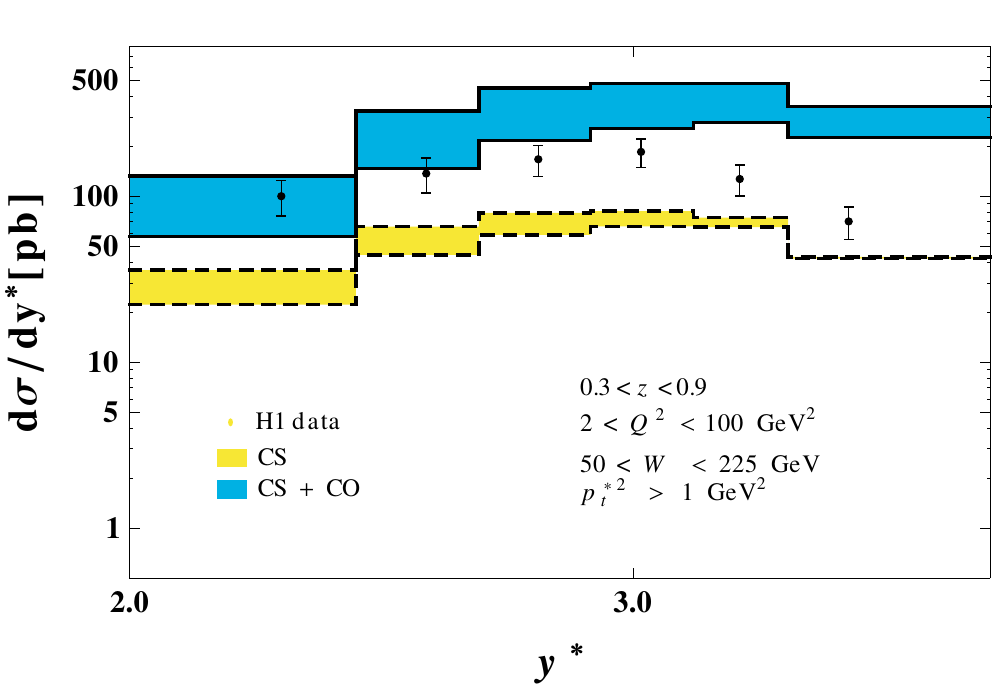}
\includegraphics[width=0.32\textwidth]{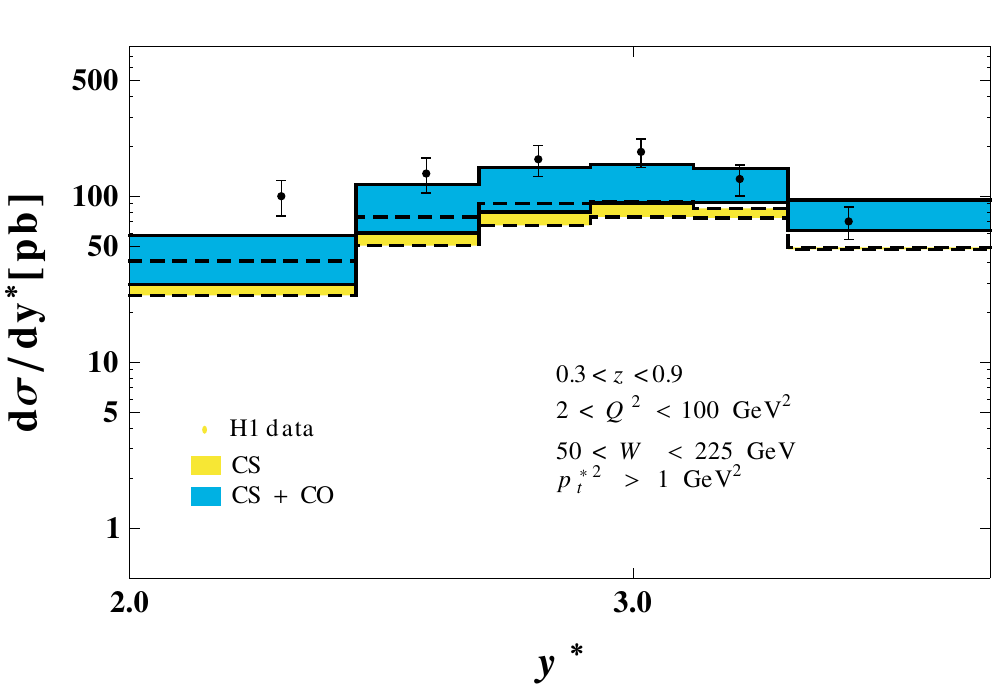}
\includegraphics[width=0.32\textwidth]{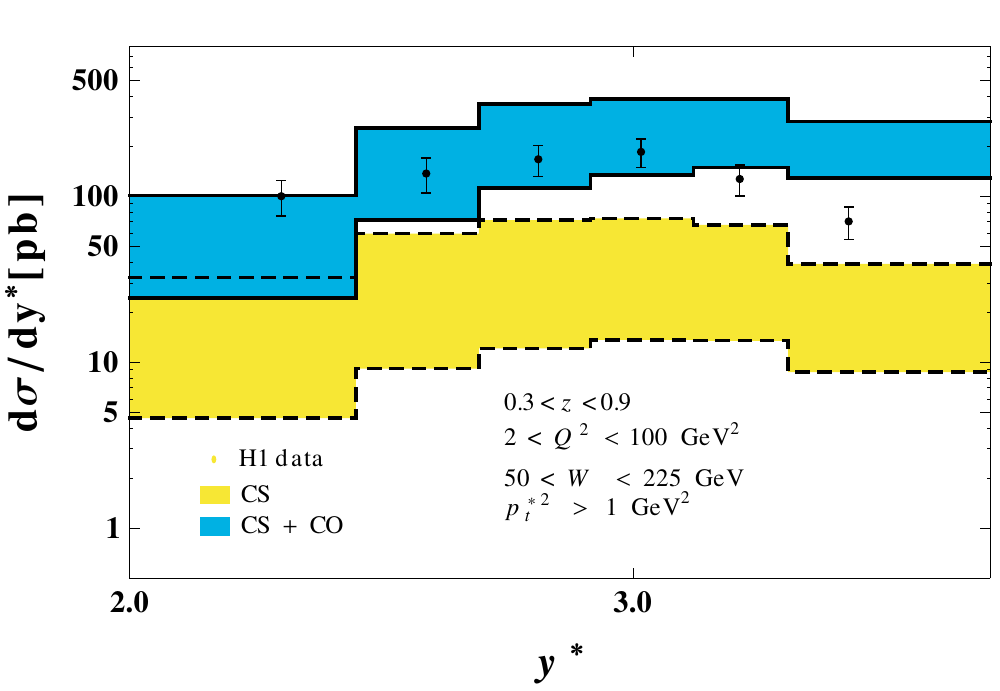}\\
\includegraphics[width=0.32\textwidth]{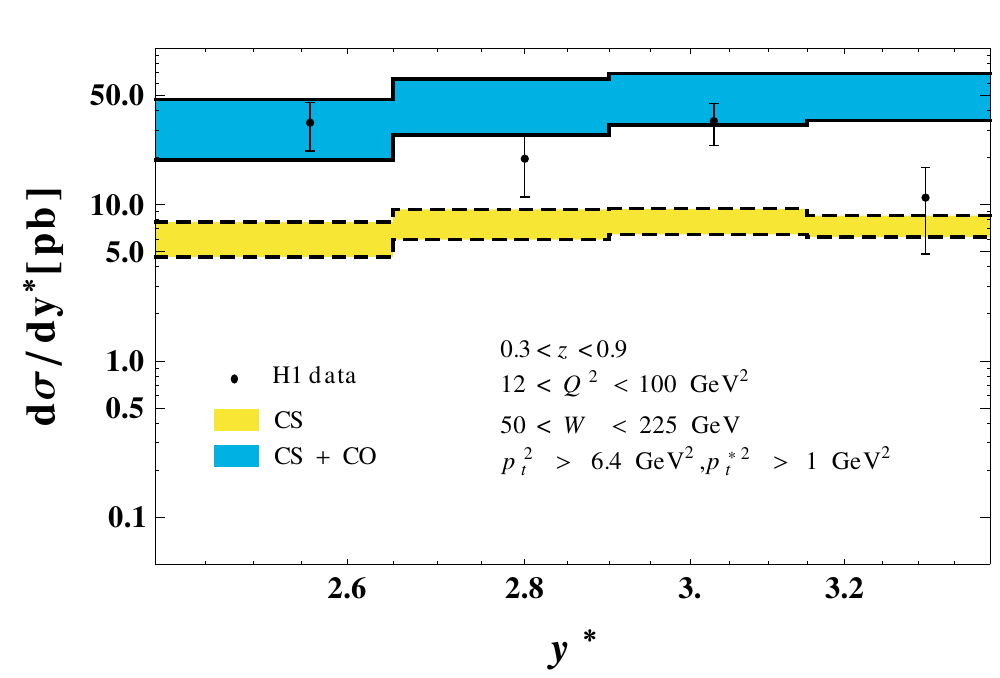}
\includegraphics[width=0.32\textwidth]{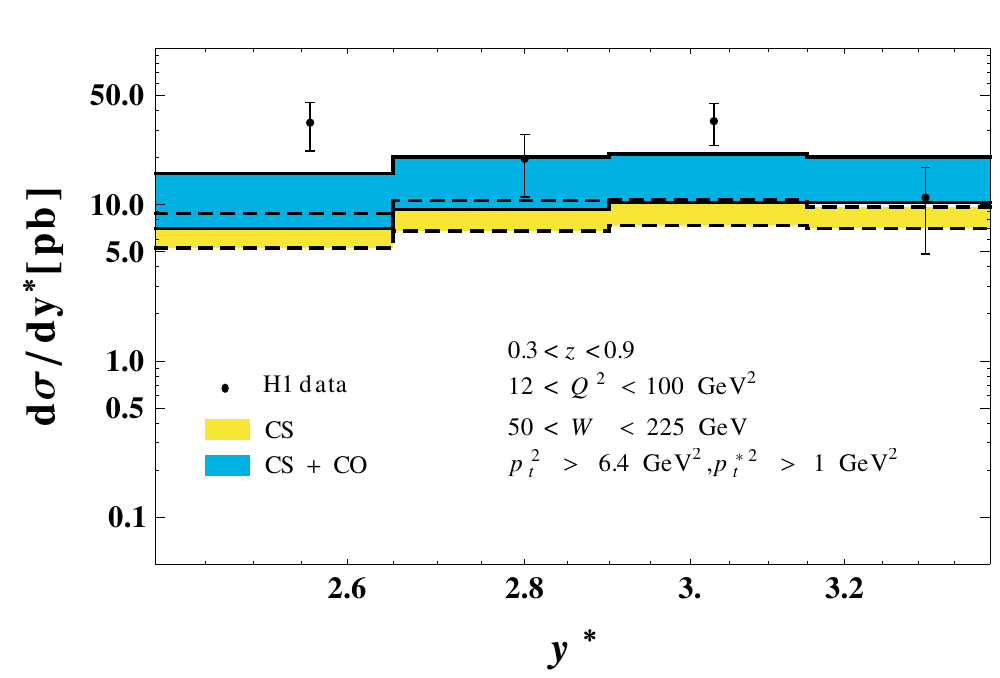}
\includegraphics[width=0.32\textwidth]{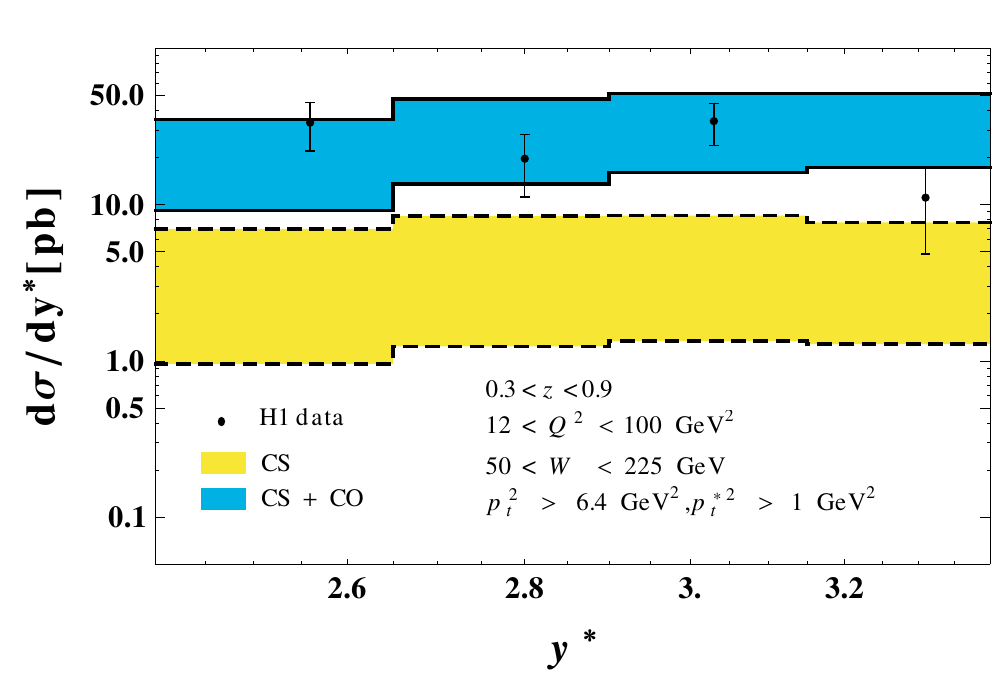}
\caption{\label{fig:ys-2002}
The differential cross sections for the $J/\psi$ production in DIS with respect to $y_\psi^\star$.
The experimental data are taken from Reference~\cite{Adloff:2002ey}.
The l.h.s., mid, and r.h.s plots correspond to the LDMEs taken in References~\cite{Chao:2012iv},~\cite{Butenschoen:2011yh}, and~\cite{Zhang:2014ybe}, respectively.
}
\end{figure}

\begin{figure}
\includegraphics[width=0.32\textwidth]{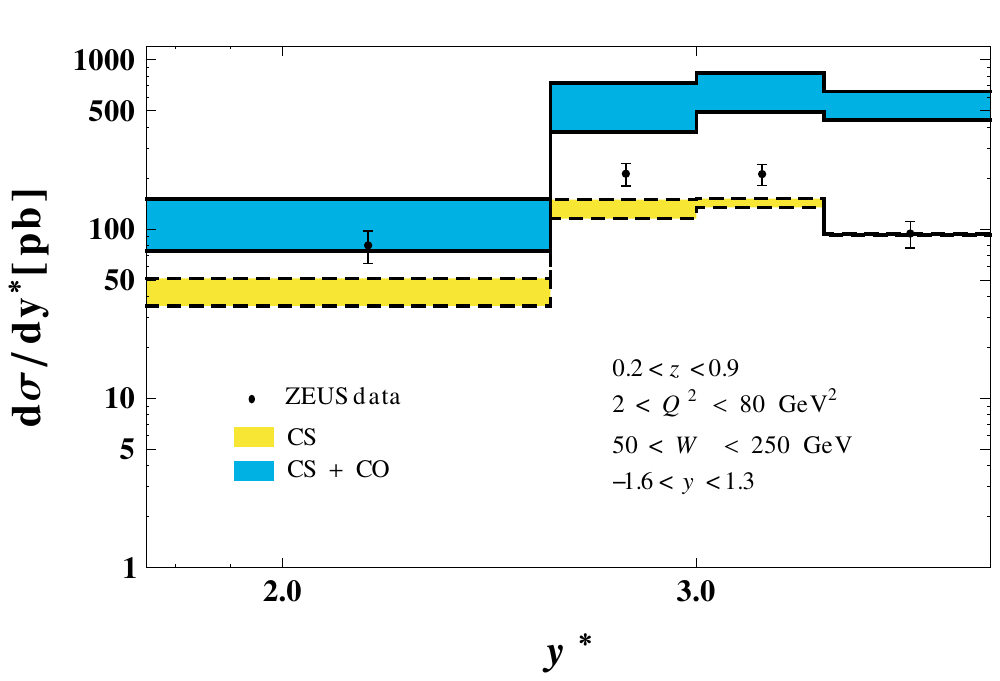}
\includegraphics[width=0.32\textwidth]{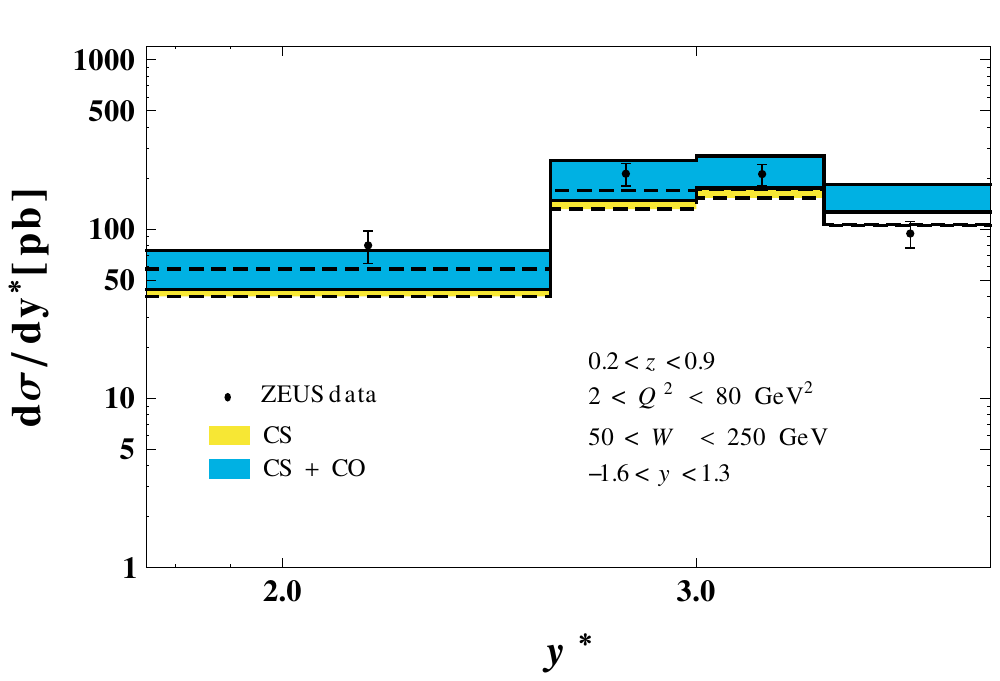}
\includegraphics[width=0.32\textwidth]{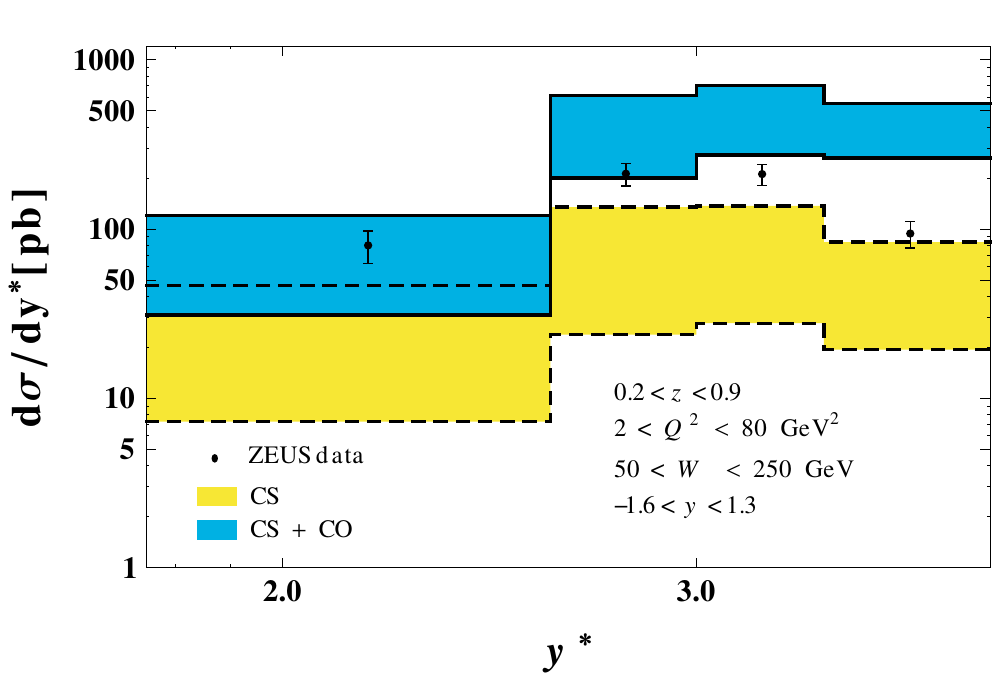}\\
\includegraphics[width=0.32\textwidth]{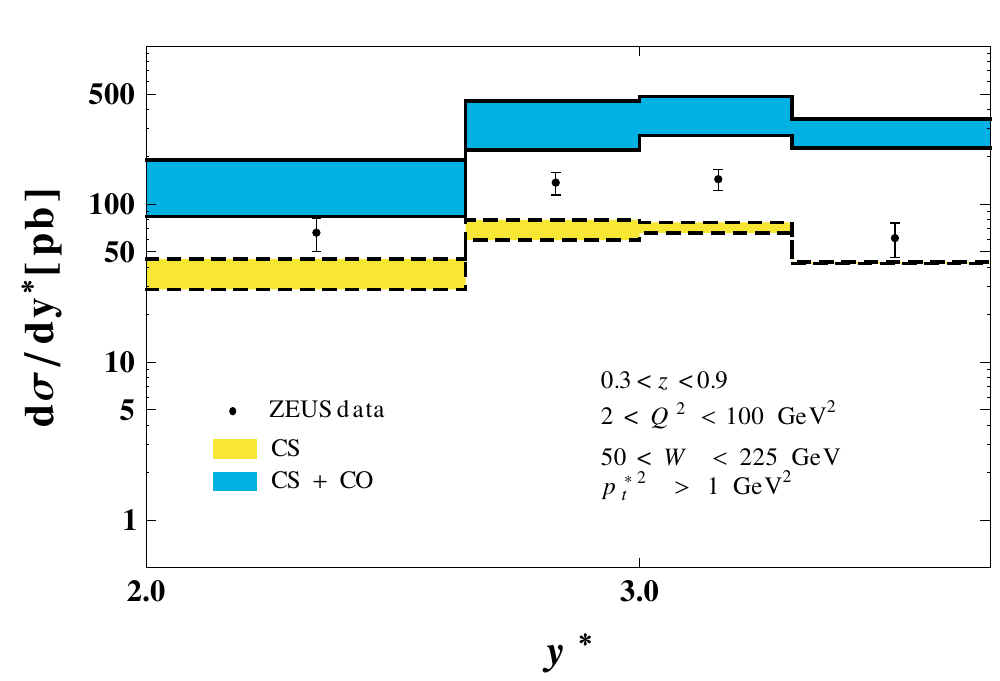}
\includegraphics[width=0.32\textwidth]{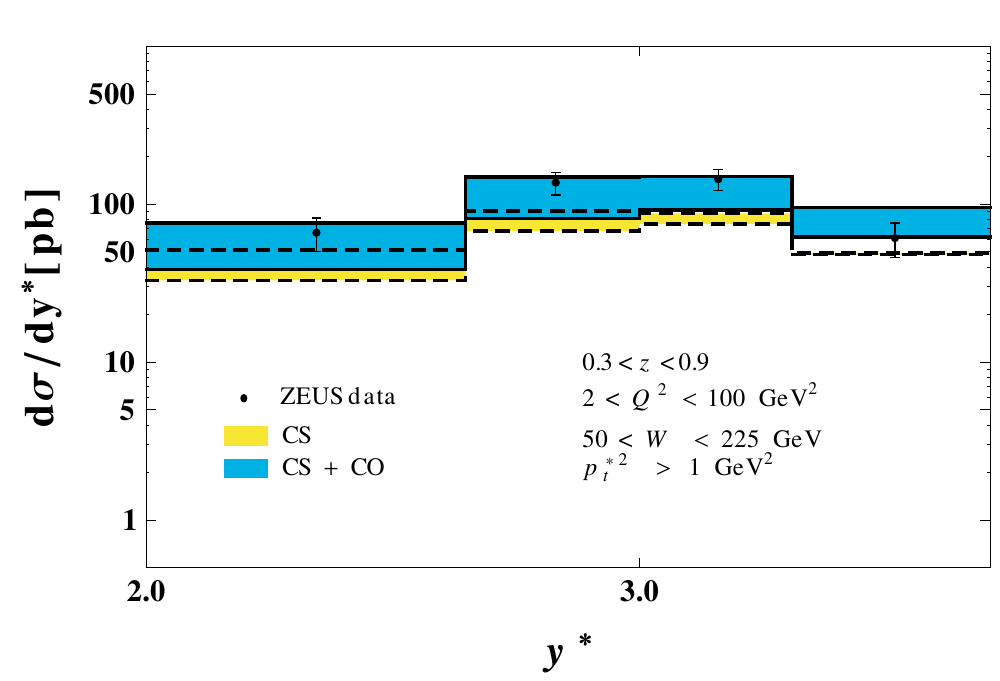}
\includegraphics[width=0.32\textwidth]{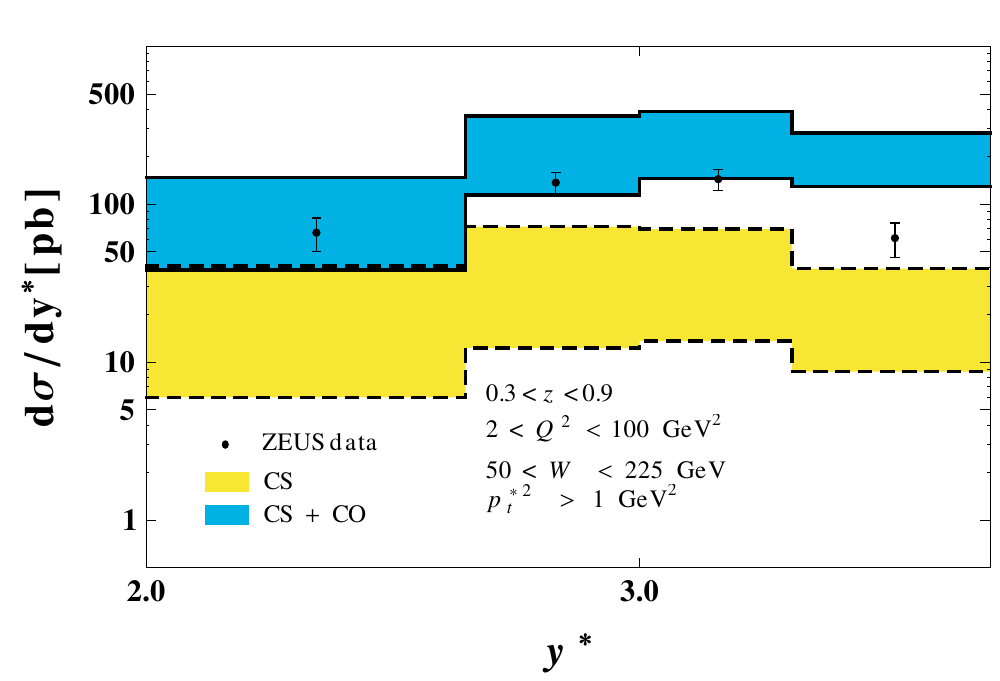}
\caption{\label{fig:ys-2005}
The differential cross sections for the $J/\psi$ production in DIS with respect to $y_\psi^\star$.
The experimental data are taken from Reference~\cite{Chekanov:2005cf}.
The l.h.s., mid, and r.h.s plots correspond to the LDMEs taken in References~\cite{Chao:2012iv},~\cite{Butenschoen:2011yh}, and~\cite{Zhang:2014ybe}, respectively.
}
\end{figure}

\begin{figure}
\includegraphics[width=0.32\textwidth]{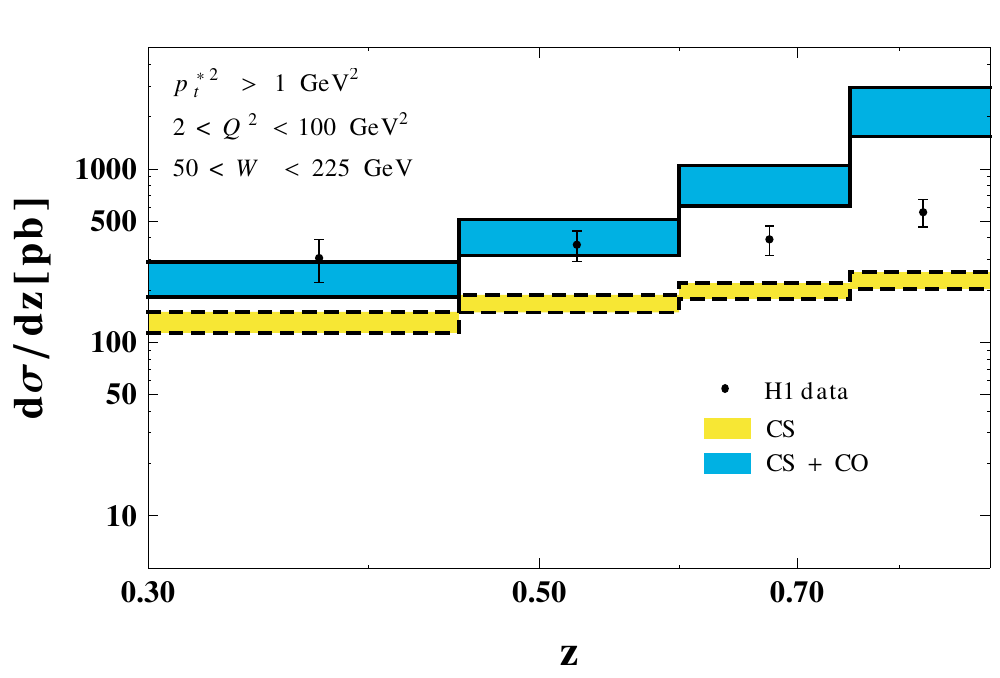}
\includegraphics[width=0.32\textwidth]{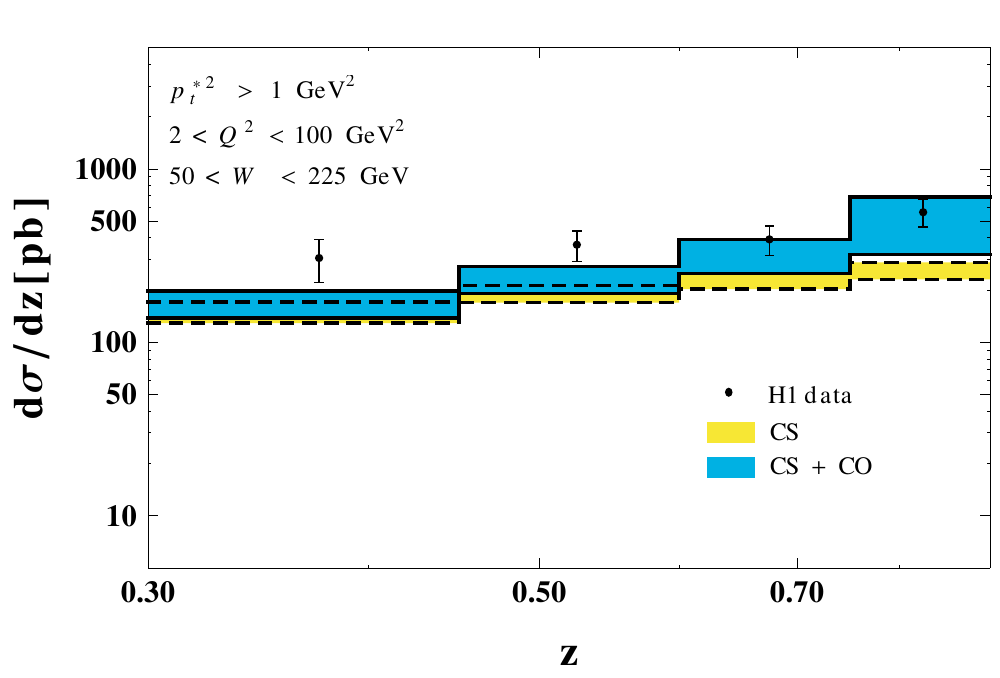}
\includegraphics[width=0.32\textwidth]{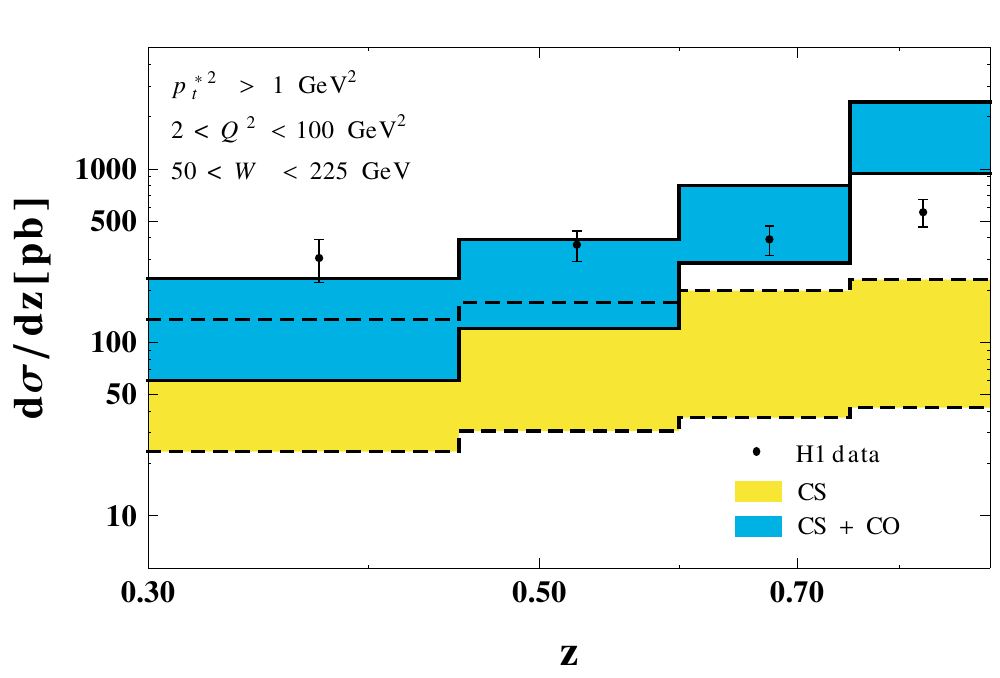}\\
\includegraphics[width=0.32\textwidth]{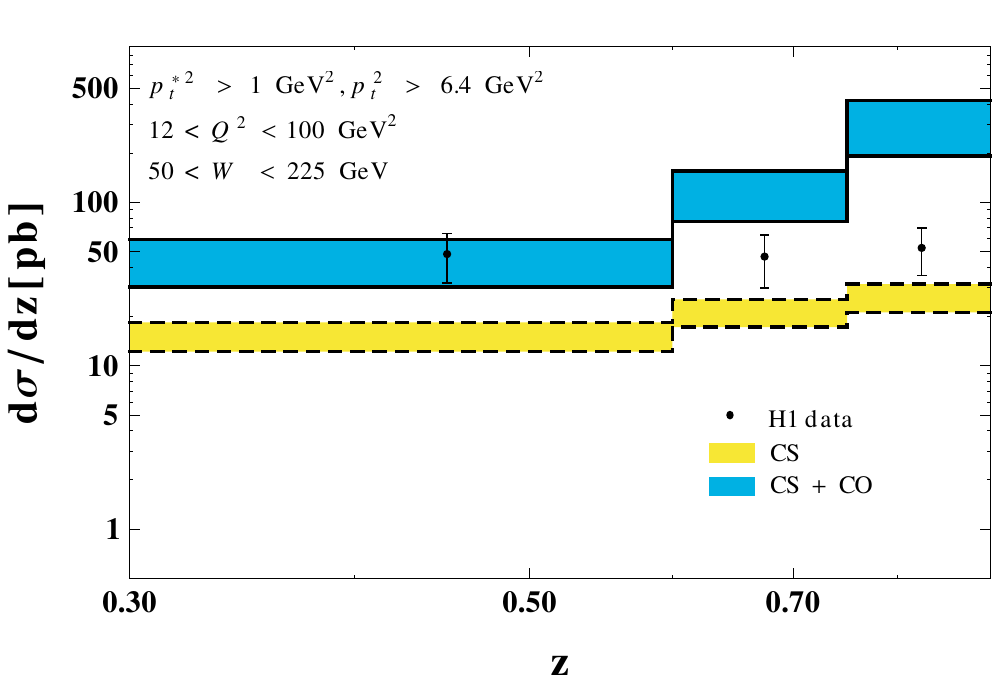}
\includegraphics[width=0.32\textwidth]{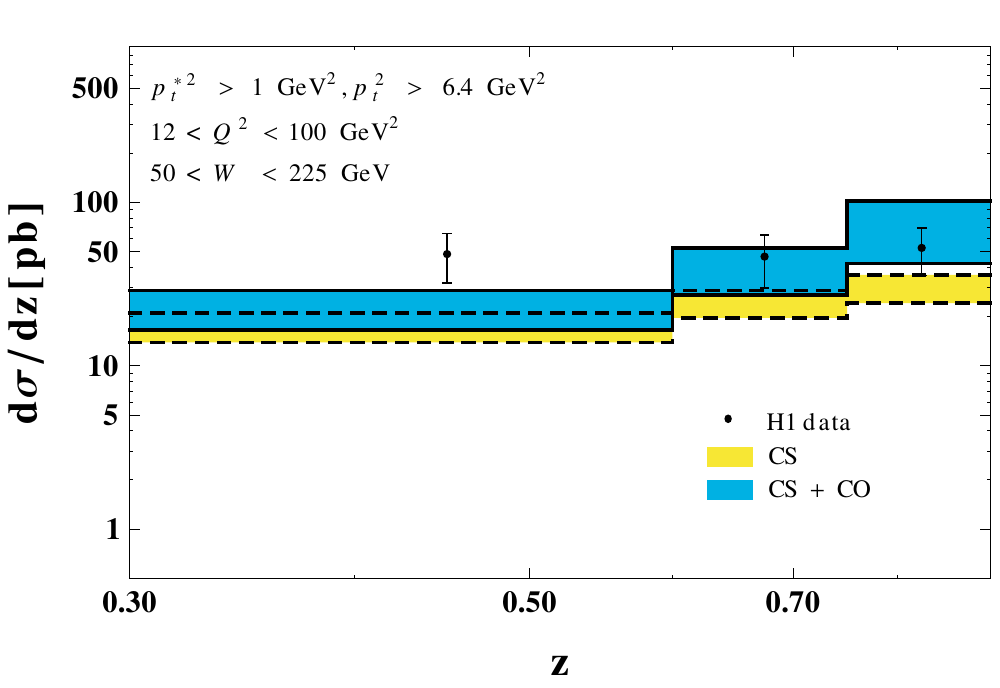}
\includegraphics[width=0.32\textwidth]{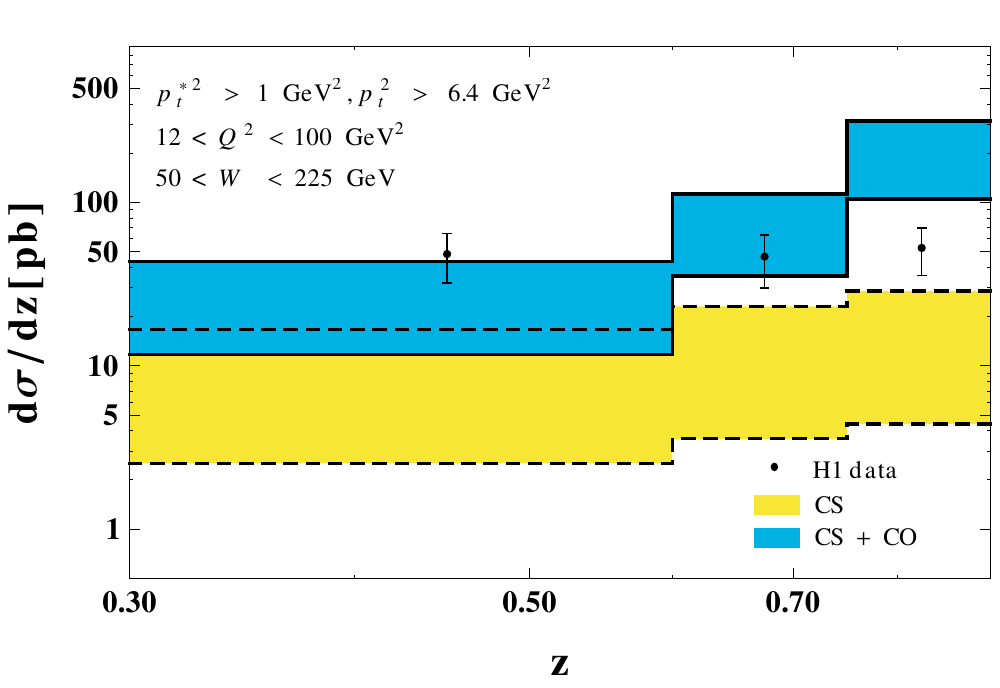}\\
\includegraphics[width=0.32\textwidth]{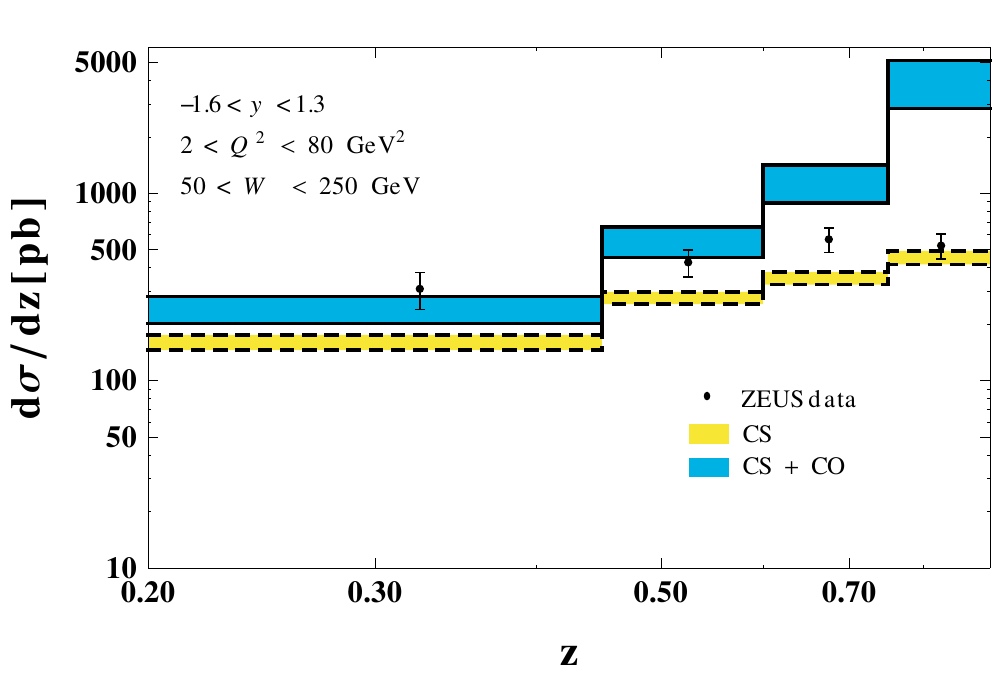}
\includegraphics[width=0.32\textwidth]{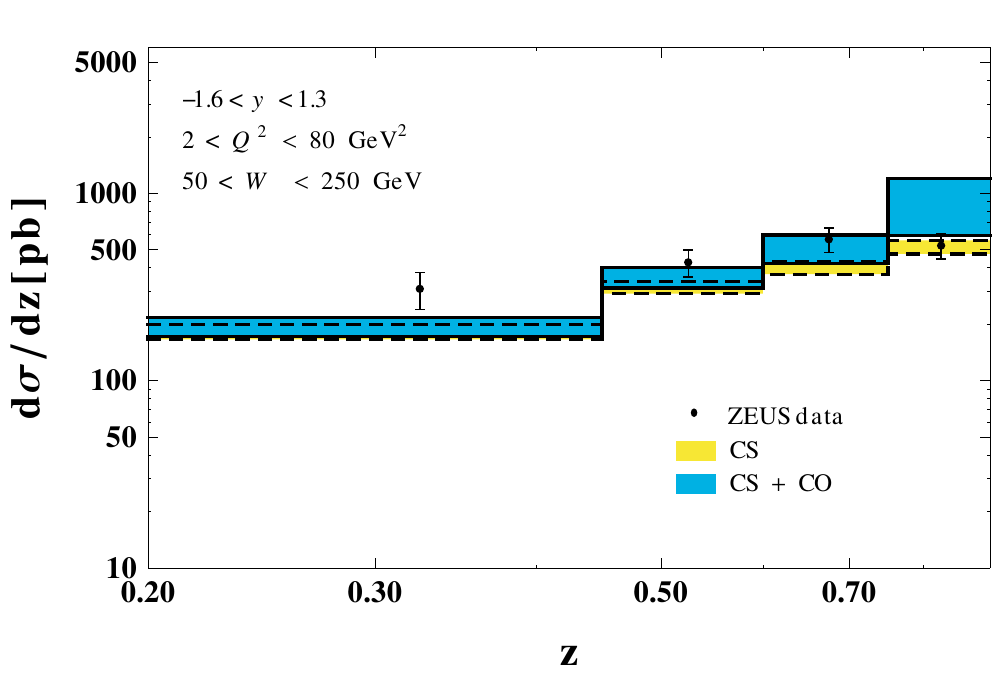}
\includegraphics[width=0.32\textwidth]{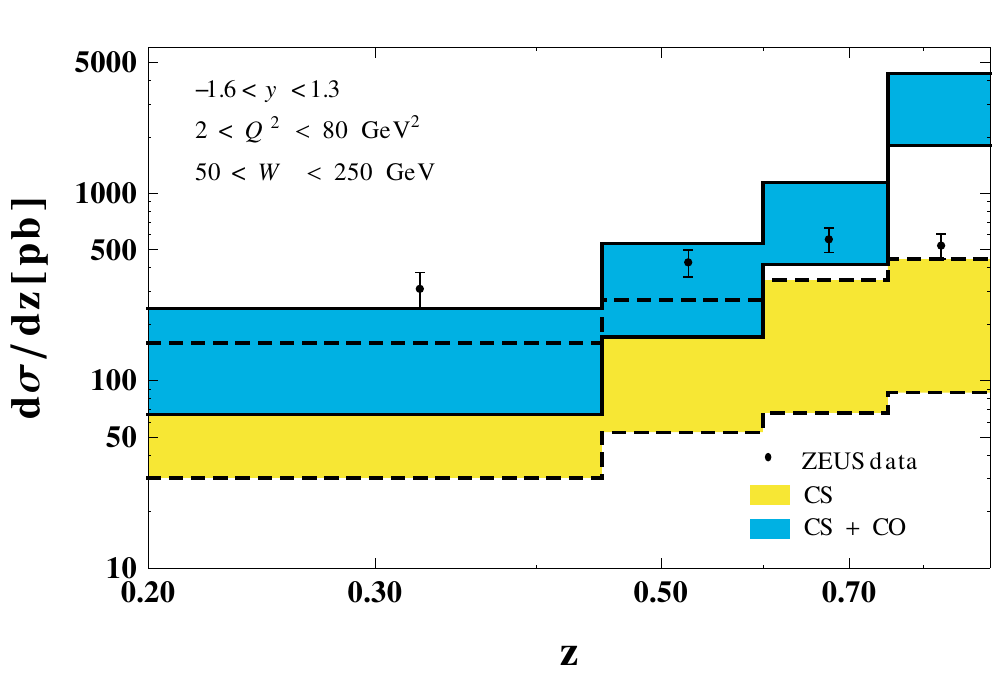}\\
\includegraphics[width=0.32\textwidth]{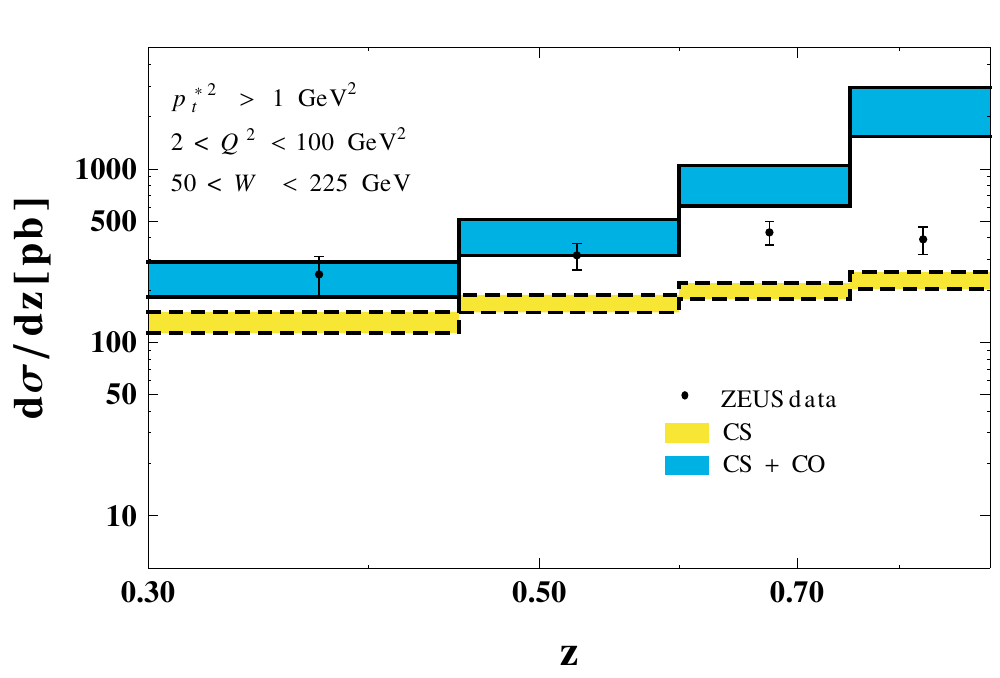}
\includegraphics[width=0.32\textwidth]{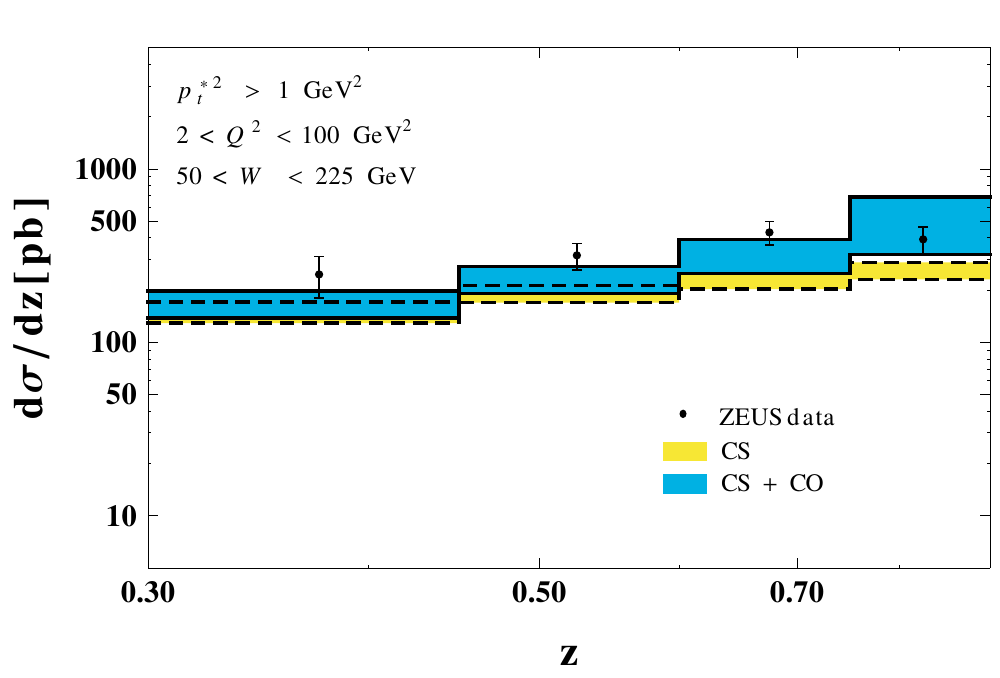}
\includegraphics[width=0.32\textwidth]{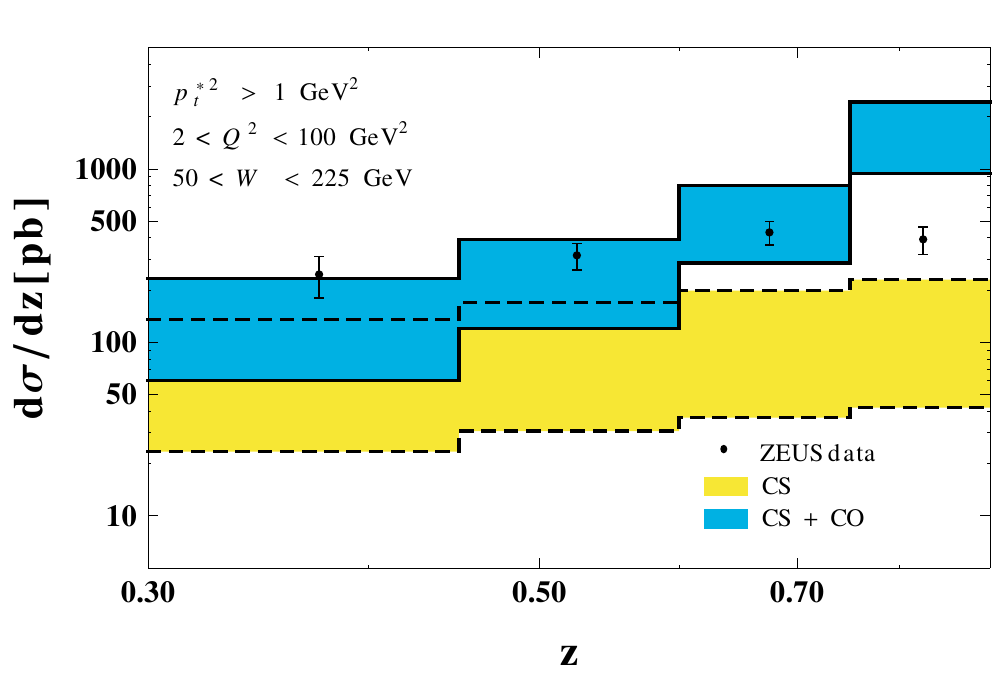}
\caption{\label{fig:z-2002}
The differential cross sections for the $J/\psi$ production in DIS with respect to $y_\psi^\star$.
The experimental data are taken from Reference~\cite{Adloff:2002ey}.
The l.h.s., mid, and r.h.s plots correspond to the LDMEs taken in References~\cite{Chao:2012iv},~\cite{Butenschoen:2011yh}, and~\cite{Zhang:2014ybe}, respectively.
}
\end{figure}

\begin{figure}
\includegraphics[width=0.32\textwidth]{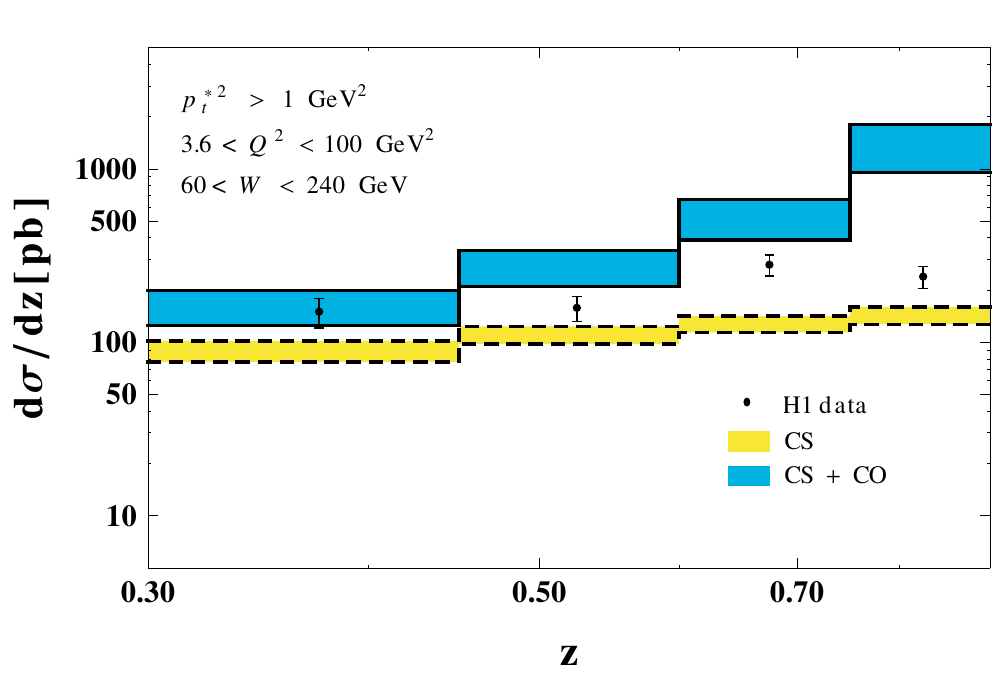}
\includegraphics[width=0.32\textwidth]{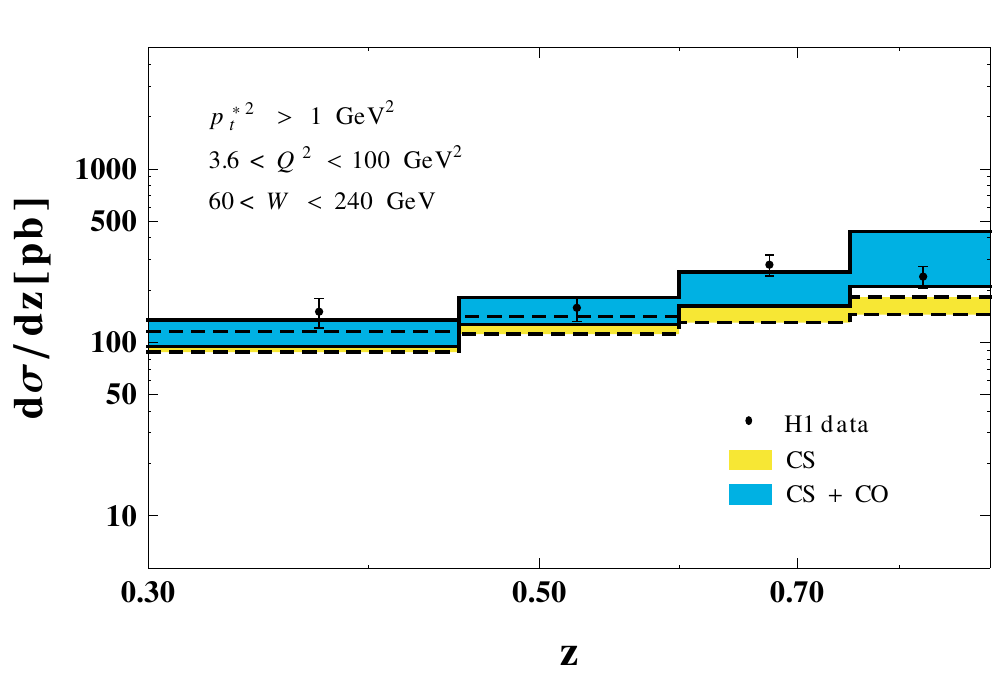}
\includegraphics[width=0.32\textwidth]{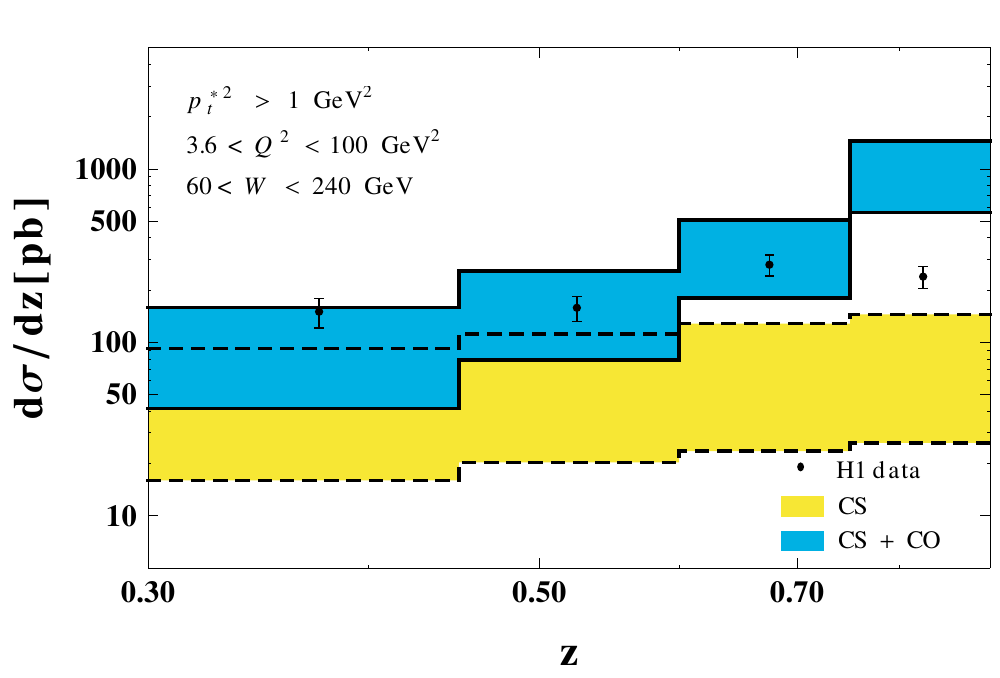}\\
\includegraphics[width=0.32\textwidth]{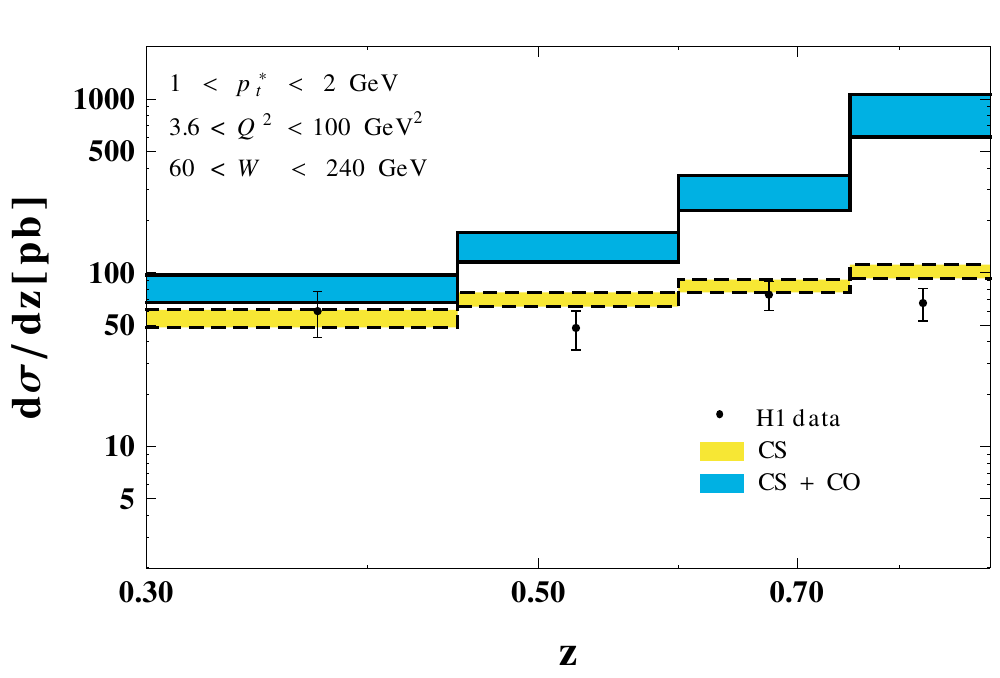}
\includegraphics[width=0.32\textwidth]{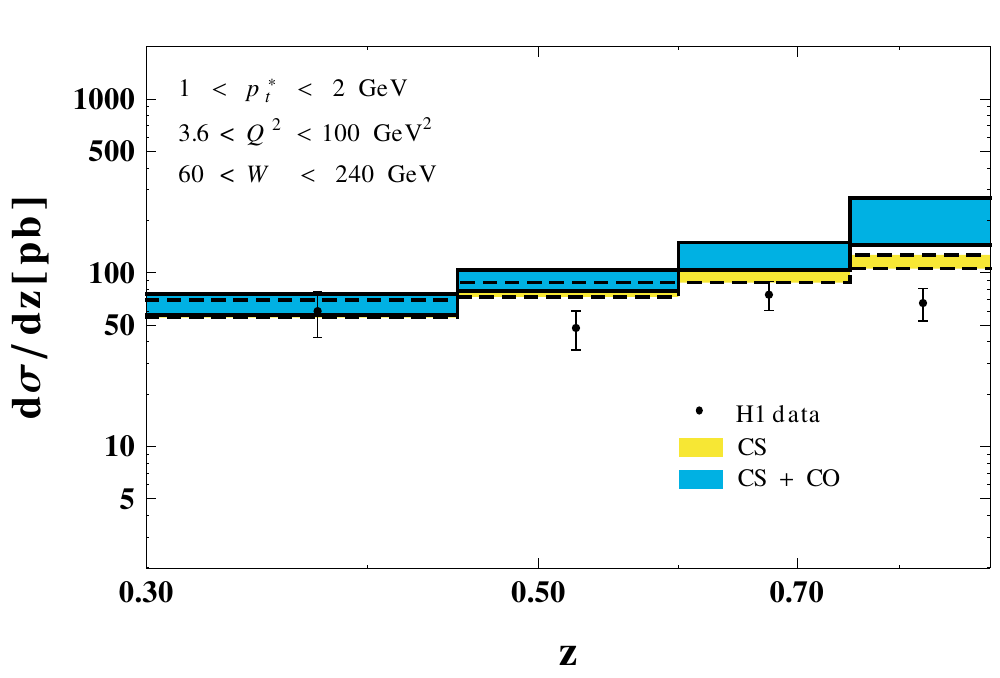}
\includegraphics[width=0.32\textwidth]{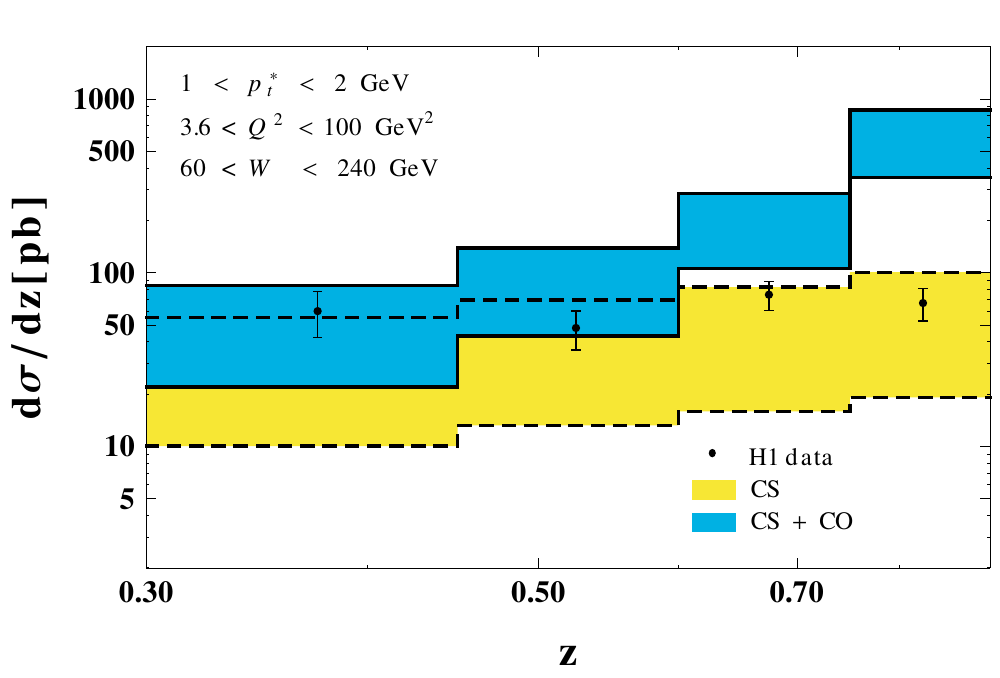}\\
\includegraphics[width=0.32\textwidth]{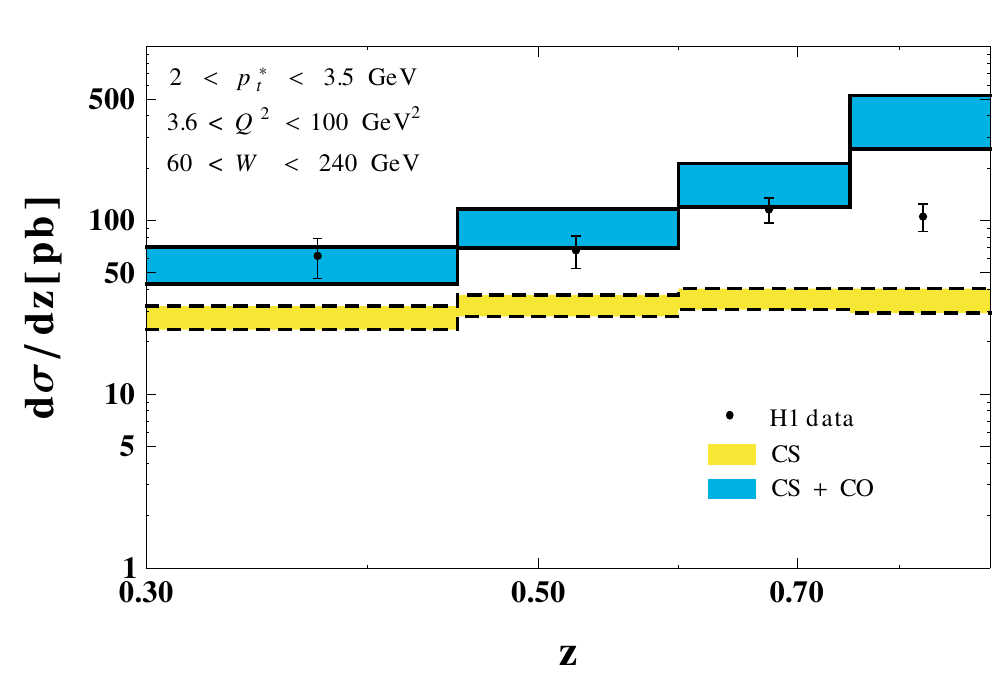}
\includegraphics[width=0.32\textwidth]{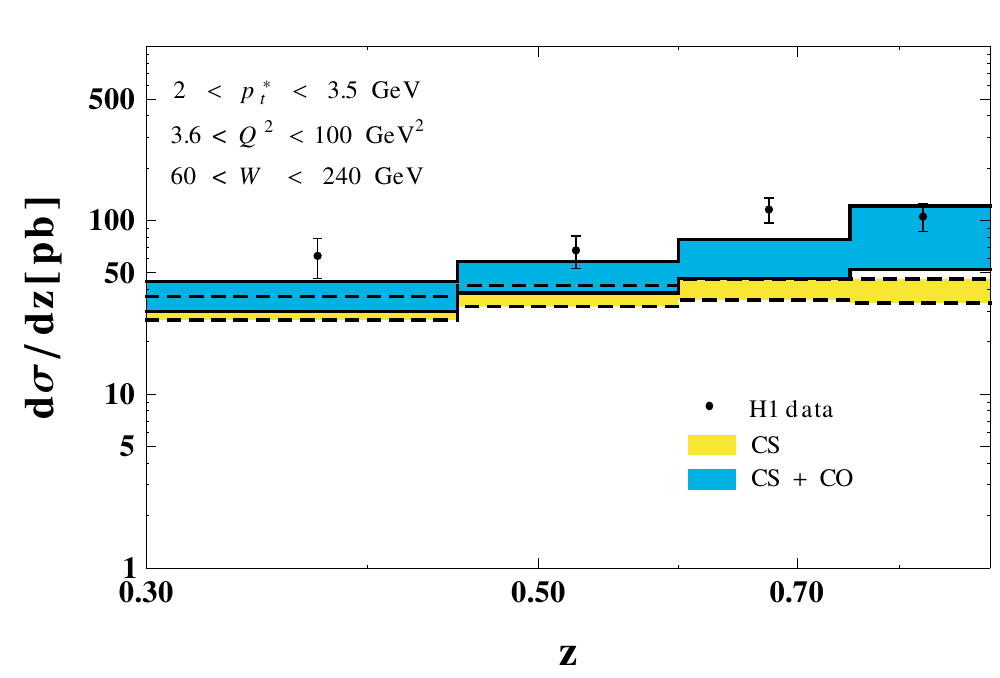}
\includegraphics[width=0.32\textwidth]{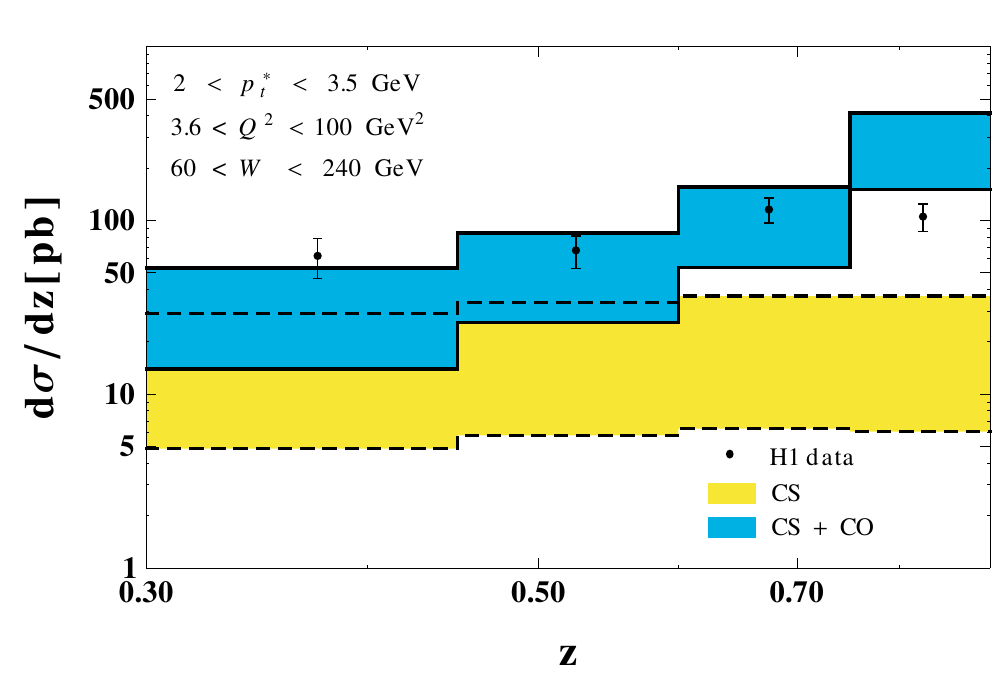}\\
\includegraphics[width=0.32\textwidth]{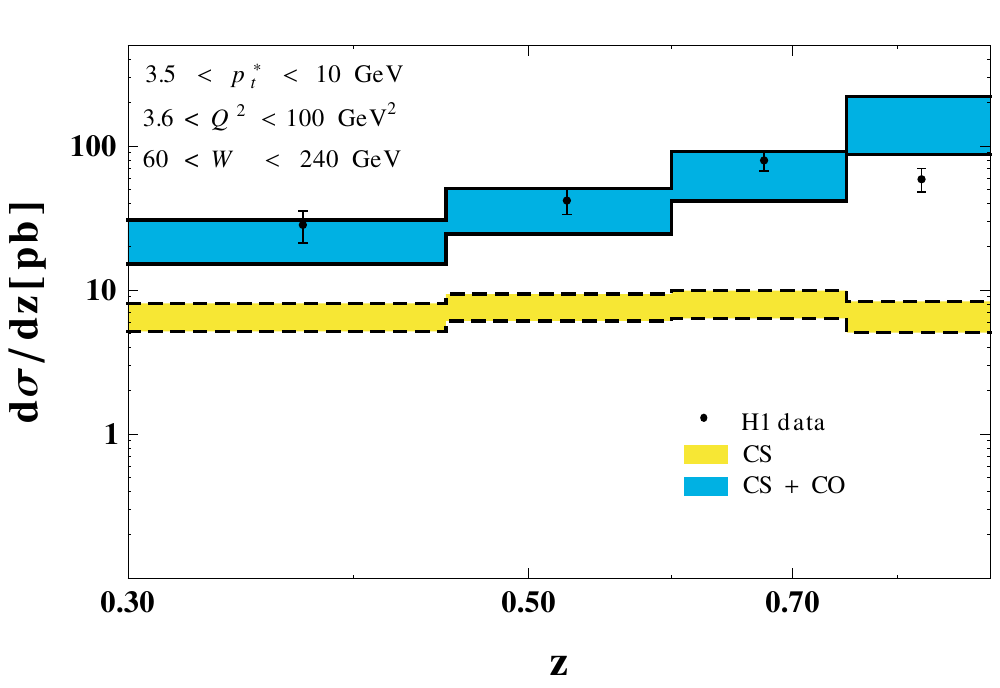}
\includegraphics[width=0.32\textwidth]{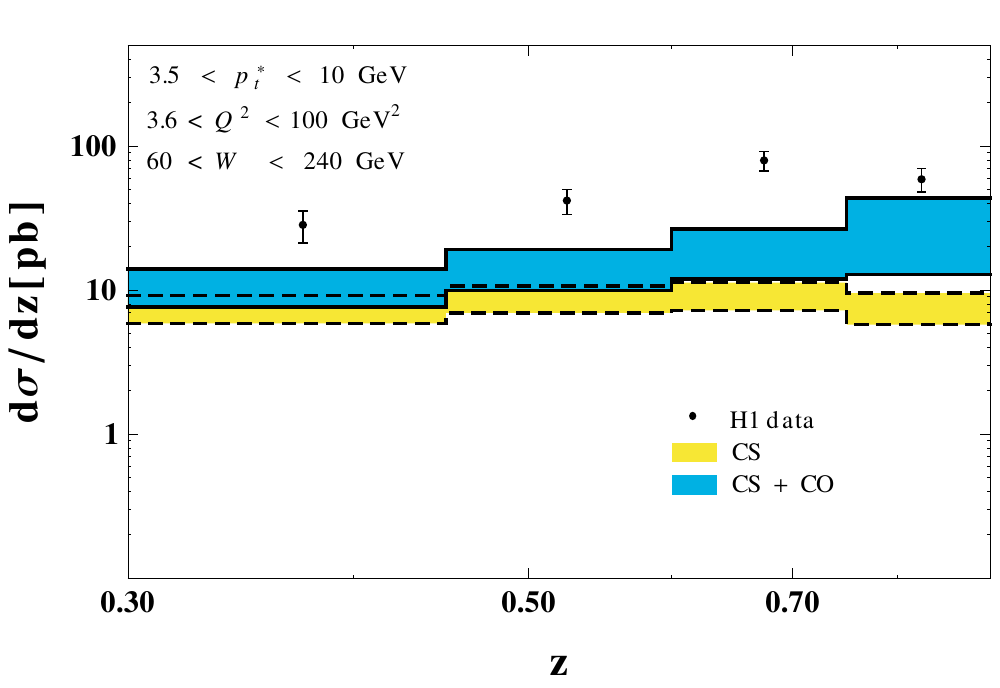}
\includegraphics[width=0.32\textwidth]{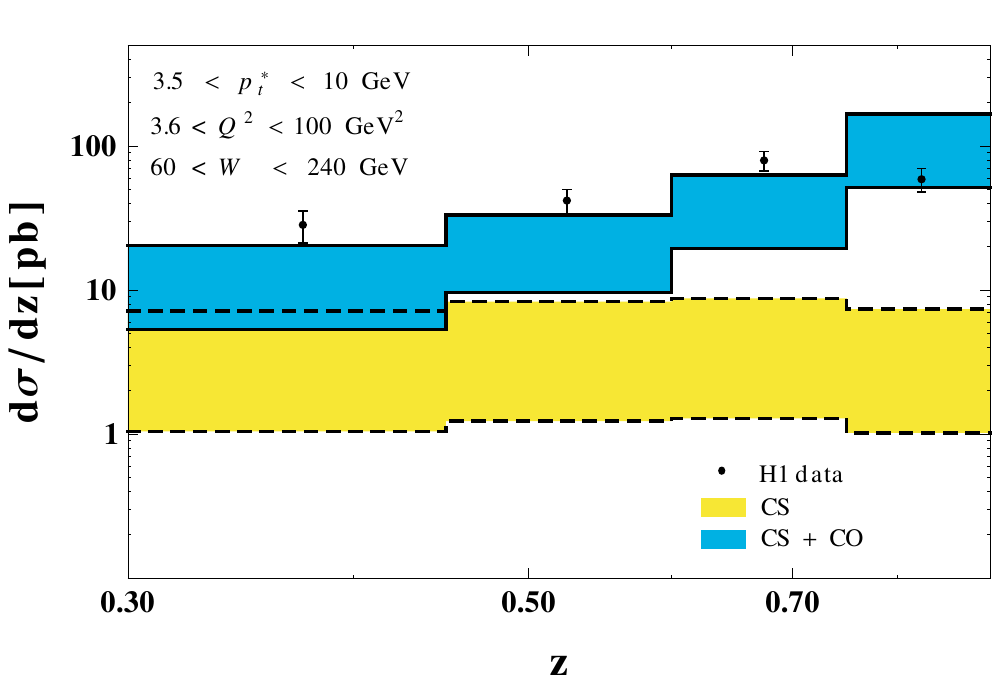}
\caption{\label{fig:z-2010}
The differential cross sections for the $J/\psi$ production in DIS with respect to $y_\psi^\star$.
The experimental data are taken from Reference~\cite{Aaron:2010gz}.
The l.h.s., mid, and r.h.s plots correspond to the LDMEs taken in References~\cite{Chao:2012iv},~\cite{Butenschoen:2011yh}, and~\cite{Zhang:2014ybe}, respectively.
}
\end{figure}

\begin{figure}
\includegraphics[width=0.32\textwidth]{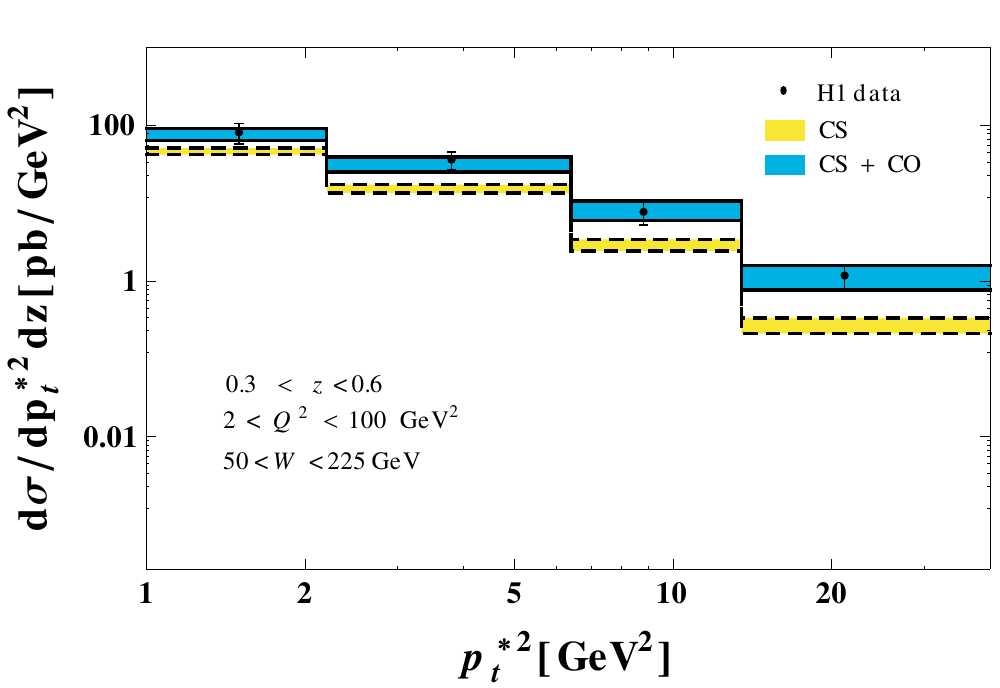}
\includegraphics[width=0.32\textwidth]{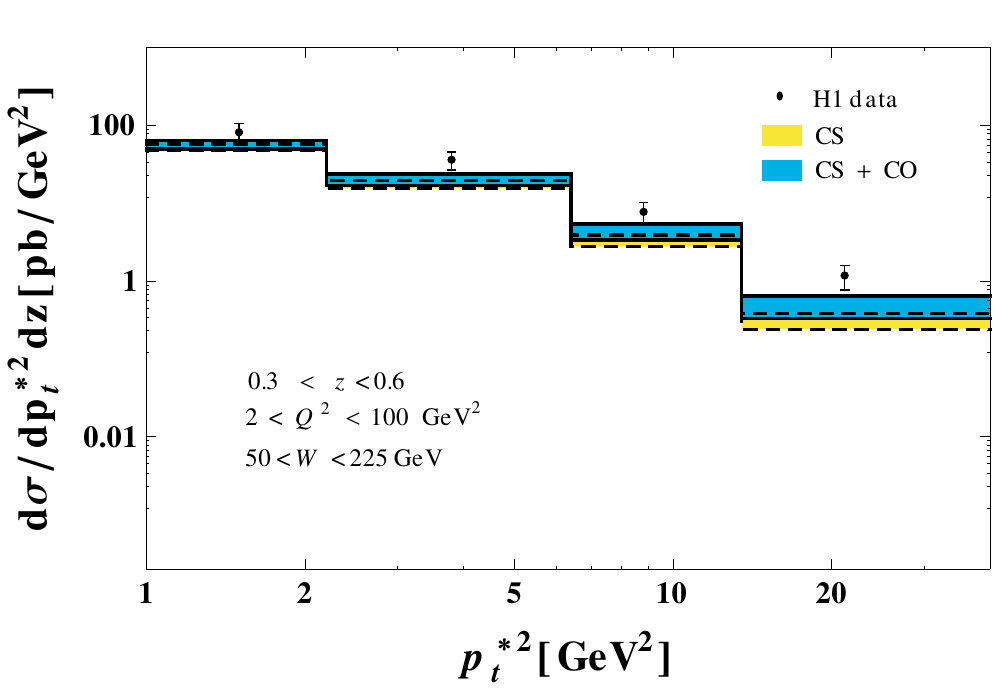}
\includegraphics[width=0.32\textwidth]{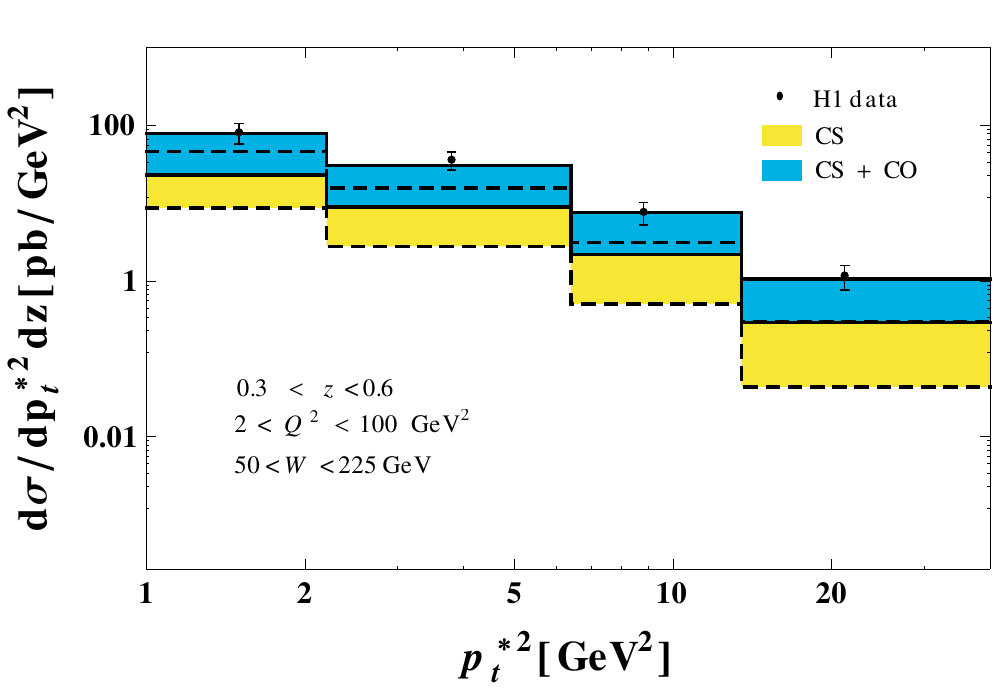}\\
\includegraphics[width=0.32\textwidth]{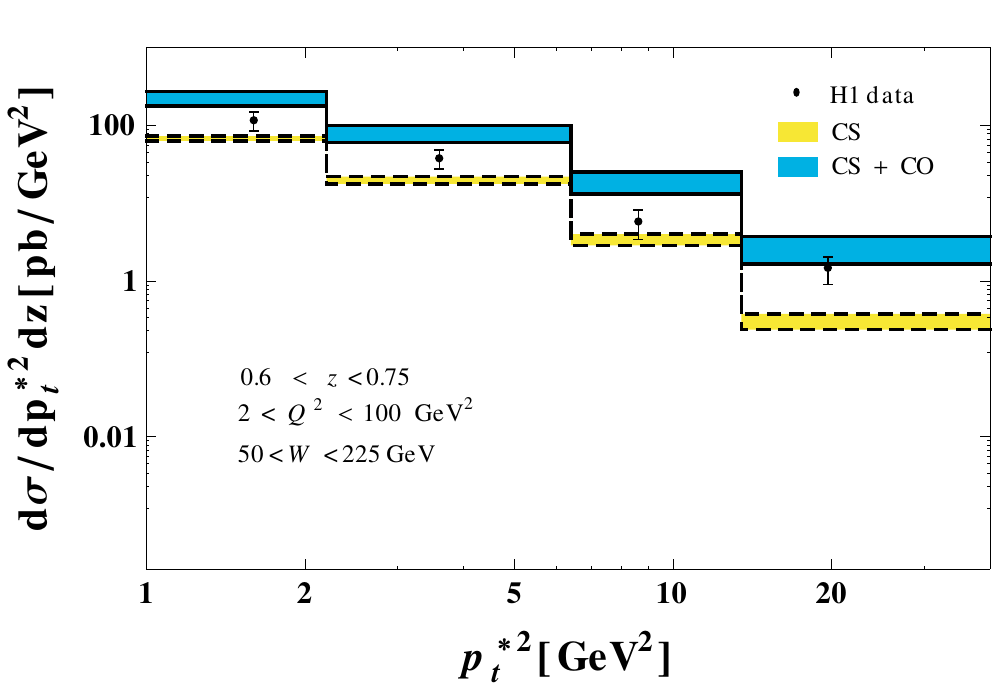}
\includegraphics[width=0.32\textwidth]{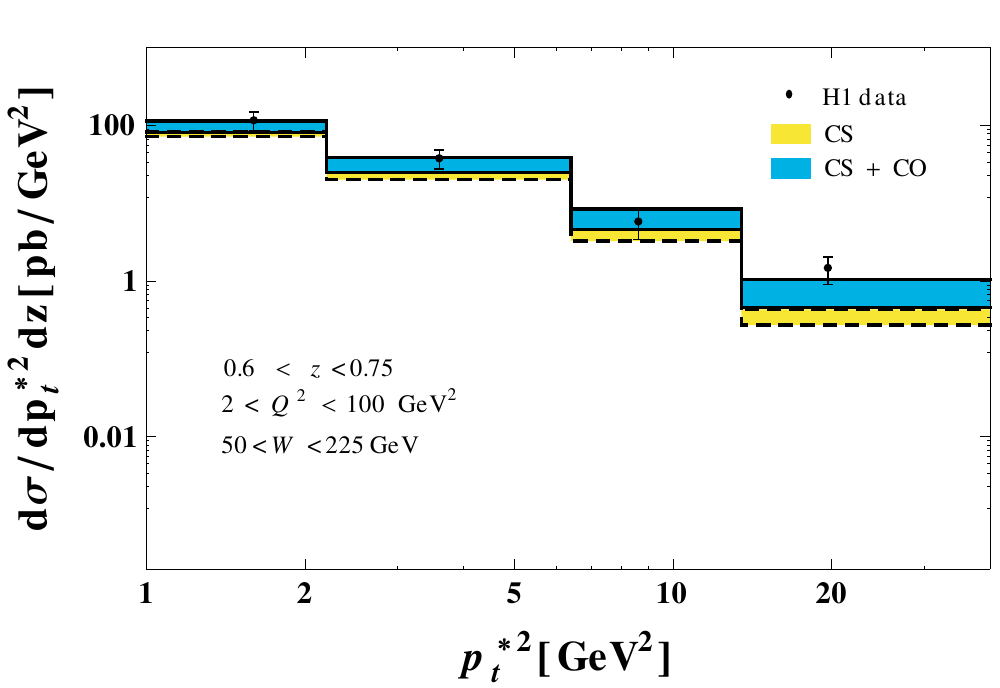}
\includegraphics[width=0.32\textwidth]{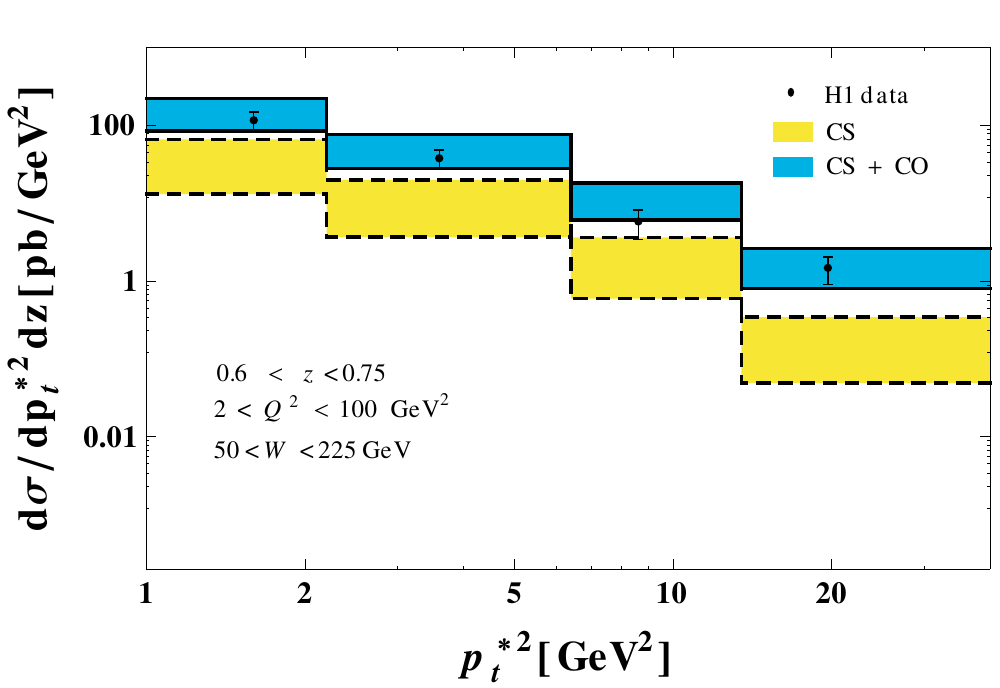}\\
\includegraphics[width=0.32\textwidth]{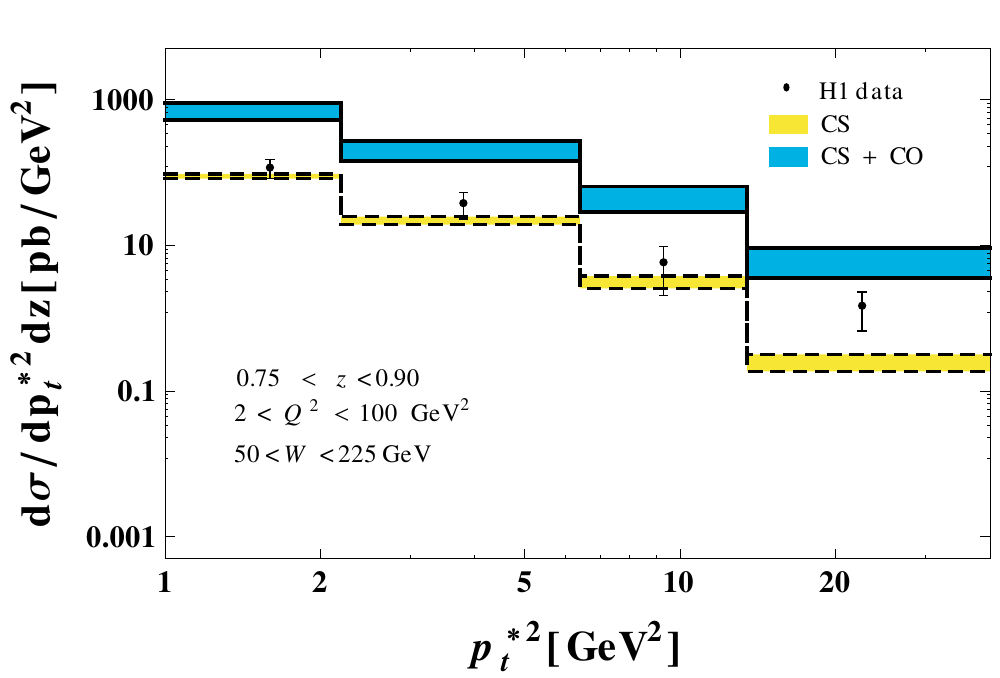}
\includegraphics[width=0.32\textwidth]{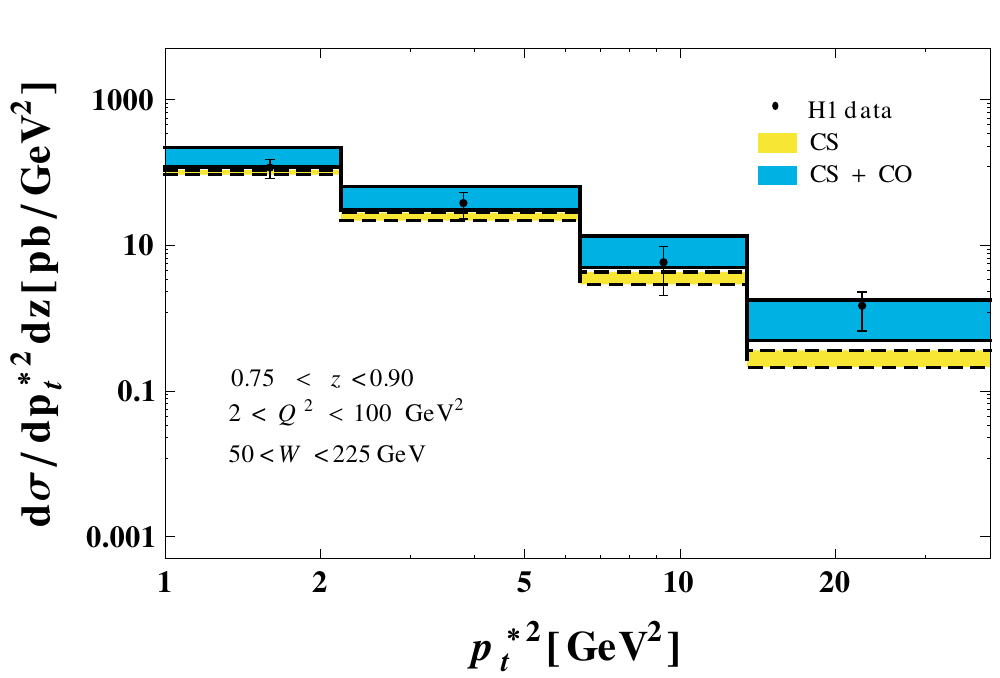}
\includegraphics[width=0.32\textwidth]{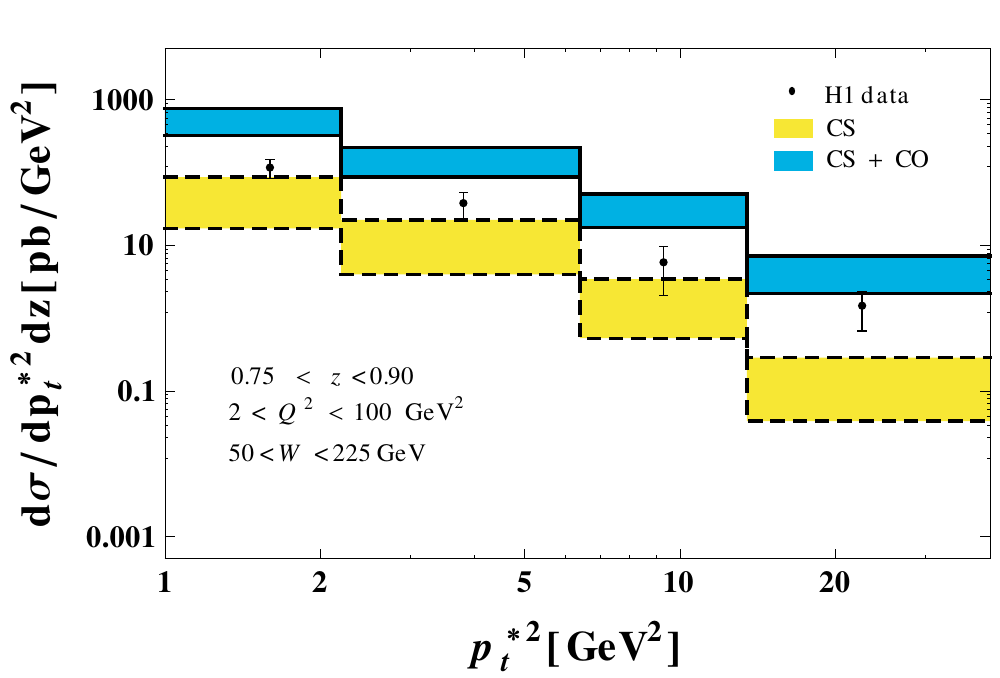}
\caption{\label{fig:ddpts2z}
The double differential cross sections for the $J/\psi$ production in DIS with respect to $p_t^{\star2}$ and $z$.
The experimental data are taken from Reference~\cite{Adloff:2002ey}.
The l.h.s., mid, and r.h.s plots correspond to the LDMEs taken in References~\cite{Chao:2012iv},~\cite{Butenschoen:2011yh}, and~\cite{Zhang:2014ybe}, respectively.
}
\end{figure}

\begin{figure}
\includegraphics[width=0.32\textwidth]{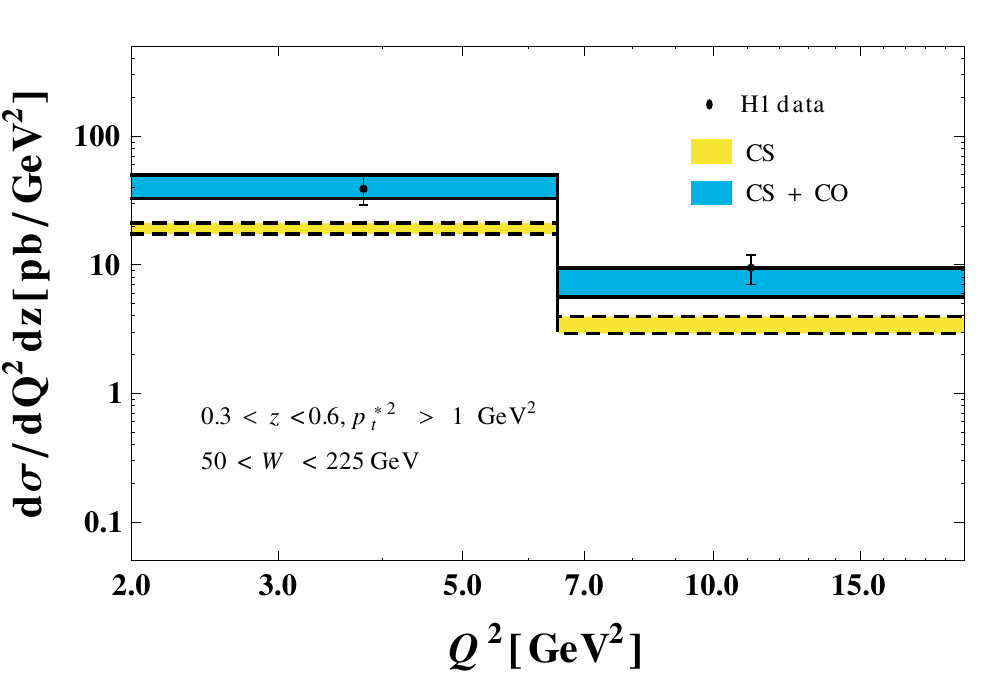}
\includegraphics[width=0.32\textwidth]{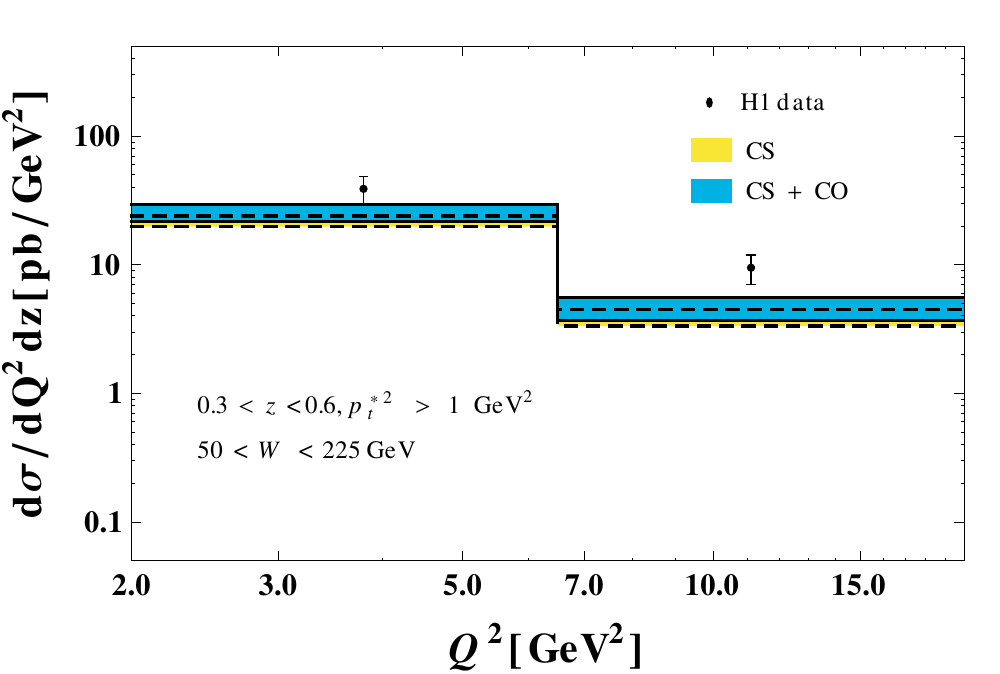}
\includegraphics[width=0.32\textwidth]{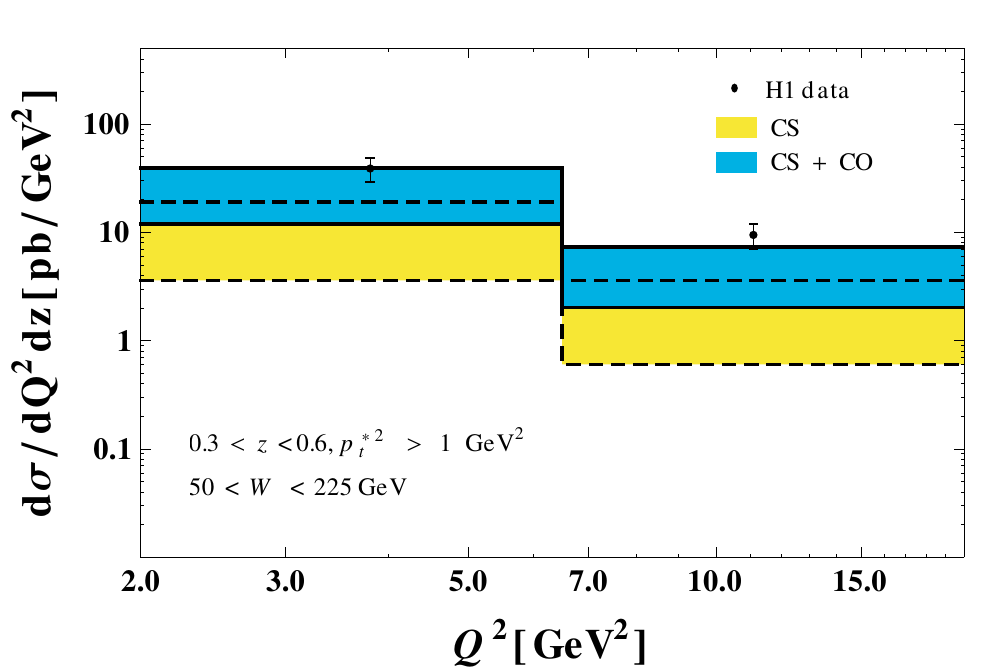}\\
\includegraphics[width=0.32\textwidth]{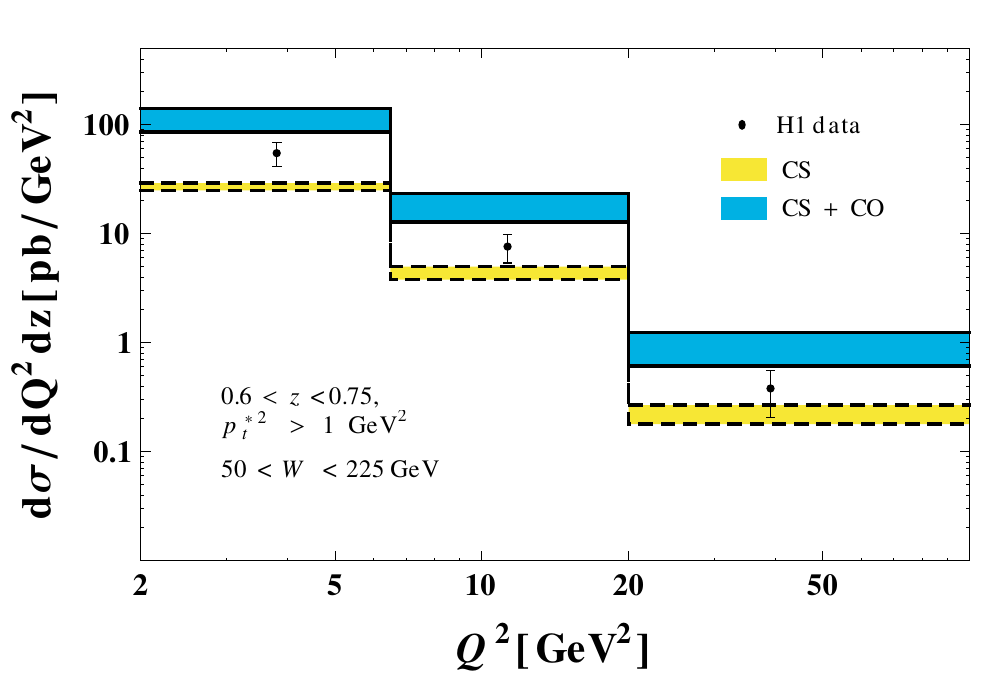}
\includegraphics[width=0.32\textwidth]{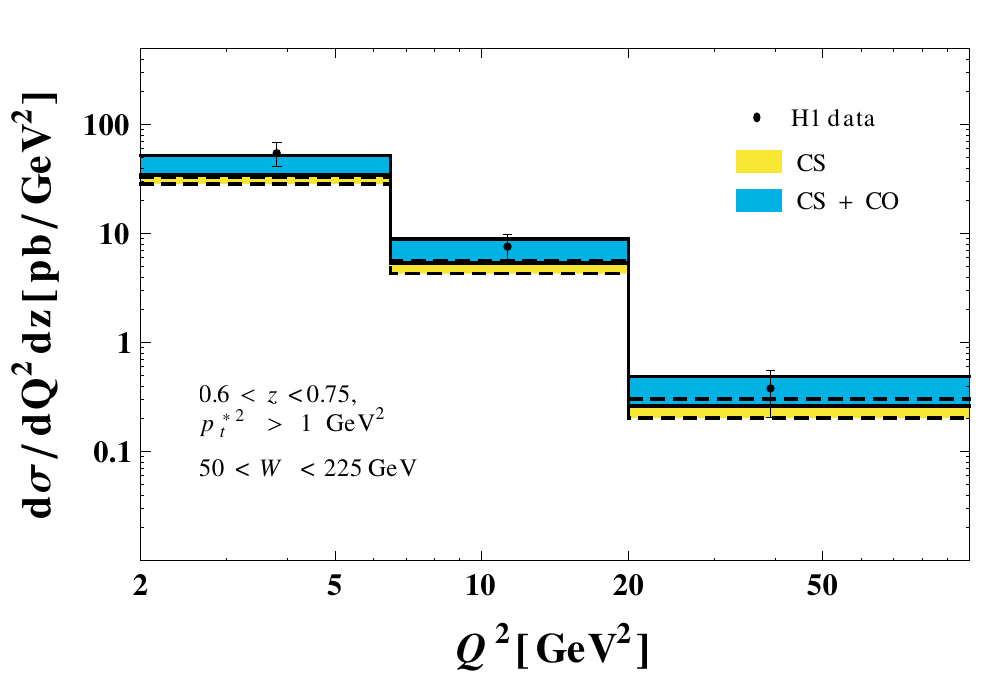}
\includegraphics[width=0.32\textwidth]{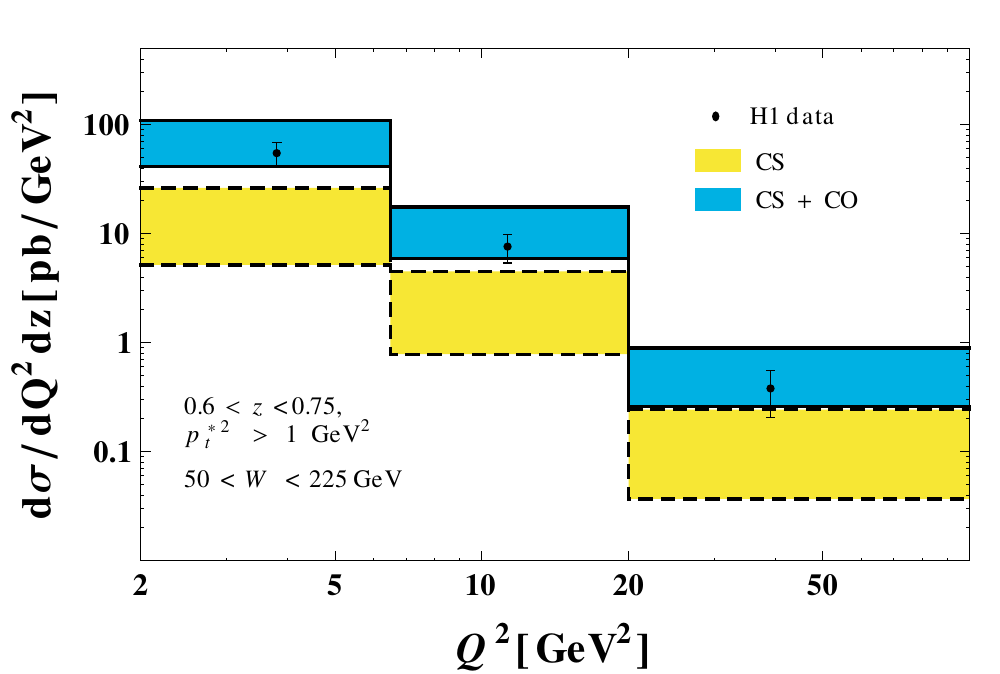}\\
\includegraphics[width=0.32\textwidth]{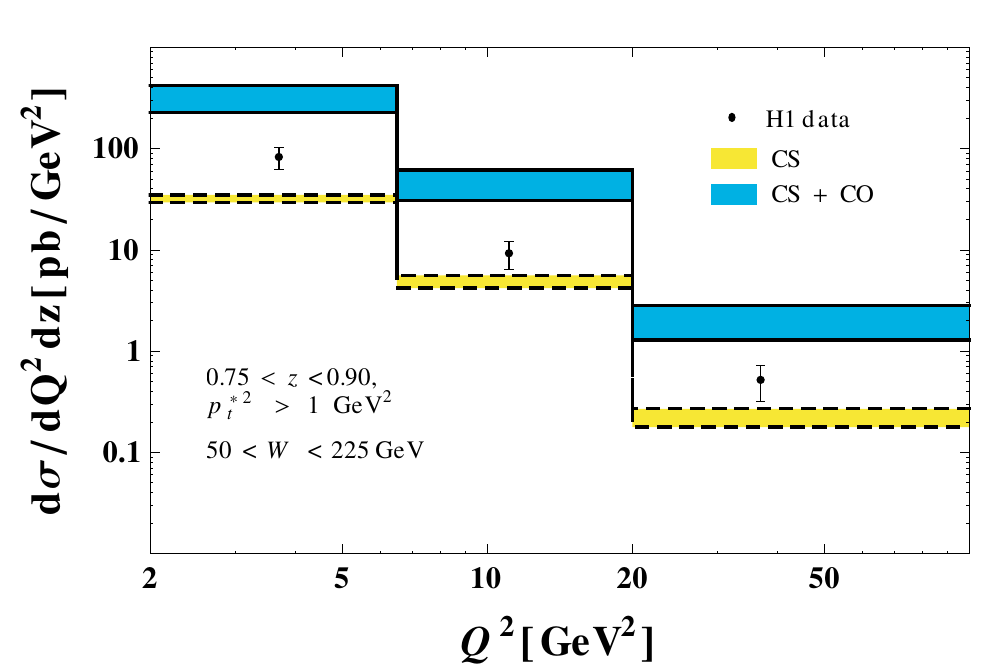}
\includegraphics[width=0.32\textwidth]{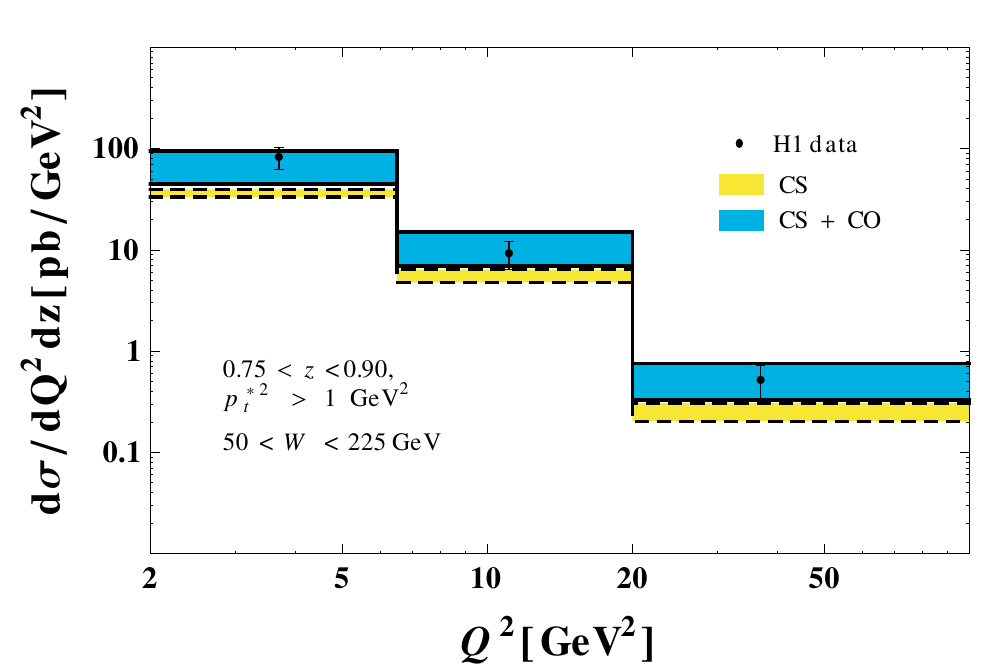}
\includegraphics[width=0.32\textwidth]{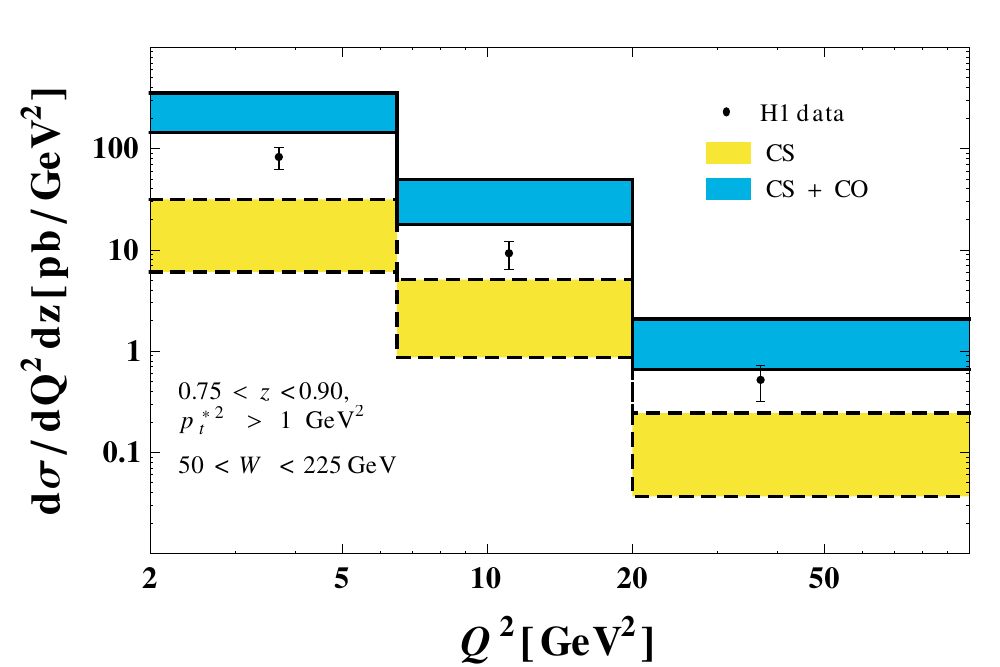}
\caption{\label{fig:ddq2z}
The double differential cross sections for the $J/\psi$ production in DIS with respect to $Q^2$ and $z$.
The experimental data are taken from Reference~\cite{Adloff:2002ey}.
The l.h.s., mid, and r.h.s plots correspond to the LDMEs taken in References~\cite{Chao:2012iv},~\cite{Butenschoen:2011yh}, and~\cite{Zhang:2014ybe}, respectively.
}
\end{figure}

As follows, we give the results for the differential cross sections with respect to $p_t^2$, $p_t^{\star2}$, $Q^2$, $W$,
$y_\psi$, $y_\psi^\star$, and $z$, and the double differential cross sections with respect to $p_t^{\star2}$ and $z$, and $Q^2$ and $z$.
The figures for each observed physical variable are grouped according to the experiment papers that the corresponding data are taken from.
All these numerical results are presented in Figures~\ref{fig:pt2-2002}-\ref{fig:ddq2z}.
For each experimental condition, there are three plots,
corresponding to the three sets of the LDMEs given in References~\cite{Chao:2012iv, Butenschoen:2011yh, Zhang:2014ybe}.

One can easily find that the theoretical results via the CS mechanism at LO are generally below the data.
However, the discrepancy is not so large as that in the $J/\psi$ hadroproduction cases.
The largest discrepancy emerges in high $p_t^2$, $p_t^{\star2}$ and $Q^2$ regions,
where the perturbative calculations are credible.
In these regions, the CS results are one order of magnitude smaller than the data.
According to Reference~\cite{Sun:2017wxk}, the QCD corrections to the CS channel is minor,
to this end, the CO contributions are necessary for understanding the $J/\psi$ leptoproduction data.

\begin{table}
\centering
\caption{\label{tab:chi2}
$\chi^2$/d.o.f. for the LDMEs taken from References~\cite{Chao:2012iv},~\cite{Butenschoen:2011yh}, and~\cite{Zhang:2014ybe}, respectively.
}
\begin{tabular}{lccccc}
\hline
Condition & Reference~\cite{Chao:2012iv} & Reference~\cite{Butenschoen:2011yh} & Reference~\cite{Zhang:2014ybe} \\
$z<0.9$ & 11.54 & 2.07 & 7.03 \\
$z<0.6$ & 2.35 & 1.90 & 1.82 \\
\hline
\end{tabular}
\end{table}

To quantitatively investigate the agreement between the theoretical results and data,
we present the $\chi^2$/d.o.f. for each set of the LDMEs in Table~\ref{tab:chi2},
where two cut conditions are considered.
For the condition, $z<0.9$, all the data points calculated in this paper are included,
while for $z<0.6$, only those satisfying $z<0.6$ are counted.
We can see that the LDMEs given in Reference~\cite{Butenschoen:2011yh} work equally well in both large $z$ and small $z$ regions,
while the other two sets fail in the region, $0.6<z<0.9$,
where a divergence factor $1/(1-z)$ will make the theoretical results larger than the data.
In the region, $z<0.6$, the LDMEs given in Reference~\cite{Zhang:2014ybe} work best.
However, the differences of the $\chi^2$/d.o.f. among the results obtained by using the three sets of the LDMEs are not significant.
Besides, the small-$x$ effects and threshold resummation are not considered in our calculation.
Therefore, it is too early to judge which of the three is better.
Once these two challenging works are finished,
the $J/\psi$ production in DIS can provide a good reference to distinguish the quarkonium production mechanisms.

\section{Summary\label{sec:summary}}

In this paper, we studied the $J/\psi$ production within the NRQCD framework at QCD LO.
Although this process has already been investigated in three previous papers~\cite{Fleming:1997fq, Yuan:2000cn, Kniehl:2001tk},
their calculations employed a form of the leptonic tensor
which will lead to wrong results when the $p_t^2$ and $y_\psi$ in the laboratory frame do not cover all values possible for them.
Since many of the existing data are measured at specific values of $p_t^2$ or $y_\psi$,
and even for the $Q^2$, $W$, $y_\psi^\star$ and $z$ distributions,
some measurements applied a $p_t^2$ cut, our renewed investigation is necessary.
The calculation in this paper are based on a formalism proposed in Reference~\cite{Zhang:2017dia},
which significantly reduces the complexity of the computation by reducing the number of the involved momenta.

We presented in this paper the numerical results comparing with all the existing data released from HERA collaborations.
Three representative sets of the LDMEs were employed for comparison.
We found that the CS contributions at QCD LO are generally below the data.
In the regions where the perturbative calculations are expected to work well,
the CS results are almost one order of magnitude smaller than the HERA data.
The NRQCD results obtained by using the three sets of the LDMEs have their own advantages in the description of the data in specific kinematical regions.
In general, the results using the LDMEs in Reference~\cite{Chao:2012iv} are the largest among the three,
while those for Reference~\cite{Butenschoen:2011yh} are the smallest.
Since the gluon saturation effect might be important,
and the behaviour of the $z$ distribution of the $^1S_0^{[8]}$ and $^3P_J^{[8]}$ SDCs might be completely changed if the $1/(1-z)$ singularities are smeared,
it is not proper to make definite conclusions until these two works are accomplished.
Besides, the LDMEs employed in our calculation are all extracted with the QCD NLO SDCs.
When the QCD corrections to all the CO channels are achieved,
the comparison of the theoretical results to data can provide better references for the universality of the LDMEs.

In any sense, the $J/\psi$ production in DIS provides an alternative device for the study of the $J/\psi$ production mechanism.
Benefiting from the various observables, they can provide new references for the theoretical studies.
They may also help to fix the LDMEs for the $J/\psi$ production when new progress in the phenomenology are made,
and new experiments in the future colliders, such as the EIC, are carried out.

\begin{acknowledgments}
This work is supported by the National Natural Science Foundation of China (Grant No. 11405268, 11647113 and 11705034).
\end{acknowledgments}

\begin{appendix}

\section{Analytical expressions for $H_1$, $H_2$, $H_3$ and $H_4$\label{app:h}}

In this appendix, we present the analytical expressions for $H_1$, $H_2$, $H_3$ and $H_4$.
The average on the initial spin and colour are not implemented.
To make the expression more compact, we define
\bea
s=2p\cdot q,~~~~t=-2p\cdot p_a,~~~~u=-2q\cdot p_a. \label{eqn:stu}
\eea
the relations between the variables defined in Equation~\ref{eqn:stu} and the ordinarily used Mandelstam variables,
\bea
\hat{s}=(p+q)^2,~~~~\hat{t}=(q-p_\psi)^2,~~~~\hat{u}=(p-p_\psi)^2, \label{eqn:stuhat}
\eea
are
\bea
s=\hat{s}+Q^2,~~~~t=\hat{t},~~~~u=\hat{u}+Q^2. \label{eqn:sturelation}
\eea
Thus we have
\bea
s+t+u=M^2+Q^2, \label{eqn:stumq}
\eea
which we have employed to eliminate $M^2$ in our expressions.
In the following, $e_q$ denotes the fractional electric charge of quark $q$.

The expressions are listed below.

\bea
e+q(\bar{q})\rightarrow c\bar{c}[^3S_1^{[1]}]+e+q(\bar{q}): \NO
\eea
\bea
H_1=H_2=H_3=H_4=0.
\eea

\bea
e+q(\bar{q})\rightarrow c\bar{c}[^1S_0^{[8]}]+e+q(\bar{q}): \NO
\eea
\bea
H_1=-\frac{32e_c^2}{Mt(s+u)^2}(2Q^2t+s^2+u^2), \NO
\eea
\bea
H_2=-\frac{16e_c^2}{M(s+u)^2}(Q^2t+su), \NO
\eea
\bea
H_3=H_4=0.
\eea

\bea
e+q(\bar{q})\rightarrow c\bar{c}[^3S_1^{[8]}]+e+q(\bar{q}): \NO
\eea
\bea
H_1&=&\frac{16e_q^2}{3M^3(Q^2-s)^2(Q^2-u)^2} \NO \\
&\times&\{2Q^6t-2Q^4[s(2t+u)+t(t+2u)] \NO \\
&+&Q^2[s^2(t+2u)+2s(t^2+4tu+u^2)+tu(2t+u)] \NO \\
&-&su(s^2+2st+2t^2+2tu+u^2)\}, \NO
\eea
\bea
H_2=-\frac{8e_q^2}{3M^3(Q^2-u)^2}[t(s+t)^2], \NO
\eea
\bea
H_3&=&-\frac{8e_q^2}{3M^3(Q^2-s)(Q^2-u)^2} \NO \\
&\times&\{Q^4t^2+Q^2t(s^2+st-2tu) \NO \\
&+&[-s^3t-2s^2t^2-st^2(t-u)+t^2u(t+u)]\}, \NO
\eea
\bea
H_4&=&\frac{16e_q^2}{3M^3(Q^2-s)^2(Q^2-u)^2} \NO \\
&\times&\{-Q^6t^2-\frac{1}{2}Q^4t(s^2-2st-2tu+u^2) \NO \\
&+&Q^2t(s^3+s^2t-3stu+tu^2+u^3) \NO \\
&-&\frac{1}{2}t[s^4+2s^3t+s^2t(t-2u)-2stu(t+u)+u^2(t+u)^2]\}.
\eea

\bea
e+q(\bar{q})\rightarrow c\bar{c}[^3P_J^{[8]}]+e+q(\bar{q}): \NO
\eea
\bea
H_1&=&\frac{128e_c^2}{M^3t(s+u)^4} \NO \\
&\times&\{24Q^6t-4Q^4(3s^2+8st+12t^2+8tu+3u^2) \NO \\
&+&2Q^2[8s^3+s^2(9t+8u)+2s(6t^2+5tu+4u^2)+8t^3+12t^2u+9tu^2+8u^3] \NO \\
&-&(s+u)[7s^3+s^2(12t+7u)+s(8t^2+16tu+7u^2)+u(8t^2+12tu+7u^2)]\}, \NO
\eea
\bea
H_2&=&-\frac{64e_c^2}{M^3t(s+u)^4} \NO \\
&\times&\{-4Q^4t[-s^2+st+t(2t+u)] \NO \\
&+&Q^2[-2s^4-4s^3t-s^2(5t^2+8tu+2u^2)+2st(6t^2+tu-2u^2)+t^2(8t^2+12tu+7u^2)] \NO \\
&+&s(s+u)[2s^3+2s^2t+s(8t^2+5tu+2u^2)+t(8t^2+12tu+7u^2)]\}, \NO
\eea
\bea
H_3&=&-\frac{128e_c^2}{Mt(s+u)^4} \NO \\
&\times&\{-4Q^4t^2+2Q^2t[s(2t-u)+4t^2+2tu+u^2] \NO \\
&+&s(s+u)[s^2+(2t+u)^2]\}, \NO
\eea
\bea
H_4&=&-\frac{128e_c^2}{Mt(s+u)^4} \NO \\
&\times&\{-8Q^4t^2+2Q^2t[s^2+2s(2t-u)+8t^2+4tu+u^2] \NO \\
&+&(s+u)[s^3+s^2u+s(4t^2+8tu+u^2)+4t^2u+u^3)]\}.
\eea

\bea
e+g\rightarrow c\bar{c}[^3S_1^{[1]}]+e+g: \NO
\eea
\bea
H_1&=&\frac{1024e_c^2}{27M(s+t)^2(s+u)^2(t+u)^2} \NO \\
&\times&\{-6Q^6t^2+2Q^4t[s(4t-3u)+4t(t+u)] \NO \\
&-&Q^2[s^2(3t^2-4tu+3u^2)+2st(t^2-tu-2u^2)+t^2(t^2+2tu+3u^2)] \NO \\
&+&2[s^3(t^2+tu+u^2)+s^2(t+u)^3+stu(t^2+3tu+u^2)+t^2u^2(t+u)]\}, \NO
\eea
\bea
H_2&=&\frac{256e_c^2}{27M(s+t)^2(s+u)^2(t+u)^2} \NO \\
&\times&\{2Q^4t^2(s^2-2t^2)-2Q^2t(s^2-2t^2)(s(t-u)+t(t+u)) \NO \\
&+&s[s^3(t^2+u^2)+2s^2t^2(t+u)+st^2(t^2+6tu+u^2)+4t^3u(t+u)]\}, \NO
\eea
\bea
H_3&=&\frac{256e_c^2}{27M(s+t)^2(s+u)^2(t+u)^2} \NO \\
&\times&\{4Q^6t^3+2 Q^4t^2[s^2-3s(t-u)-t(5t+3u)] \NO \\
&-&2Q^2t[s^3(t-u)+2s^2u(2t-u) \NO \\
&-&st(4t^2-3tu-3u^2)-t^2(t+u)(3t+u)] \NO \\
&+&s[s^3(t^2+u^2)+s^2(2t^3+3t^2u-2tu^2+u^3) \NO \\
&+&st(t^3+8t^2u+tu^2-2u^3)+t^2u(t+u)(5t+u)]\}, \NO
\eea
\bea
H_4&=&\frac{256e_c^2}{27M(s+t)^2(s+u)^2(t+u)^2} \NO \\
&\times&\{8Q^6t^3+2Q^4t^2(s^2-6s(t-u)-10t^2-6tu+u^2) \NO \\
&+&2Q^2t[s^3(-t+u)+s^2(t^2-7tu+4u^2) \NO \\
&+&s(8t^3-6t^2u-7tu^2+u^3)+(t+u)(6t^2+2tu-u^2)] \NO \\
&+&[s^4(t^2+u^2)+2s^3(t^3+2t^2u-2tu^2+u^3) \NO \\
&+&s^2(t^4+14t^3u+2t^2u^2-4tu^3+u^4) \NO \\
&+&2 s t^2u(t+u)(5t+2u)+t^2u^2(t+u)^2]\}.
\eea

\bea
e+g\rightarrow c\bar{c}[^1S_0^{[8]}]+e+g: \NO
\eea
\bea
H_1&=&\frac{384e_c^2}{Mt(s+t)^2(s+u)^2(t+u)^2} \NO \\
&\times&\{Q^2t[s^4+2s^3(t+u)+2s^2(t+u)^2+2su(t+u)^2+u^2(2t^2+2tu+u^2)] \NO \\
&+&su[s^4+2s^3(t+u)+3s^2(t+u)^2+2s(t+u)^3+(t^2+tu+u^2)^2]\}, \NO
\eea
\bea
H_2&=&\frac{96e_c^2}{M(s+t)^2(s+u)^2(t+u)^2} \NO \\
&\times&\{2Q^4t^2u^2+2Q^2(s^2tu(t+u)+stu(t^2+tu+2u^2)) \NO \\
&+&s^2[s^2(t+u)^2+2s(t+u)(t^2+tu+u^2) \NO \\
&+&(t^2+u^2)(t^2+2tu+2u^2)]\}, \NO
\eea
\bea
H_3&=&\frac{96e_c^2}{M(s+t)^2(s+u)^2(t+u)^2} \NO \\
&\times&\{2Q^4t^2u^2-Q^2t[s^3(t+u)+s^2(t-2u)(t+u)-su(2t^2+tu+3u^2)+tu^2(t+u)] \NO \\
&+&s[s^3t(t+u)+s^2(t+u)(2t^2+u^2) \NO \\
&+&s(t-u)(t^3+t^2u-u^3)-tu(t+u)(t^2+tu+u^2)]\}, \NO
\eea
\bea
H_4&=&\frac{96e_c^2}{M^3(s+t)^2(s+u)^2(t+u)^2} \NO \\
&\times&\{-2Q^6t^2(s^2+u^2)+2Q^4t[s^3(2t-u)+2s^2(t^2-u^2)-s u (2 t^2+u^2)+2tu^2(t+u)] \NO \\
&-&Q^2[s^4(3t^2-2tu+u^2)+2s^32s^3(3t-u)(t^2-u^2) \NO \\
&-&s^2(-3t^4+6t^3u+10t^2u^2+6tu^3-u^4) \NO \\
&-&2stu(t+u)(3t^2+u^2)+3t^2u^2(t+u)^2] \NO \\
&+&(s+t+u)[s^4(t^2+u^2)+2s^3(t+u)(t^2-tu+u^2) \NO \\
&+&s^2(t^4-2t^3u-2t^2u^2+u^4)-2st^3u(t+u)+t^2u^2(t+u)^2]\}.
\eea

\bea
e+g\rightarrow c\bar{c}[^3S_1^{[8]}]+e+g: \NO
\eea
\bea
H_i[^3S_1^{[8]}]=\frac{15}{8}H_i[^3S_1^{[1]}],~~~~(i=1,~2,~3,~4).
\eea

\bea
e+g\rightarrow c\bar{c}[^3P_J^{[8]}]+e+g: \NO
\eea
\bea
H_1&=&-\frac{1536e_c^2}{M^3t(s+t)^3(s+u)^4(t+u)^3} \NO \\
&\times&\{24Q^8t^2(s^2-tu)(st-u^2) \NO \\
&-&2Q^6t[6s^5(t+u)+4s^4(7t^2+3u^2) \NO \\
&+&s^3(19t^3+25t^2u-10tu^2+24u^3) \NO \\
&-&s^2(3t^4+31t^3u+10t^2u^2+10tu^3-12u^4) \NO \\
&+&su(-22t^4-31t^3u+25t^2u^2+6u^4) \NO \\
&-&-tu^2(t+u)(3t^2-22tu-6u^2)] \NO \\
&+&2Q^4[2s^6(4t-3u)(t+u)+3s^5(9t^3+t^2u+4tu^2-4u^3) \NO \\
&+&s^4(24t^4+11t^3u-19t^2u^2+24tu^3-18u^4) \NO \\
&+&3s^3(t^5-7t^4u-8t^3u^2-6t^2u^3+8tu^4-4u^5) \NO \\
&-&s^2(4t^6+39t^5u+76t^4u^2+24t^3u^3+19t^2u^4-12tu^5+6u^6) \NO \\
&+&stu(-10t^5-39t^4u-21t^3u^2+11t^2u^3+3tu^4+2u^5) \NO \\
&+&t^2u^2(-4t^4+3t^3u+24t^2u^2+27tu^3+8u^4)] \NO \\
&+&Q^2[s^7(-(7t-16u))(t+u)+s^6(-25t^3+9t^2u+34tu^2+48u^3) \NO \\
&+&s^5(-27t^4+5t^3u+60t^2u^2+60tu^3+80u^4) \NO \\
&+&s^4(-9t^5+47t^4u+120t^3u^2+134t^2u^3+78tu^4+80u^5) \NO \\
&+&s^3(t^6+51t^5u+180t^4u^2+180t^3u^3+134t^2u^4+60tu^5+48u^6) \NO \\
&+&s^2(t^7+19t^6u+112t^5u^2+180t^4u^3+120t^3u^4+60t^2u^5+34tu^6+16u^7) \NO \\
&+&stu(2t^6+19t^5u+51t^4u^2+47t^3u^3+5t^2u^4+9tu^5+9u^6) \NO \\
&+&t^2u^2(t+u)(t^4-9t^2u^2-18tu^3-7u^4)] \NO \\
&-&(s+u)[7s^7u(t+u)+s^6u(25t^2+38tu+21u^2) \NO \\
&+&s^5(t+u)(2t^3+45t^2u+43tu^2+35u^3) \NO \\
&+&s^4(4t^5+63t^4u+132t^3u^2+156t^2u^3+98tu^4+35u^5) \NO \\
&+&s^3(t+u)(2t^5+45t^4u+91t^3u^2+99t^2u^3+57tu^4+21u^5) \NO \\
&+&s^2u(13t^6+70t^5u+136t^4u^2+132t^3u^3+88t^2u^4+38tu^5+7u^6) \NO \\
&+&stu^2(t+u)(13t^4+34t^3u+29t^2u^2+18tu^3+7u^4)+2t^4u^3(t+u)^2]\}, \NO
\eea
\bea
H_2&=&\frac{768e_c^2}{M^3t(s+t)^3(s+u)^4(t+u)^3} \NO \\
&\times&\{4Q^6t^2(st-u^2)(s^4-2s^2t^2+stu(t+u)+t^2u(2t+u)) \NO \\
&+&stu(2t^3+t^2u-tu^2-u^3)-t^2u^3(2t+u)] \NO \\
&-&Q^4t[2s^7(t+u)+4s^6(2t^2+u^2)-s^5(t-2u)(t^2+tu+4u^2) \NO \\
&+&s^4(-21t^4-13t^3u-18t^2u^2+2tu^3+4u^4) \NO \\
&-&2s^3(6t^5+10t^4u+t^3u^2+4t^2u^3-6tu^4-u^5) \NO \\
&+&2s^2t(t^5+10t^4u+11t^3u^2+9t^2u^3+6tu^4+5u^5)] \NO \\
&+&st^2u(12t^4+26t^3u+2t^2u^2+tu^3+u^4) \NO \\
&+&t^3u^2(t+u)(2t^2-8tu-7u^2)] \NO \\
&+&2Q^2[s^8(t^2-u^2)+s^7(3t^3+tu^2-2u^3) \NO \\
&-&s^6(2t^4+2t^3u+2t^2u^2-tu^3+3u^4) \NO \\
&-&s^5(12t^5+16t^4u+16t^3u^2-t^2u^3+tu^4+2u^5) \NO \\
&-&s^4(10t^6+28t^5u+28t^4u^2+13t^3u^3-7t^2u^4+3tu^5+u^6) \NO \\
&-&s^3t(t^2+6tu-u^2)(t^4+3t^3u+t^2u^2-2tu^3-2u^4) \NO \\
&+&s^2t^2(t^6+7t^5u+13t^4u^2+18t^3u^3+35t^2u^4+24tu^5+6u^6) \NO \\
&+&st^3u(t+u)(2t^4+8t^3u+8t^2u^2+12tu^3+7u^4)+t^6u^2(t+u)^2]  \NO \\
&+&s(s+u)[2s^7u(t+u)+2s^6u(3t^2+3tu+2u^2) \NO \\
&+&s^5(t+u)(5t^3+10t^2u+8tu^2+6u^3) \NO \\
&+&s^4(15t^5+38t^4u+53t^3u^2+40t^2u^3+22tu^4+4u^5) \NO \\
&+&s^3(t+u)(15t^5+37t^4u+51t^3u^2+30t^2u^3+17tu^4+2u^5) \NO \\
&+&s^2t(5t^6+32t^5u+78t^4u^2+90t^3u^3+68t^2u^4+34tu^5+9u^6) \NO \\
&+&st^2u(t+u)(7t^4+24t^3u+23t^2u^2+16tu^3+7u^4)+2t^5u^2(t+u)^2]\}, \NO
\eea
\bea
H_3&=&\frac{768e_c^2}{M^3t(s+t)^3(s+u)^4(t+u)^3} \NO \\
&\times&\{8Q^8t^3(s^2-tu)(st-u^2) \NO \\
&-&2Q^6t^2[2s^5u+2s^4(5t^2-tu+3u^2) \NO \\
&+&s^3(11t^3+9t^2u-4tu^2+10u^3) \NO \\
&-&s^2(t^4+13t^3u+8t^2u^2+4tu^3-4u^4) \NO \\
&+&su(-12t^4-13t^3u+11t^2u^2+2u^4) \NO \\
&+&tu^2(-t^3+11t^2u+12tu^2+2u^3)] \NO \\
&-&Q^4t[2s^7(t+u)+2s^6(2t^2+tu+5u^2) \NO \\
&+&s^5(-19t^3+13t^2u-8tu^2+20u^3) \NO \\
&+&s^4(-42t^4-17t^3u+7t^2u^2-18tu^3+24u^4) \NO \\
&-&2s^3(8t^5+5t^4u-5t^3u^2-3t^2u^3+4tu^4-7u^5) \NO \\
&+&s^2(5t^6+54t^5u+76t^4u^2+30t^3u^3+37t^2u^4+4tu^5+6u^6) \NO \\
&+&stu(22t^5+60t^4u+12t^3u^2-3t^2u^3+13tu^4+2u^5) \NO \\
&+&t^2u^2(5t^4-10t^3u-36t^2u^2-25tu^3-4u^4)] \NO \\
&+&Q^2[2s^8(t^2-u^2)+6s^7(t^3+t^2u+tu^2-u^3) \NO \\
&-&2s^6(4t^4-8t^3u-4t^2u^2-9tu^3+5u^4) \NO \\
&+&s^5(-33t^5-9t^4u+11t^3u^2+19t^2u^3+30tu^4-10u^5) \NO \\
&+&s^4(-25t^6-37t^5u+7t^4u^2+33t^3u^3+38t^2u^4+26tu^5-6u^6) \NO \\
&+&s^3(-t^7+3t^6u+30t^5u^2+72t^4u^3+89t^3u^4+37t^2u^5+12tu^6-2u^7) \NO \\
&+&s^2t(3t^7+27t^6u+76t^5u^2+102t^4u^3+133t^3u^4+91t^2u^5+24tu^6+4u^7) \NO \\
&+&st^2u(t+u)(6t^5+27t^4u+26t^3u^2+33t^2u^3+28tu^4+6u^5) \NO \\
&+&t^4(3t-4u)u^2(t+u)^3] \NO \\
&+&s(s+u)[2s^7u(t+u)+2s^6u(3t^2+4tu+3u^2) \NO \\
&+&s^5(5t^4+11t^3u+20t^2u^2+16tu^3+10u^4) \NO \\
&+&s^4(15t^5+24t^4u+40t^3u^2+34t^2u^3+21tu^4+10u^5) \NO \\
&+&s^3(15t^6+32t^5u+50t^4u^2+48t^3u^3+35t^2u^4+18tu^5+6u^6) \NO \\
&+&s^2(5t^7+18t^6u+35t^5u^2+34t^4u^3+35t^3u^4+28t^2u^5+11tu^6+2u^7) \NO \\
&+&stu(t+u)(3t^5+6t^4u-2t^3u^2+3t^2u^3+7tu^4+2u^5) \NO \\
&-&2t^3u^2(t+u)^3(t+2u)]\}, \NO
\eea
\bea
H_4&=&\frac{768e_c^2}{M^3t(s+t)^3(s+u)^4(t+u)^3} \NO \\
&\times&\{16Q^8t^3(s^2-tu)(st-u^2) \NO \\
&-&4Q^6t^2[s^5(t+2u)+s^4(11t^2-tu+5u^2) \NO \\
&+&s^3(11t^3+10t^2u-4tu^2+10u^3) \NO \\
&-&s^2(t^4+13t^3u+8t^2u^2+4tu^3-5u^4) \NO \\
&+&su(-12t^4-13t^3u+10t^2u^2-tu^3+2u^4) \NO \\
&+&tu^2(-t^3+11t^2u+11tu^2+u^3)] \NO \\
&-&2Q^4t[s^7(t+u)+2s^6u(t+4u) \NO \\
&+&s^5(-22t^3+13t^2u-2tu^2+17u^3) \NO \\
&+&s^4(-39t^4-10t^3u+22t^2u^2-13tu^3+24u^4) \NO \\
&+&s^3(-13t^5+t^4u+20t^3u^2+6t^2u^3-13tu^4+17u^5) \NO \\
&+&s^2(5t^6+57t^5u+76t^4u^2+20t^3u^3+22t^2u^4-2tu^5+8u^6) \NO \\
&+&su(22t^6+57t^5u+t^4u^2-10t^3u^3+13t^2u^4+2tu^5+u^6) \NO \\
&+&tu^2(t+u)(5t^4-18t^3u-21t^2u^2-tu^3+u^4)] \NO \\
&+&2Q^2[s^8(t^2-u^2)+s^7(3t^3+6t^2u+5tu^2-4u^3) \NO \\
&+&s^6(-6t^4+25t^3u+16t^2u^2+15tu^3-8u^4) \NO \\
&+&s^5(-21t^5+26t^4u+51t^3u^2+28t^2u^3+28tu^4-10u^5) \NO \\
&+&s^4(-14t^6+11t^5u+70t^4u^2+61t^3u^3+38t^2u^4+28tu^5-8u^6) \NO \\
&+&s^3(2t^7+28t^6u+66t^5u^2+72t^4u^3+61t^3u^4+28t^2u^5+15tu^6-4u^7) \NO \\
&+&s^2(3t^8+30t^7u+76t^6u^2+66t^5u^3+70t^4u^4+51t^3u^5+16t^2u^6+5tu^7-u^8) \NO \\
&+&st^2u(t+u)(6t^5+24t^4u+4t^3u^2+7t^2u^3+19tu^4+6u^5) \NO \\
&+&t^2u^2(t+u)^2(3t^4-4t^3u-9t^2u^2+tu^3+u^4)] \NO \\
&+&(s+u)[2s^8u(t+u)+2s^7u(3t^2+5tu+4u^2) \NO \\
&+&s^6(5t^4+7t^3u+29t^2u^2+27tu^3+16u^4) \NO \\
&+&s^5(15t^5+10t^4u+50t^3u^2+62t^2u^3+39tu^4+20u^5) \NO \\
&+&s^4(15t^6+14t^5u+51t^4u^2+83t^3u^3+70t^2u^4+39tu^5+16u^6) \NO \\
&+&s^3(5t^7+8t^6u+39t^5u^2+68t^4u^3+83t^3u^4+62t^2u^5+27tu^6+8u^7) \NO \\
&+&s^2u(t+u)(t^6+17t^5u+22t^4u^2+29t^3u^3+21t^2u^4+8tu^5+2u^6) \NO \\
&+&stu^2(t+u)^2(t^4+6t^3u+t^2u^2+2tu^3+2u^4)+5t^4u^3(t+u)^3]\}.
\eea

\end{appendix}


\end{document}